\documentclass[aps,pre,reprint,superscriptaddress,longbibliography]{revtex4-2}

\usepackage{amsmath}
\usepackage{amsthm}
\usepackage{graphicx}
\usepackage{fancyhdr, color}
\usepackage{bm}
\usepackage{accents}
\usepackage{amssymb}
\usepackage{url}
\usepackage{setspace}
\usepackage{hyperref}
\usepackage[utf8]{inputenc}
\usepackage[T1]{fontenc}

\newtheorem{prop}{Proposition}

\newcommand{\Tr}{\text{Tr}}

\newcommand{\Var}{\text{Var}}

\newcommand{\revision}[1]{{{#1}}}

\begin{document}


\title{Hierarchy of entropy production and thermodynamic trade-off relations in non-Markovian systems}



\author{Ken Funo}
\affiliation{Department of Applied Physics, The University of Tokyo, 7-3-1 Hongo, Bunkyo-ku, Tokyo 113-8656, Japan}
\email{funo@ap.t.u-tokyo.ac.jp}

\author{Tan Van Vu}
\affiliation{Center for Gravitational Physics and Quantum Information, Yukawa Institute for Theoretical Physics, Kyoto University,
Kitashirakawa Oiwakecho, Sakyo-ku, Kyoto 606-8502, Japan}

\author{Keiji Saito}
\affiliation{Department of Physics, Kyoto University, Kyoto 606-8502, Japan}

\date{\today}

\begin{abstract}
\revision{Non-Markovian dynamics arise when a system is coupled to a bath with finite correlation time, producing memory effects that can modify finite-time thermodynamic behavior. We study this problem using a Markovian embedding of generalized Langevin dynamics generated by a Gaussian bath model. 
We show that the entropy production of the reduced non-Markovian description decomposes into the Markovian entropy production of the enlarged Markovian dynamics and a nonnegative correlation contribution associated with the system-auxiliary modes, establishing a hierarchy of entropy production under Markovian embedding.
The resulting hierarchy yields embedding-independent thermodynamic trade-off relations, including entropic bounds, power-efficiency trade-off relations, thermodynamic uncertainty relations, and thermodynamic speed limits. These results clarify how bath memory modifies finite-time constraints on operational speed, engine performance, and current precision. We further show that the entropy-production hierarchy extends beyond Gaussian baths to more general classical and quantum bath models under appropriate thermodynamic conditions.}
\end{abstract}


\maketitle


{\it Introduction.---} Many open systems of experimental and theoretical interest across biology, chemistry, and physics are coupled to structured environments whose response is not instantaneous~\cite{vanKampen2007StochasticProcesses, klippenstein2021introducing, Memory_and_Friction, das2023enhanced, PhysRevX.14.021032}. 
The resulting non-Markovian dynamics, in which fluctuations and dissipation depend on the system's past evolution, reflects bath memory and information backflow from the environment~\cite{RevModPhys.88.021002,RevModPhys.89.015001}. Within the framework of stochastic thermodynamics~\cite{seifert2012stochastic,ohkuma2007fluctuation,speck2007jarzynski,PhysRevLett.116.020601,PhysRevX.7.011008,RevModPhys.92.041002,kanazawa2025stochastic}, entropy production---which quantifies irreversibility---is always nonnegative, yet memory effects can render its rate transiently negative~\cite{PhysRevE.99.012120,RevModPhys.93.035008}, indicating that bath memory has the potential to reduce irreversibility.
This suggests that memory is a potential physical resource, as explored for channel capacities~\cite{RevModPhys.86.1203}, metrology~\cite{yang2024control}, information erasure~\cite{bylicka2016thermodynamic}, and work extraction~\cite{PhysRevE.106.014114,krishnamurthy2023overcoming, ginot2025energy, muruga2026extractingworkhiddendegrees}. 

\begin{figure*}[t]
    \centering
    \includegraphics[width=\linewidth]{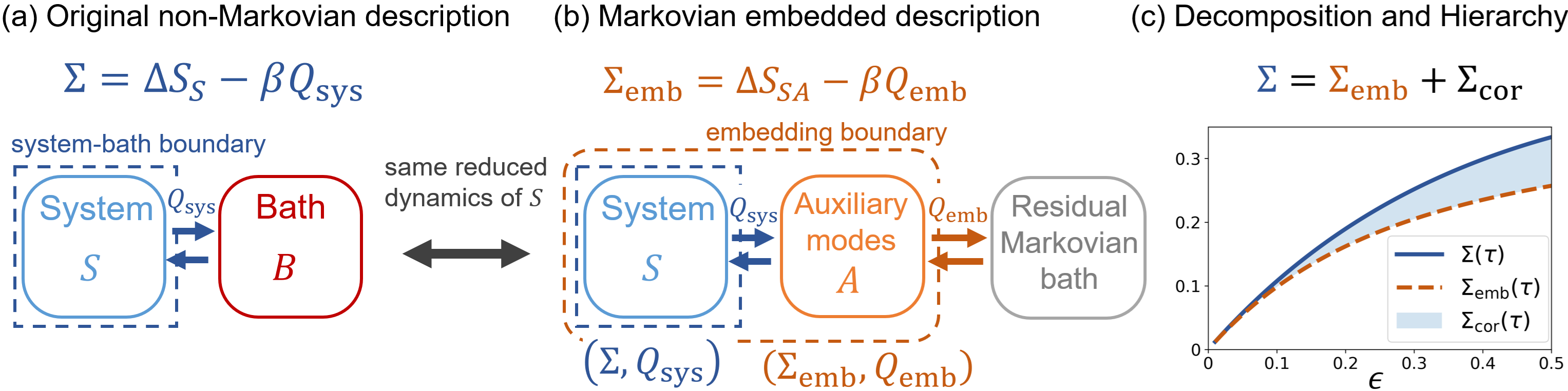}
    \caption{Hierarchy of entropy production under Markovian embedding. (a) Original non-Markovian description~\eqref{eq: GLE} with the entropy production $\Sigma$ and heat $Q_{\rm sys}$ based on system observables. (b) Equivalent Markovian embedded description~\eqref{eq: Markovian-embedded model Gauss}, in which the bath is given by auxiliary modes $A$ and a residual Markovian bath. The inner system-bath boundary recovers $(\Sigma,Q_{\rm sys})$, whereas the outer embedding boundary gives $(\Sigma_{\rm emb},Q_{\rm emb})$. 
    \revision{(c) Entropy productions for the example in Fig.~\ref{fig: Example}, by varying the inverse memory time $\epsilon$.
    Their difference is the nonnegative contribution $\Sigma_{\rm cor}$, implying the hierarchy $\Sigma\geq\Sigma_{\rm emb}$~\eqref{eq: relation Sigma}.
    Here, $\Sigma_{\rm emb}(\tau)$ and $\Sigma(\tau)$ scale as $O(\epsilon)$ (see also Fig.~\ref{fig: Example}).
    }
    }
    \label{fig: model}
\end{figure*}

A powerful framework for analyzing such dynamics is the Markovian embedding technique~\cite{klippenstein2021introducing,Memory_and_Friction,PhysRevResearch.2.033442,10.1021/ct900563s,PhysRevResearch.6.023270,kanazawa2025stochastic,PhysRevA.55.2290,PhysRevLett.120.030402,Nazir2018,strasberg2016nonequilibrium, PhysRevX.10.031040, PhysRevLett.127.250404, PhysRevE.110.014125, kanazawa2025stochastic, Andreas_Kanazawa}, which represents a non-Markovian bath using auxiliary modes that encode memory and a residual Markovian bath that accounts for irreversible dissipation. 
Beyond reproducing the reduced system dynamics, this embedding reveals two distinct thermodynamic system-bath boundaries, one for the original non-Markovian description and the other for the enlarged Markovian system (see Fig.~\ref{fig: model}).
These two boundaries naturally define two entropy productions: the former quantifies irreversibility in the original non-Markovian description, including memory effects, whereas the latter quantifies dissipation into the residual Markovian bath. 
Although entropy production in embedded Markovian systems has been studied in several contexts~\cite{strasberg2016nonequilibrium, PhysRevE.110.014125, kanazawa2025stochastic, Andreas_Kanazawa}, its precise relation to that of the original non-Markovian system has remained largely unexplored. 
Clarifying this relation is essential for disentangling memory-induced contributions from irreversible dissipation.

Entropy production also underlies finite-time thermodynamic trade-off relations, including thermodynamic uncertainty relations (TURs)~\cite{TUR2,TUR1,PhysRevLett.114.158101,PhysRevLett.116.120601}, speed limits~\cite{aurell2012refined, PhysRevLett.121.070601,PhysRevX.13.011013}, entropic bounds~\cite{PhysRevE.97.062101}, and power-efficiency bounds~\cite{PhysRevLett.117.190601, PhysRevLett.120.190602}.
These relations constrain current precision, operational speed, and engine performance, through entropy production and are well established in Markovian systems.
In the non-Markovian regime, however, \revision{bath memory modifies the entropy production structure, and how this affects} such finite-time trade-off relations remains unclear. A precise relation between the non-Markovian entropy production and the embedded Markovian one is therefore needed.

Here we establish a hierarchy between the entropy production of the non-Markovian system and that of the embedded Markovian system, and apply it to derive finite-time thermodynamic trade-off relations in the non-Markovian regime.
\revision{We further obtain a decomposition of the entropy production into a contribution associated with dissipation into the residual Markovian bath and a nonnegative system-auxiliary correlation contribution (see Fig.~\ref{fig: model}). We apply this hierarchy to the Gaussian bath model, and}
derive entropic bounds on currents and power-efficiency trade-off relations for underdamped generalized Langevin systems, and thermodynamic speed limits and TURs in the overdamped regime.  
In the underdamped case, we identify spectral conditions under which the prefactor of the entropic bound diverges, allowing non-vanishing heat exchanges at vanishingly small entropy production, and conditions under which it remains finite, prohibiting the achievement of Carnot efficiency at finite power. In the overdamped case, we show that the precision-to-dissipation ratio can exceed its Markovian value, highlighting the potential usefulness of bath memory for improving thermodynamic performance.

{\it Gaussian bath model.---}
\revision{We focus on a Gaussian bath model, in which the bath consists of harmonic oscillators linearly coupled to the system~\cite{zwanzig1973nonlinear} (see Appendix A).}  
The system Hamiltonian is given by $H_{S}^{\lambda_{t}}=\frac{P^{2}}{2M}+V^{\lambda_{t}}_{S}(X)$, where $\lambda_{t}$ describes the time-dependence of the potential. 
Starting from the system-bath Hamiltonian dynamics, the reduced dynamics of the system is described by the generalized Langevin equation~\cite{zwanzig1973nonlinear,Memory_and_Friction,klippenstein2021introducing}:
\begin{align}
\frac{dX}{dt} &= \frac{P}{M} \nonumber \\
\frac{dP}{dt} &= -\partial_{X}V^{\lambda_{t}}_{S} - \int^{t}_{0}ds K(t-s) \frac{P(s)}{M} + \eta_{t}  , \label{eq: GLE}
\end{align}
where $K(t):= \frac{2}{\pi}\int_{0}^{\infty} d\omega \frac{J(\omega)}{\omega}\cos\omega t$ 
is the memory kernel, $J(\omega)$ is the bath spectral density that reflects the properties of the bath, and $\eta_{t}$ is the colored Gaussian noise satisfying $\mathbb{E}[\eta_{t}]=0$ and the fluctuation-dissipation relation $\mathbb{E}[\eta_{t}\eta_{s}]=K(t-s)/\beta$. Here, $\beta$ is the inverse temperature of the bath. 

The entropy production $\Sigma$ from $t=0$ to $t=\tau$ for the non-Markovian system described by Eq.~\eqref{eq: GLE} is defined as~\cite{ohkuma2007fluctuation,speck2007jarzynski,PhysRevLett.116.020601,PhysRevX.7.011008,RevModPhys.92.041002}
\begin{align}
    \Sigma(\tau) := \Delta S_{S} - \beta Q_{\rm sys}\geq 0, \label{eq: EP nonMarkov}
\end{align}
where $\Delta S_{S}$ is the change in the Shannon entropy of the system and $Q_{\rm sys}:=\int^{\tau}_{0}dt \dot{Q}_{\rm sys}$. Here, the heat current is defined based on the first law of thermodynamics as $\dot{Q}_{\rm sys}:=\frac{d}{dt}\langle H_{S}^{\lambda_{t}}\rangle_{t} - \dot{W}$, where $\dot{W}:=\dot{\lambda}_{t}\langle \partial_{\lambda_{t}}H_{S}^{\lambda_{t}}\rangle_{t}$ is the work rate. 
The entropy production quantifies the total entropy generated over the entire system and serves as a measure of thermodynamic irreversibility. 
Crucially, the entropy production~\eqref{eq: EP nonMarkov} depends only on the probability distribution $f^{S}_{t}$ and the observables of the system.

{\it Markovian embedding.---}
Let us now focus on the Markovian embedding of Eq.~\eqref{eq: GLE} in which the system $S$ interacts with auxiliary modes $A$, such that the joint dynamics of $SA$ is Markovian. 
Specifically, we assume that the time-evolution equation of the joint system $SA$ is given by the Fokker-Planck equation 
\begin{align}
    \partial_{t}f^{SA}_{t}= \{H_{SA}^{\rm tot},f^{SA}_{t}\} + \mathcal{D}f^{SA}_{t}, \label{eq: Markovian-embedded model Gauss}
\end{align}
where $\{A,B\}$ denotes the Poisson bracket, $\mathcal{D}$ describes the dissipative part of the Fokker-Planck operator \revision{acting on $A$, and the instantaneous stationary state of Eq.~\eqref{eq: Markovian-embedded model Gauss} is given by the thermal state $\pi^{SA}_{\lambda_{t}}\propto\exp(-\beta H_{SA}^{\rm tot})$, where the total Hamiltonian reads $H_{SA}^{\rm tot}=H_{S}^{\lambda_{t}}+H_{\rm int}+H_{A}$. 
We further assume the initial distribution to take the form $f^{SA}_{0}=f^{S}_{0}\pi^{A|S}$, where $\pi^{A|S}\propto \exp(-\beta (H_{\rm int}+H_{A}))$ is the conditional thermal state of the auxiliary modes.}

\revision{
Following a standard finite-dimensional approach to Markovian embedding, we consider memory kernels whose one-sided Laplace transform $\widehat K(s):=\int^{\infty}_{0}dt\, e^{-st}K(t)$ 
is a real, strictly proper, and stable rational function, and satisfies $J(\omega)/\omega\geq 0$ for all $\omega\geq 0$. 
In the Supplemental Material~\cite{supplement}, we show that for any memory kernel of the above class, we can construct a Markovian embedding of the form given in Eq.~\eqref{eq: Markovian-embedded model Gauss} using a finite set of quadratic auxiliary modes $A$. 
A commonly used subclass, which includes exponentially decaying and damped-oscillatory memory kernels, is $K(t)=\sum_{k=1}^{M}e^{-\Gamma_k|t|}[a_k\cos(\Omega_k t)+        b_k\sin(\Omega_k|t|)]$, with $\Gamma_k>0$, $\Omega_k\geq0$, and coefficients chosen such
that $J(\omega)/\omega \geq0$. Beyond this finite-dimensional rational class, we show in the Supplemental Material that any continuous real, even, positive-definite kernel with finite $K(0)$ can be approximated uniformly on any prescribed finite time interval by finite sums of damped oscillatory kernels.
}



{\it Hierarchy of entropy production.---}
For the Markovian embedded model~\eqref{eq: Markovian-embedded model Gauss}, the standard definitions of thermodynamic quantities, such as work, heat, and entropy production, apply to the {\it joint} system $SA$. In particular, the entropy production rate for the joint system is defined as
\begin{align}
    \dot{\Sigma}_{\rm emb} := \dot{S}_{SA} - \beta \dot{Q}_{\rm emb} \geq 0,
\end{align}
where $S_{SA}$ is the Shannon entropy of the joint system and $\dot{Q}_{\rm emb} := \frac{d}{dt}\langle H_{SA}^{\rm tot} \rangle -\dot{W}$ is the heat flux from the residual Markovian bath. 
We emphasize that the entropy production for the embedded Markovian system $\Sigma_{\rm emb}(\tau):=\int^{\tau}_{0}dt \dot{\Sigma}_{\rm emb}$ and the non-Markovian entropy production $\Sigma$ in Eq.~\eqref{eq: EP nonMarkov} are defined based on different system-bath boundaries, and as a result, the two definitions are not identical (see Fig.~\ref{fig: model}). Furthermore, while $\Sigma_{\rm emb}$ depends on specific choices of Markovian embedding, $\Sigma$ is invariant under different choices of embedding. 
A natural question is therefore how these two entropy productions are related to each other.

Our first main result is the hierarchy of entropy production under Markovian embedding, i.e., the non-Markovian entropy production $\Sigma$ is {\it larger} than or equal to that of the Markovian embedded model $\Sigma_{\rm emb}$:
\begin{align}
    \Sigma \geq \Sigma_{\rm emb}. \label{eq: relation Sigma}
\end{align}
\revision{We refer to this ordering as a hierarchy because it establishes an ordering of the entropy productions associated with the nested thermodynamic boundaries introduced by the embedding (see Fig.~\ref{fig: model}).}
This hierarchy is obtained from the decomposition of $\Sigma$ into two nonnegative contributions as
\begin{align}
        \Sigma &= \Sigma_{\rm emb} + \Sigma_{\rm cor},\  
    \Sigma_{\rm cor}(\tau):=D(f^{SA}_{\tau} \Vert  f^{S}_{\tau}  \pi^{A|S})\geq 0, \label{eq: Relation between EP cond eq}
\end{align}
under the initial condition $f^{SA}_{0}=f^{S}_{0}\pi^{A|S}$. 
Here, $D(f\Vert g)$ is the relative entropy, and equality of~\eqref{eq: relation Sigma} is achieved if and only if $f^{SA}_{\tau}=f^{S}_{\tau}\pi^{A|S}$.
\revision{Here, Eqs.~\eqref{eq: relation Sigma} and~\eqref{eq: Relation between EP cond eq} are derived by assuming a Gaussian bath model. However, we can generalize them to a generic bath model in both the classical and quantum regimes, as discussed in Appendix~B. 
}

\revision{Although $\Sigma_{\rm emb}$ and $\Sigma_{\rm cor}$ explicitly depend on the choice of embedding, the decomposition~\eqref{eq: Relation between EP cond eq} gives important physical insights into the entropy production.
Recalling that the Markovian embedding technique effectively separates the bath into the residual Markovian bath and the auxiliary modes, Eq.~\eqref{eq: Relation between EP cond eq} yields a corresponding decomposition of the entropy production: $\Sigma_{\rm emb}$ describes dissipation into the residual Markovian bath, while $\Sigma_{\rm cor}$ captures system-auxiliary correlations and deviations of the auxiliary modes from conditional equilibrium.   
This decomposition also shows that whenever $\dot{\Sigma}(t)$ becomes temporarily negative, the negative entropy production rate must originate from a decrease in $\Sigma_{\rm cor}(t)$, as $\dot{\Sigma}_{\rm emb}(t)\geq 0$ is always satisfied.} 

\revision{
To further support the above intuition, we derive an embedding-independent lower bound on $\Sigma_{\rm cor}$ for the Gaussian bath model, which reads
\begin{align}
    \Sigma_{\rm cor}\geq \Sigma_{F^{\rm NM}}:=\frac{\beta}{2K(0)}\langle (F_{\tau}^{\rm NM})^{2}\rangle_{\tau}, \label{eq: Sigma mem lower bound}
\end{align}
where $F_{t}^{\rm NM}(X,P):=\mathbb{E}[\mathcal{F}_{t}^{\rm GLE}|X(t)=X,P(t)=P]$ with $\mathcal{F}_{t}^{\rm GLE}:=-\int^{t}_{0}dsK(t-s)\frac{P(s)}{M}+\eta_{t}$ is the bath force that includes bath memory effects and enters the reduced evolution of the system as $\partial_{t}f_{t}^{S}=\{ H_{S}^{\lambda_{t}},f_{t}^{S}\}-\partial_{P}(F_{t}^{\rm NM}f_{t}^{S})$. 
Therefore, although $\Sigma_{\rm cor}$ itself depends on the choice of embedding, Eq.~\eqref{eq: Sigma mem lower bound} provides an embedding-independent lower bound based on the bath force. We further note that in the Markovian limit $K(t)\propto \delta(t)$, the lower bound $\Sigma_{F^{\rm NM}}$ vanishes. 
}

From the first main result~\eqref{eq: relation Sigma}, $\Sigma$ is bounded from below by the embedded Markovian entropy production $\Sigma_{\rm emb}$. This allows us to apply techniques from stochastic thermodynamics for Markovian systems to derive thermodynamic trade-off relations in the non-Markovian regime based on $\Sigma$, which makes the results independent of the choice of embedding. In what follows, we derive the entropic bound~\eqref{eq: time-dep TUR}, power-efficiency trade-off relation~\eqref{eq: power-efficiency trade-off}, and TUR~\eqref{eq: nonMarkov TUR Koyuk Seifert}. 
\revision{We also derive thermodynamic speed limits in Appendix C. See Supplemental Material~\cite{supplement} for the derivation of these results.} 

{\it Entropic bound.---} Let us introduce the bath-induced current $\mathcal{J}_{\mathcal{O}}(t)$, which is part of the change in the observable $\mathcal{O}_{t}(X,P)$ induced by the bath. 
As a special case, by choosing $\mathcal{O}_{t}=H^{\lambda_{t}}_{S}$, the bath-induced current reduces to the heat current: $\mathcal{J}_{\mathcal{O}}=\dot{Q}_{\rm sys}$.
Our second main result is a non-Markovian entropic bound:\revision{ 
    \begin{align}
    \frac{1}{\Theta \Sigma}\Bigl( \int^{\tau}_{0}dt |\mathcal{J}_{\mathcal{O}}(t)| \Bigr)^{2} &\leq 1-\frac{\Sigma_{F^{\rm NM}}}{\Sigma} \leq 1   ,\label{eq: time-dep TUR} 
\end{align}
}where 
\begin{align}
    \Theta := \frac{\mathcal{S}_{K}}{\beta}\int^{\tau}_{0}dt \int dXdP (\partial_{P}\mathcal{O}_{t})^{2} f_{t}^{S}(X,P), \label{eq: bound theta}
\end{align}
and the coefficient $\mathcal{S}_{K}$ is defined as~\cite{footnote_EP}\revision{
\begin{align}
    \mathcal{S}_{K} &:=     \sup_{\omega\geq 0} \frac{|\widehat{K}(i\omega)|^{2}}{\mathrm{Re}\widehat{K}(i\omega)} = \sup_{\omega\geq 0}\frac{\frac{J(\omega)^{2}}{\omega^{2}}+[\int^{\infty}_{0}dt K(t)\sin\omega t]^{2}}{\frac{J(\omega)}{\omega}}. 
\label{eq: bound S} 
\end{align}
We note that $\mathcal{S}_{K}$ is bounded from below as $\mathcal{S}_{K}\geq \int^{\infty}_{0}dt K(t)$, and this equality holds for the overdamped Drude-Lorentz memory kernel $K^{\rm od}(t)$ introduced later.

This inequality~\eqref{eq: time-dep TUR} shows a non-Markovian thermodynamic trade-off relation between the bath-induced current and the entropy production for generalized Langevin dynamics. 
The bound is controlled by $\Theta$, which is determined by the bath-side property $\mathcal{S}_{K}/\beta$ and the system-side observable $\int^{\tau}_{0}dt\langle (\partial_{P}\mathcal{O}_{t})^{2}\rangle $, and is independent of the choice of embedding. 
Therefore, the dependence of $\mathcal{S}_{K}$ on the memory kernel allows us to characterize how modifying the bath spectrum changes the current-dissipation trade-off in Eq.~\eqref{eq: time-dep TUR}, as follows.}

\begin{figure}
    \centering
    \includegraphics[width=0.99\linewidth]{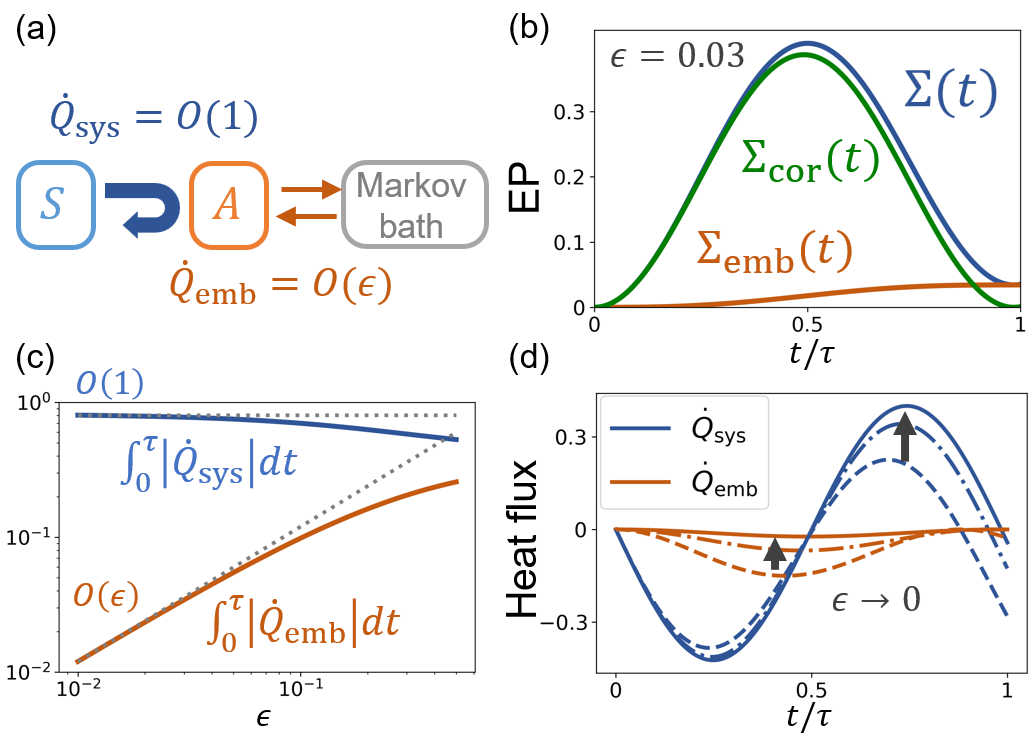}
    \caption{Memory-assisted heat exchange with suppressed dissipation. (a) Schematic of the model. \revision{(b) Plot of $\Sigma(t)$ (blue curve), $\Sigma_{\rm cor}(t)$ (green curve), and $\Sigma_{\rm emb}(t)$ (orange curve) for $\epsilon=0.03$.} (c) Numerical plot showing the scaling of $\int^{\tau}_{0}dt|\dot{Q}_{\rm sys}|=O(1)$ (blue curve) and $\int^{\tau}_{0}dt|\dot{Q}_{\rm emb}|=O(\epsilon)$ (orange curve).  (d) Plot of $\dot{Q}_{\rm sys}(t)$ (blue curve) and $\dot{Q}_{\rm emb}(t)$ (orange curve), for \revision{$\epsilon=0.3,\,0.1,\,0.03$} (dashed, dashdot, and solid curves, respectively).
    A near-reversible exchange of heat between the system and auxiliary modes is realized in the limit of $\epsilon\rightarrow 0$, while dissipation is strongly suppressed, as $\int^{\tau}_{0}dt|\dot{Q}_{\rm emb}|$ and $\Sigma(\tau)$ scale as $O(\epsilon)$ (see also Fig.~\ref{fig: model}(c)).}
    \label{fig: Example}
\end{figure}

{\it Case 1: finite $\Theta$.} It follows from Eq.~\eqref{eq: time-dep TUR} that whenever the prefactor $\Theta$ remains finite as $\Sigma\rightarrow 0$,  the time-integrated absolute value of the current $\int^{\tau}_{0}dt |\mathcal{J}_{\mathcal{O}}(t)|$ also vanishes. Note that this property is the key to showing the unattainability of Carnot efficiency with finite power in heat engines, as we discuss later. 
\revision{In the Supplemental Material~\cite{supplement}, we show that $\mathcal{S}_{K}<\infty$ if $0<J(\omega)/\omega<\infty$ for every finite $\omega\geq 0$ and $0<\lim_{\omega\rightarrow\infty} \omega J(\omega)<\infty$. 

{\it Case 2: divergent $\Theta$.} Conversely, when $\Theta$ diverges, the inequality no longer forbids finite currents at vanishing entropy production~\cite{Tajima21}. 
As a concrete example, we take $K_{\epsilon}(t)=ae^{-\epsilon|t|},a>0$, where $\epsilon>0$ is the inverse memory time of the kernel. Equation~\eqref{eq: bound S} gives $\mathcal{S}_{K_{\epsilon}}=a/\epsilon$, which diverges as $\epsilon\rightarrow 0$. Therefore, the scaling of $\Sigma(\tau)=O(\epsilon)$ and $\int^{\tau}_{0}dt|\dot{Q}_{\rm sys}|=O(1)$ is allowed by the trade-off relation~\eqref{eq: time-dep TUR}, which is numerically confirmed in Fig.~\ref{fig: model}(c) and Fig.~\ref{fig: Example}.
Below, we discuss its underlying mechanism in terms of the embedding using a single overdamped harmonic auxiliary coordinate whose relaxation rate is $\epsilon$. 
In the $\epsilon\rightarrow 0$ limit, the memory time diverges, and the auxiliary mode can store energy and subsequently return it to the system, while suppressing dissipation into the residual bath as $\int^{\tau}_{0}dt|\dot{Q}_{\rm emb}|=O(\epsilon)$ and
$\Sigma_{\rm emb}=O(\epsilon)$.
This enables the oscillatory heat exchange with negligible dissipation shown in Fig.~\ref{fig: model}(c) and Fig.~\ref{fig: Example}, highlighting a unique feature of memory effects.    

{\it Case 3: Markovian limit.} Let us consider the memory kernel $K(t)=ae^{-\Gamma|t|}$ and take the Markovian limit: $\Gamma,a\rightarrow \infty$ with $a/\Gamma\rightarrow M\gamma$. Then, we have $K(t)\rightarrow 2M\gamma\delta(t)$, $\Sigma_{F^{\rm NM}}\rightarrow 0$, and $\mathcal{S}_{K}\rightarrow M\gamma$.  Therefore, Eq.~\eqref{eq: time-dep TUR} reproduces the results obtained in Refs.~\cite{PhysRevLett.117.190601, PhysRevE.97.062101} in the Markovian limit. 
}

{\it Power-efficiency trade-off relation.---}
Let us apply the entropic bound to a heat engine setup in which the system is in contact with hot and cold baths with inverse temperatures $\beta_{H}$ and $\beta_{C}$, respectively. 
We now present a trade-off relation between the power $\mathcal{P}$ and efficiency $\eta$ of non-Markovian heat engines:
    \begin{align}
    \mathcal{P} \leq 
    \beta_{C} \bar{\Theta}_{\rm m} \eta (\eta_{\rm Car}-\eta), \label{eq: power-efficiency trade-off}
\end{align}
where $\eta_{\rm Car}:=1-\beta_{H}/\beta_{C}$ is the Carnot efficiency and $\bar{\Theta}_{\rm m}:=\Theta_{\rm m}/\tau$, with $\Theta_{\rm m}$ the multi-bath generalization of Eq.~\eqref{eq: bound theta} for $\mathcal{O}_{t}=H_{S}^{\lambda_{t}}$ and $\tau$ the duration of one engine cycle (see Supplemental Material~\cite{supplement} for details). This quantity $\bar{\Theta}_{\rm m}$ captures the non-Markovian bath property and controls the prefactor of the bound.  
When $\bar{\Theta}_{\rm m}$ remains finite as $\eta\rightarrow \eta_{\rm Car}$, the bound~\eqref{eq: power-efficiency trade-off} implies $\mathcal{P}\rightarrow 0$, i.e., the Carnot efficiency at finite power cannot be achieved.

{\it Overdamped systems and TUR.---}
We next consider the overdamped limit, in which thermodynamic speed limits and TURs are most naturally formulated. The reduced dynamics of the system is described by the overdamped generalized Langevin equation~\cite{nguyen2018small}
\begin{align}
    \frac{1}{\mu}\frac{dX}{dt}
    =\frac{\xi_{t}}{\mu}
    -\partial_{X}V_{S}^{\lambda_{t}}
    -\int_{0}^{t}dsK^{\rm od}(t-s)\dot{X}(s)
    +\eta_{t}^{\rm od}.
    \label{eq: overdamped GLE}
\end{align}
Here, \revision{$K^{\rm od}(t):=\sum_{k}a_{k} e^{-\Gamma_{k}|t|}$ with $\Gamma_{k}>0,\, a_{k}\geq 0$} is the memory kernel corresponding to the overdamped Drude-Lorentz spectral density, and $\eta_{t}^{\rm od}$ is the associated colored noise. 
The additional white noise $\xi_t$ satisfies $\mathbb{E}[\xi_t\xi_s]=(2\mu/\beta)\delta(t-s)$ and represents the direct Markovian bath contribution required to ensure relaxation to the correct thermal equilibrium. 
In Appendix C, we extend the hierarchy relation~\eqref{eq: relation Sigma} to Eq.~\eqref{eq: hierarchy Sigma vs Markov2} in this overdamped regime.

\begin{figure}[t]
    \centering
    \includegraphics[width=0.7\linewidth]{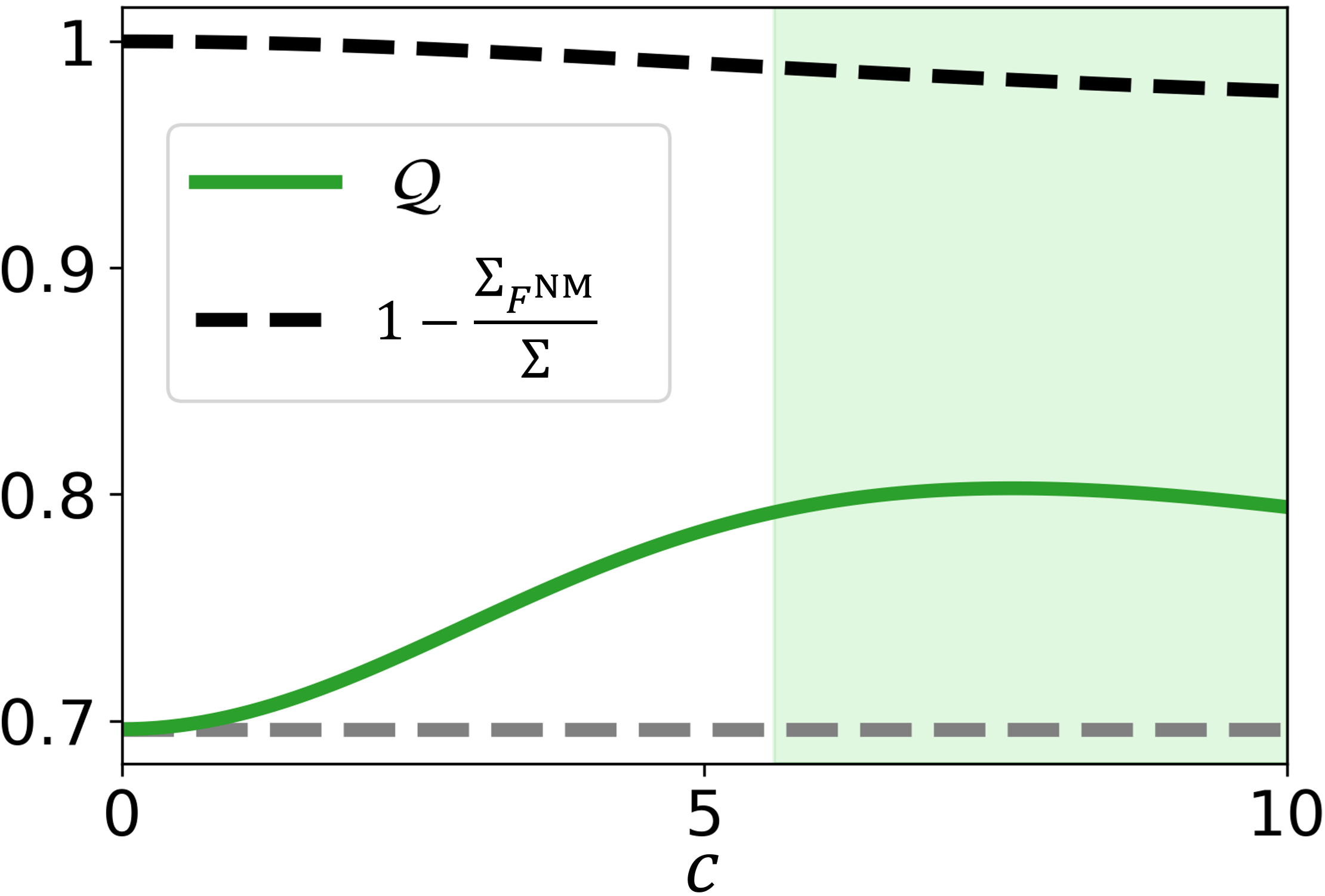}
    \caption{Effect of memory on the transient TUR~\eqref{eq: nonMarkov TUR Koyuk Seifert}. Precision-to-dissipation ratio $\mathcal{Q}$ (green curve) and the bound $1-\Sigma_{F^{\rm NM}}/\Sigma$ (black dashed curve) as a function of the coupling $c$ to the non-Markovian bath. The ratio can be enhanced relative to the corresponding value in the Markovian limit $c=0$ (gray dashed line). The green shaded region marks the range of $c$ for which memory effects suppress both $\Sigma$ and $\Var[\mathcal{J}_{\tau}^{S}]$ relative to the Markovian limit $c=0$ (see also Appendix C and Fig.~\ref{fig: TUR2}).}
    \label{fig: TUR}
\end{figure}

We now derive a transient TUR for non-Markovian overdamped dynamics. \revision{We consider the protocol of the form $\lambda_{t}^{v}:=\lambda_{vt}$, where $v$ is the speed parameter, and a time-averaged current $\mathcal{J}_{\tau}^{S}=\tau^{-1}\int^{\tau}_{0}dt \psi[X(t),\lambda_{t}^{v}]\circ \dot{X}(t)$, where $\psi$ is an arbitrary function and $\circ$ is the Stratonovich product.} Applying the TUR~\cite{PhysRevLett.125.260604} to the embedded dynamics and then using the decomposition~\eqref{eq: Relation between EP cond eq}, we obtain\revision{
\begin{align}
     \mathcal{Q}:=
     \frac{2\bigl[\langle \mathcal{J}^{S}_{\tau}\rangle +\Delta \langle \mathcal{J}^{S}_{\tau} \rangle\bigr]^{2}}{ \Sigma  \Var[\mathcal{J}^{S}_{\tau}]}
     \leq 1-\frac{\Sigma_{F^{\rm NM}}}{\Sigma} \leq 1 ,
     \label{eq: nonMarkov TUR Koyuk Seifert}
\end{align}
where $\Delta:= \tau\partial_{\tau}-v\partial_{v}$ and $\mathcal{Q}$ characterizes the precision-to-dissipation ratio. The bound~\eqref{eq: nonMarkov TUR Koyuk Seifert} shows that higher entropy production is required to increase the precision of the current. Importantly, all quantities that appear in Eq.~\eqref{eq: nonMarkov TUR Koyuk Seifert} are embedding-independent and can be evaluated from the reduced dynamics~\eqref{eq: overdamped GLE}.
In particular, the weaker bound $\mathcal Q\leq 1$ allows one to place
a lower bound on the entropy production using reduced-system current
statistics and their response to the observation time and driving
speed, without reconstructing a specific Markovian embedding.
Figure~\ref{fig: TUR} plots $\mathcal{Q}$ as a function of the coupling $c$ to the non-Markovian bath (green solid curve). By incorporating $\Sigma_{F^{\rm NM}}$, we obtain a tighter upper bound on $\mathcal{Q}$ (black dashed curve). We also find that $\mathcal{Q}$ exceeds its Markovian value at $c=0$, demonstrating that bath memory can improve the precision-to-dissipation ratio (see also Appendix C). 
}

{\it Conclusion.---} We have established a hierarchy of entropy production under Markovian embedding~\eqref{eq: relation Sigma}, decomposing the non-Markovian entropy production into the \revision{Markovian entropy production of the enlarged system and a nonnegative correlation contribution}~\eqref{eq: Relation between EP cond eq}. We further extend this structure to generic bath models and the quantum regime~\eqref{eq: EP decomp quantum}. 
Applying this hierarchy to generalized Langevin systems, we derive entropic bounds~\eqref{eq: time-dep TUR} and power-efficiency trade-off relations~\eqref{eq: power-efficiency trade-off}, and identify spectral conditions under which Carnot efficiency at finite power is unattainable, and conditions under which finite heat currents with vanishing entropy production are allowed. 
In the overdamped regime, we derive thermodynamic speed limits~\eqref{eq: nonMarkov TSL} and TURs~\eqref{eq: nonMarkov TUR Koyuk Seifert} for non-Markovian systems and show that bath memory \revision{can enhance the precision-to-dissipation ratio.

An important consequence of the hierarchy relation is that, although the Markovian embedding is used in the derivation, the resulting thermodynamic trade-off relations can be formulated in terms of embedding-independent quantities.
This provides a thermodynamically consistent framework for extending finite-time thermodynamic constraints to reduced non-Markovian dynamics.

We note that the derivation of thermodynamic trade-off relations presented in this paper assumes a Gaussian bath model, which is a specific but standard class of thermodynamically consistent descriptions of non-Markovian dynamics. 
Extending these trade-off relations beyond the Gaussian-bath setting to broader classes of environments, including those arising from more general coarse-graining procedures~\cite{vroylandt2022derivation,10.1063/5.0345751}, is an important direction for future work.
}


\begin{acknowledgments} 
We are grateful to Takahiro Sagawa, Andreas Dechant, Kay Brandner, Paul Menczel, and Graeme Pleasance for valuable discussions.
We acknowledge that the divergence of $\Theta$ was also discussed with Hiroyasu Tajima and Kiyoshi Kanazawa, and the discussion was based on their proposal (JSPS KAKENHI Grant No.~JP25K24774).
We also acknowledge the use of ChatGPT 5.6 and 6 for assistance with manuscript preparation, the improvement of earlier results, and the construction of approximation schemes for general memory kernels.

K.F. is supported by JST ERATO Grant No.~JPMJER2302, Japan, and JSPS KAKENHI Grant Nos.~JP23K13036, JP24H00831 and JP26H02012.
T.V.V. is supported by JSPS KAKENHI Grant Nos.~JP23K13032, JP26K00019, JP26K00022, and JP26H02015.
K.S. is supported by JSPS KAKENHI Grant Nos.~JP23K25796, JP26H02015, and JP26H00388.
\end{acknowledgments}

%


\section*{End Matter}

\revision{
{\it Appendix A: Gaussian bath model.---}
The Hamiltonian of the Gaussian bath model (Zwanzig model) including the interaction is given by 
\begin{align}
    H_{\rm int}+H_{B} = \sum_{k}\Bigl[ \frac{p_{k}^{2}}{2m_{k}}+\frac{m_{k}\omega_{k}^{2}}{2} \Bigl( x_{k}-\frac{c_{k}X}{m_{k}\omega_{k}^{2}}\Bigr)^{2} \Bigl]. \label{eq: Zwanzig Htot}
\end{align}
The reduced system dynamics is exactly given by Eq.~\eqref{eq: GLE} with the spectral density $J(\omega)=\pi\sum_{k}(c_{k}^{2}/2m_{k}\omega_{k})\delta(\omega-\omega_{k})$, under the assumption of the initial conditional equilibrium of the bath: $f^{SB}_{0}=f^{S}_{0}\pi^{B|S}$ with $\pi^{B|S}\propto e^{-\beta(H_{\rm int}+H_{B})}$. 
Under this setting, the entropy production is defined as~\cite{PhysRevE.99.012120}
\begin{align}
    \Sigma(\tau) = D(f_{\tau}^{SB} || f_{\tau}^{S}\pi^{B|S}) \geq 0, \label{eq: EP Zwanzig}
\end{align}
 which is equivalent to the expression given in Eq.~\eqref{eq: EP nonMarkov}. This can be checked by using the expression $D(f_{\tau}^{SB} || f_{\tau}^{S}\pi^{B|S})=\Delta S_{S}+\beta \int^{\tau}_{0}dt \int ( H_{\rm int}+H_{B})\partial_{t}f_{t}^{SB}$. 
 Further using the Liouville equation $\partial_{t}f_{t}^{SB}=\{H_{S}^{\lambda_{t}}\!+\!H_{\rm int}+H_{B},f_{t}^{SB}\}$ and $\int H_{S}^{\lambda_{t}}\partial_{t}f_{t}^{SB}=\dot{Q}_{\rm sys}$ shows the equivalence between Eqs.~\eqref{eq: EP nonMarkov} and~\eqref{eq: EP Zwanzig}.}

{\it Appendix B: Generalization of the hierarchy.---} 
The trade-off relations above were derived for generalized Langevin systems. However, the hierarchy relation itself extends to generic bath models in both the classical and quantum regimes.  
In the quantum regime, non-Markovian open-system modeling is often essential in solid-state quantum devices~\cite{odeh2025non}, complex molecular systems~\cite{lambert2013quantum, tanimura2020numerically}, and photonic systems~\cite{liu2011experimental}, where coherence and entanglement between the system and bath can play a crucial role~\cite{PhysRevA.104.052408, PRXQuantum.4.020353}.

We use the framework based on the Hamiltonian of mean force~\cite{PhysRevLett.116.020601} together with a Markovian embedding based on the Gorini-Kossakowski-Sudarshan-Lindblad (GKSL) master equation~\cite{Breuer} 
\begin{align}
    \partial_{t}\rho^{SA}_{t}=-\frac{i}{\hbar}[H_{SA}^{\rm tot},\rho^{SA}_{t}] + \mathcal{D}_{t}[\rho^{SA}_{t}]. \label{appendix: GKLS}
\end{align} 
Here, $H_{SA}^{\rm tot}=H_{S}^{\lambda_{t}}+H_{\rm int}^{g_{t}}+H_{A}$, where $g_{t}$ is a time-dependent coupling, and we assume that the instantaneous stationary state of Eq.~\eqref{appendix: GKLS} is given by $\pi^{SA}_{\lambda_{t},g_{t}}\propto\exp(-\beta H_{SA}^{\rm tot})$. 
\revision{
We note that derivation of the hierarchy relation~\eqref{eq: EP decomp quantum} itself does not require the explicit closed form of the reduced non-Markovian equation. We therefore define the non-Markovian dynamics of the system as the reduced dynamics $\rho_{t}^{S}=\Tr_{A}[\rho_{t}^{SA}]$ obtained from Eq.~\eqref{appendix: GKLS}. 
For concreteness, we refer to the reaction-coordinate mapping method~\cite{strasberg2016nonequilibrium, Nazir2018}, in which the Caldeira-Leggett model is rewritten by promoting a collective degree of freedom of the original bath to an auxiliary reaction coordinate $A$, while the remaining bath degrees of freedom constitute a residual bath. When this residual bath can be treated within an appropriate Markovian approximation, the enlarged system $SA$ is described by a GKSL master equation of the form~\eqref{appendix: GKLS}. 

For a given thermodynamically consistent Markovian dynamics of the joint system $SA$ mentioned above, we show that the entropy productions associated with the reduced dynamics and enlarged dynamics satisfy  
\begin{align}
    \Sigma
    =
    \Sigma_{\rm emb}
    +\mathcal{C}_{\tau}
    -\mathcal{C}_{0},
    \label{eq: EP decomp general}
\end{align}
where
\begin{align}
    \mathcal{C}_{t}
    &:=
    D(\rho^{SA}_{t}\Vert\pi^{SA}_{\lambda_t,g_t})
    -    D(\rho^{S}_{t}\Vert\pi^{S}_{\lambda_t,g_t})
    \geq0,
\end{align}
and $\pi^{S}_{\lambda_{t},g_{t}}:=\Tr_{A}[\pi^{SA}_{\lambda_{t},g_{t}}]$. For initial states satisfying $\mathcal{C}_{0}=0$, this relation reduces to
\begin{align}
    \Sigma &= \Sigma_{\rm emb}+\Sigma_{\rm cor} ,\quad     \Sigma_{\rm cor}:= \mathcal{C}_{\tau}\geq 0, \label{eq: EP decomp quantum}
\end{align}
which shows the hierarchy relation in the quantum regime.

The condition $\mathcal{C}_{0}=0$ is the equality condition for the monotonicity of the relative entropy under partial trace and can equivalently be characterized by the Petz recovery map~\cite{10.1093/qmath/39.1.97}. This initial condition includes, for example, a product state $\rho^{SA}_{0}=\rho^{S}_{0}\otimes \frac{e^{-\beta H_A}}{Z_A}$ with the interaction initially switched off and the thermal state of the joint system $\rho^{SA}_{0}=\pi^{SA}_{\lambda_{0},g_{0}}$.
In the classical regime, it reduces to the conditional thermal initial state $f^{SA}_{0}=f^{S}_{0}\pi^{A|S}_{g_{0}}$. 
Other initial-state conventions, including a product state with nonvanishing interaction, are discussed in the Supplemental Material~\cite{supplement}.
}

{\it Derivation of the hierarchy.---}
Here, we derive Eqs.~\eqref{eq: EP decomp general} and~\eqref{eq: EP decomp quantum}. Note that Eq.~\eqref{eq: Relation between EP cond eq} follows as a special case of this derivation.   
In the strong coupling regime, the internal energy $\langle\mathcal{E}_{S}\rangle$ and the entropy rate $\dot{S}_{S}^{*}$ of the system are modified by the interaction as $ \langle \mathcal{E}_{S}\rangle := \langle \partial_{\beta}(\beta H^{*}_{\lambda_{t},g_{t}})\rangle $ and $ \dot{S}_{S}^{*}:= \dot{S}_{S} + \beta^{2}\frac{d}{dt} \langle \partial_{\beta}H^{*}_{\lambda_{t},g_{t}}\rangle$~\cite{PhysRevLett.116.020601,PhysRevX.7.011008,RevModPhys.92.041002}. Here, $S_{S}$ is the von Neumann entropy and $H^{*}_{\lambda_{t},g_{t}} := -\beta^{-1} \ln ( \Tr_{A}[ e^{-\beta H_{SA}^{\rm tot}}/Z_{A}])$ 
is the Hamiltonian of mean force with $Z_{A}:=\Tr_{A}[e^{-\beta H_{A}}]$. 
The heat current is then defined via the first law of thermodynamics as $    \dot{Q}_{\rm sys}:=\frac{d}{dt}\langle \mathcal{E}_{S} \rangle -\dot{W}$, where the work rate is defined as $\dot{W} := \langle \dot{H}_{SA}^{\rm tot}\rangle$. The entropy production is defined as $\Sigma:= \int^{\tau}_{0}dt (\dot{S}^{*}_{S} - \beta \dot{Q}_{\rm sys})$. On the other hand, for embedded dynamics~\eqref{appendix: GKLS}, the heat flux and the entropy production rate are defined as $\dot{Q}_{\rm emb}:=\frac{d}{dt}\langle H_{SA}^{\rm tot}\rangle-\dot{W}$ and $\dot{\Sigma}_{\rm emb}:=\dot{S}_{SA}-\beta \dot{Q}_{\rm emb}$. 
Using the relation 
\begin{align}
\dot{Q}_{\rm sys}=\dot{Q}_{\rm emb}+\frac{1}{\beta}\frac{d}{dt}\langle \ln \pi^{SA}_{\lambda_{t},g_{t}}-\ln\pi^{S}_{\lambda_{t},g_{t}} +\beta^{2}\partial_{\beta}H^{*}_{\lambda_{t},g_{t}}\rangle,    
\end{align}
we obtain Eq.~\eqref{eq: EP decomp general}, which reduces to Eq.~\eqref{eq: EP decomp quantum} for $\mathcal{C}_{0}=0$.


The corresponding classical result follows by replacing density
operators with phase-space distributions.
For the Gaussian bath model~\eqref{eq: Zwanzig Htot}, the Hamiltonian of mean force reduces to the system Hamiltonian $H_{S}^{\lambda_{t}}$, and Eq.~\eqref{eq: EP decomp quantum} reduces to Eq.~\eqref{eq: Relation between EP cond eq}. 

\revision{
{\it Appendix C: Overdamped hierarchy and thermodynamic trade-off relations.---}
We further discuss the hierarchy and thermodynamic trade-off relations for the overdamped generalized Langevin dynamics~\eqref{eq: overdamped GLE}. We first introduce the conditional non-Markovian force $F_t^{\rm NM}(X):=
\mathbb{E}[-\int_0^t ds\,K^{\rm od}(t-s)\dot X(s)+\eta_t^{\rm od}
\,|\,X(t)=X]$. Using this definition, Eq.~\eqref{eq: Sigma mem lower bound} also holds for the overdamped dynamics with the memory kernel $K^{\rm od}$.} The reduced probability distribution obeys 
\begin{align}
    \partial_{t} f_{t}^{S}(X)=-\partial_{X} \Bigl[\Bigl(\nu_{t}^{\rm M}(X)+\mu F_{t}^{\rm NM}(X)\Bigr)f_{t}^{S}(X)\Bigr], \label{eq: continuity}
\end{align}
where $\nu_{t}^{\rm M}:=-\mu\partial_{X}V_{S}^{\lambda_{t}}-\frac{\mu}{\beta}\partial_{X}\ln f_{t}^{S}$ is the conventional Markovian mean local velocity. The entropy production can be decomposed as $\Sigma=\Sigma_{\rm M}+\Sigma_{\rm NM}$, where $\Sigma_{\rm M}:=\frac{\beta}{\mu} \int^{\tau}_{0}dt \langle (\nu_{t}^{\rm M})^{2}\rangle $ is the apparent Markovian entropy production, and $\Sigma_{\rm NM}:=\beta \int^{\tau}_{0}dt \langle \nu_{t}^{\rm M}F_{t}^{\rm NM}\rangle$ is a {\it negative} non-Markovian correction term. As shown in the Supplemental Material~\cite{supplement}, these quantities satisfy the extended hierarchy
\begin{align}
    \Sigma_{\rm emb}\leq 
    \Sigma
    =
    \Sigma_{\rm M}+\Sigma_{\rm NM}
    \leq
    \Sigma_{\rm M},
    \quad
    \Sigma_{\rm NM}\leq 0.
    \label{eq: hierarchy Sigma vs Markov2}
\end{align}
The inequality $\Sigma_{\rm NM}\leq0$ shows that the time-integrated correlation between the Markovian mean local velocity and the conditional non-Markovian force is nonpositive, thereby reducing $\Sigma$ below the apparent Markovian entropy production $\Sigma_{\rm M}$.

\revision{
{\it Thermodynamic speed limit.---}
Combining the contiuity equation~\eqref{eq: continuity} with the Benamou-Brenier formula~\cite{benamou2000computational} and the hierarchy relation~\eqref{eq: hierarchy Sigma vs Markov2}, together with the lower bound~\eqref{eq: Sigma mem lower bound} (see Supplemental Material~\cite{supplement}), we obtain the non-Markovian thermodynamic speed limit
\begin{align}
\frac{\beta}{\mu\Sigma}\frac{\mathcal{W}_{2}(f^{S}_{0},f^{S}_{\tau})^{2}}{\tau}  \leq 1-\frac{\Sigma_{F^{\rm NM}}}{\Sigma}\leq 1 , 
\label{eq: nonMarkov TSL}
\end{align}
where $\mathcal{W}_{2}(f,g)$ denotes the 2-Wasserstein distance~\cite{villani2009optimal}. This bound shows that the rate at which the probability distribution can evolve is fundamentally constrained by the entropy production. 
All quantities appearing in Eq.~\eqref{eq: nonMarkov TSL} are embedding-independent and can be evaluated directly from the reduced non-Markovian dynamics. 
Although the right-most bound of Eq.~\eqref{eq: nonMarkov TSL} takes exactly the same form as that of the Markovian case, incorporating $\Sigma_{F^{\rm NM}}$ tightens the bound.  
}

\begin{figure}[!t]
    \centering
    \includegraphics[width=0.99\linewidth]{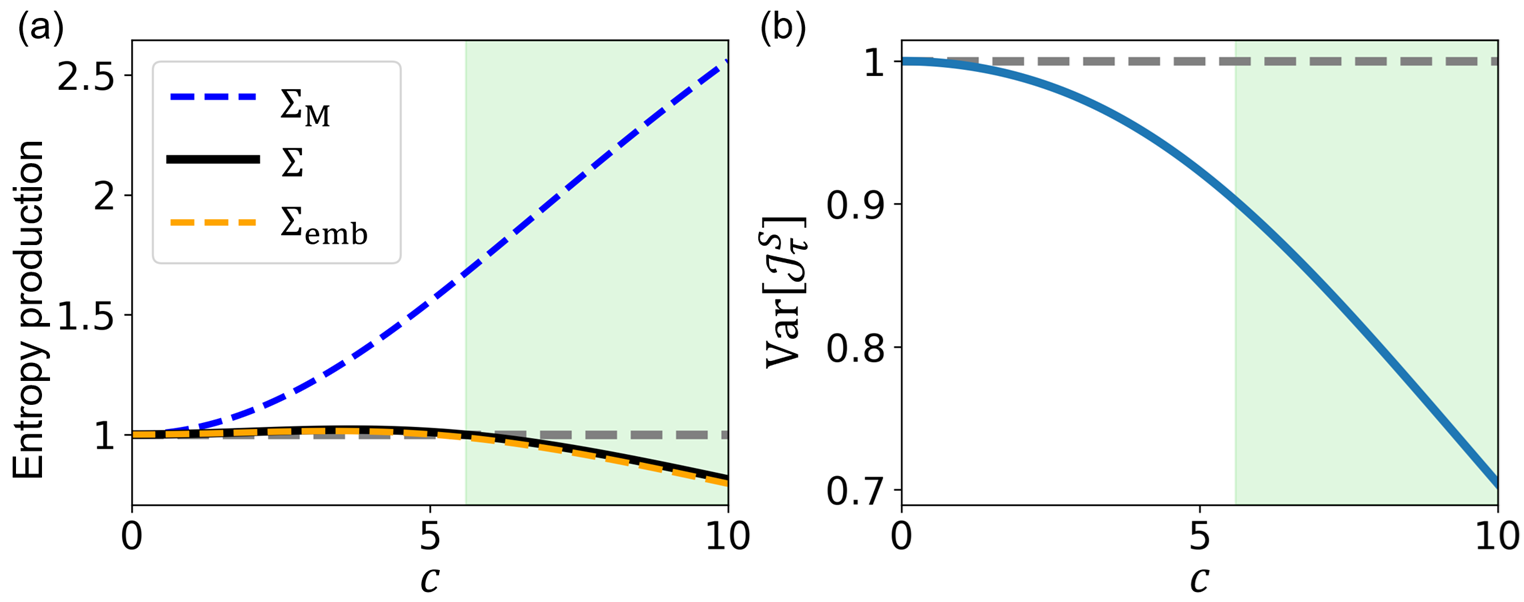}
    \caption{(a) Entropy production $\Sigma$ (black solid curve), apparent Markovian entropy production $\Sigma_{\rm M}$ (blue dashed curve), and embedded entropy production $\Sigma_{\rm emb}$ (orange dashed curve). The difference between $\Sigma$ and $\Sigma_{\rm M}$ is the negative non-Markovian correction $\Sigma_{\rm NM}\leq 0$. 
    (b) Variance of the current $\Var[\mathcal{J}_{\tau}^{S}]$. All quantities are normalized relative to the corresponding values in the Markovian limit $c=0$.}
    \label{fig: TUR2}
\end{figure}

{\it Origin of the TUR enhancement.---} 
The enhancement of $\mathcal{Q}$ for $c>0$ in Fig.~\ref{fig: TUR} originates from the suppression of both $\Sigma$ and $\Var[\mathcal{J}_{\tau}^{S}]$, as shown in Fig.~\ref{fig: TUR2}. 
From $\Sigma_{\rm NM}\leq0$, the non-Markovian force lowers $\Sigma$ relative to $\Sigma_{\rm M}$, and, for the parameters used in Fig.~\ref{fig: TUR2}(a), even below its value at $c=0$. At the same time, temporal correlations induced by the non-Markovian bath suppress $\Var[\mathcal{J}_{\tau}^{S}]$, as shown in Fig.~\ref{fig: TUR2}(b). 
For Figs.~\ref{fig: TUR} and~\ref{fig: TUR2}, we choose $V^{S}_{\lambda^{v}_{t}}(X)=\kappa (X-vt)^2/2$ and $\mathcal{J}_{\tau}^{S}=[X(\tau)-X(0)]/\tau$.

\clearpage
\appendix 
\widetext


\section{\label{sec:Markov_embedding}Markovian embedding of the generalized Langevin equation}

Here, we introduce thermal linear
Gaussian embeddings of the generalized Langevin equation.

\subsection{\label{sec:Markov_embedding_Langevin}Thermal linear Gaussian embedding}

The auxiliary Hamiltonian reads 
\begin{align}
    H_{A}+H_{\rm int}=\frac{1}{2}(\bm{y}-\bm{c}X)^{T}(\bm{y}-\bm{c}X), \label{sup: Haux_general}
\end{align}
where $\bm{y}\in \mathbb{R}^{n}$ and $\bm{c}\in \mathbb{R}^{n}$ are the auxiliary coordinates and system-auxiliary coupling vector, respectively.
Here, we take
$H_A=\bm y^T\bm y/2$ and
$H_{\rm int}=-X\bm c^T\bm y+X^2\bm c^T\bm c/2$,
so that the counterterm is included in $H_{\rm int}$.

We now introduce a reversible interaction matrix $J^{T}=-J$ and an irreversible dissipation matrix $D=D^{T}\succeq0$. We assume that $A:=J-D$ is Hurwitz, that is, every eigenvalue $\lambda_{i}$ satisfies $\mathrm{Re}(\lambda_{i})<0$. 
The auxiliary Langevin equation reads
\begin{align}
    \dot{\bm y}
    =A(\bm y-\bm cX)
    +\sqrt{\frac{2D}{\beta}}\,\bm\xi_{t},
    \label{eq: note y Langevin}
\end{align}
where $\mathbb E[\bm\xi_{t}]=0$ and $    \mathbb E[\bm\xi_{t}\bm\xi_{s}^T]=I_n\delta(t-s)$.

To simplify the notation, we introduce the shifted coordinate
\begin{align}
    \bm z:=\bm y-\bm cX,
    \label{eq: note zdef}
\end{align}
where Eq.~\eqref{sup: Haux_general} reads $H_A+H_{\rm int}=\frac{1}{2}\bm z^T\bm z$. Also, Langevin equation for the system and auxiliary modes read 
\begin{align}
    \dot X&=\frac{P}{M}, \\
    \dot P&=-\partial_XV_{S}^{\lambda_t}(X)+\bm c^T\bm z, \label{eq: note P Langevin} \\
    \dot{\bm z}&=A\bm z-\bm c\frac{P}{M}
    +\sqrt{\frac{2D}{\beta}}\,\bm\xi_{t}.
    \label{eq: note z Langevin}
\end{align}

In what follows, we eliminate the auxiliary modes and derive the generalized Langevin equation. By formally solving Eq.~\eqref{eq: note z Langevin}, we have 
\begin{align}
    \bm z(t)
    ={}&e^{At}\bm z(0)
    -\int_0^t ds\,e^{A(t-s)}\bm c\frac{P(s)}{M}+\int_0^t ds\,e^{A(t-s)}
    \sqrt{\frac{2D}{\beta}}\,\bm\xi_{s}.
    \label{eq: note formal z}
\end{align}
Substituting Eq.~\eqref{eq: note formal z} into Eq.~\eqref{eq: note P Langevin} yields 
\begin{align}
    \dot{P}= -\partial_XV_{S}^{\lambda_t}(X)-\int_0^t ds\,K(t-s)\frac{P(s)}{M}+\eta_{t}, \label{sup: GLE}
\end{align}
where
\begin{align}
    K(t)=\bm c^T e^{A|t|}\bm c,
    \label{eq: note K general}
\end{align}
and
\begin{align}
    \eta_t
    :={}&\bm c^Te^{At}\bm z(0)
    +\bm c^T\int_0^t ds\,e^{A(t-s)}
    \sqrt{\frac{2D}{\beta}}\,\bm\xi_{s}.
    \label{eq: note eta general}
\end{align}

In what follows, we show the fluctuation-dissipation relation. We consider solution of the free auxiliary dynamics  $\bm{z}^{\rm f}(t)=e^{At}\bm{z}(0)+\int^{t}_{0}ds e^{A(t-s)}\sqrt{2D/\beta}\bm{\xi}_{s}$, and also define $    \Xi_t := \mathbb E[\bm z^{\rm f} (t)\bm z^{\rm f} (t)^{T}]$. We use Eq.~\eqref{eq: note z Langevin} to obtain
\begin{align}
    \frac{d}{dt}\Xi_t
    =
    A\Xi_t+\Xi_tA^T+\frac{2D}{\beta}.
    \label{eq: note covariance evolution}
\end{align}
The stationary solution $\Xi_{\rm eq}$ satisfies
\begin{align}
    A\Xi_{\rm eq}
    +
    \Xi_{\rm eq}A^T
    +
    \frac{2D}{\beta}
    =
    0,
    \label{eq: note stationary Lyapunov}
\end{align}
and the solution to Eq.~\eqref{eq: note stationary Lyapunov} is given by $    \Xi_{\rm eq}
    =
    \frac{1}{\beta}I_n$, by noting that  
$A=J-D$, $J^T=-J$, and $D^T=D$. 

By assuming the initial state of the auxiliary modes as the conditional thermal state
\begin{align}
    \pi^{A|S}(\bm z|X)
    \propto\exp\left[-\beta (H_{A}+H_{\rm int}) \right],
    \label{eq: note conditional thermal}
\end{align}
we have $\mathbb E[\bm z (0)\bm z (0)^{T}]=\frac{1}{\beta}I_{n}=\Xi_{\rm eq}$. Therefore, $\mathbb E[\bm z^{\rm f} (t)\bm z^{\rm f} (t)^{T}]=\Xi_{\rm eq}$ for $t\geq 0$.  
For $t\geq s$, Eq.~\eqref{eq: note formal z} gives
\begin{align}
    \mathbb E[\bm z^{\rm f}(t)\bm z^{\rm f}(s)^T]
    &=
    e^{A(t-s)}
    \mathbb E[\bm z^{\rm f}(s)\bm z^{\rm f}(s)^T]
    =
    e^{A(t-s)}\Xi_{\rm eq}=
    \frac{1}{\beta}e^{A(t-s)},
    \label{eq: note OU two time covariance}
\end{align}
where we use the fact that $\bm{z}^{\rm f}(t)=e^{A(t-s)}\bm{z}^{\rm f}(s)+\int^{t}_{s}du e^{A(t-u)}\sqrt{\frac{2D}{\beta}}\bm{\xi}(u)$ and $\mathbb{E}[\bm{\xi}(u)\bm{z}^{\rm f}(s)^{T}]=0 $ for $u>s$. When $t\leq s$, a similar argument leads to $\mathbb E[\bm z^{\rm f}(t)\bm z^{\rm f}(s)^T]=\frac{1}{\beta}e^{A^{T}(s-t)}$. Therefore, we obtain the fluctuation-dissipation relation
\begin{align}
    \mathbb E[\eta_t\eta_{s}]=
    \frac{1}{\beta}
    \bm c^T
    e^{A|t-s|}
    \bm c
    =
    \frac{1}{\beta}K(t-s).
\end{align}


\subsection{Properties of memory kernels generated by thermal embeddings}
\label{sec:finite rational kernels}

The one-sided Laplace transform of the memory kernel~\eqref{eq: note K general} is given by
\begin{align}
    \widehat K(s)
    =\bm c^T\left[sI_n-A\right]^{-1}\bm c = \frac{\bm{c}^{T}\mathrm{adj}(sI_{n}-A)\bm{c}}{\mathrm{det}(sI_{n}-A)}.
    \label{eq: note K Laplace}
\end{align}
In the following, we show the properties of Eq.~\eqref{eq: note K Laplace}.

We start by noting that because $A$ and $\bm{c}$ are real, Eq.~\eqref{eq: note K Laplace} is a ratio of two polynomial functions of $s$ with real coefficients. 
Therefore, $\widehat{K}(s)$ is a real rational function and satisfies $\widehat K(\bar{s})=\overline{\widehat K(s)}$.
We also note that $\mathrm{deg}\, \mathrm{det}(sI_{n}-A)=n$ and $\mathrm{deg}[ \bm{c}^{T}\mathrm{adj}(sI_{n}-A)\bm{c}]\leq n-1$, and thus $\widehat{K}(s)$ is strictly proper. 

Next, we note that the poles of $\widehat{K}(s)$ can only occur at points satisfying $\mathrm{det}(sI_{n}-A)=0$. Since $A$ is Hurwitz, all poles $p$ of $\widehat{K}(s)$ satisfy $\operatorname{Re} p<0$. Therefore, $\widehat{K}(s)$ is stable.  
 
Finally, for $\operatorname{Re}s>0$, we define $\bm x_{s}:=(s I_n-A)^{-1}\bm c$. Then, we have $\widehat{K}(s)=\bm{c}^{T}\bm{x}_{s}=\bm{x}_{s}^{\dagger}(\bar{s}I_{n}-A^{T})\bm{x}_{s}$. By taking the real part, we obtain
\begin{align}   2\operatorname{Re}\widehat K(s) = \bm x_s^\dagger [(s+\bar{s})I_{n}-(A+A^{T}) ] \bm x_s = 2(\operatorname{Re}s)||\bm{x}_{s}||^{2} - \bm{x}_{s}^{\dagger}(A+A^{T})\bm{x}_{s}.
\end{align}
Using $A=J-D$, with $J^T=-J$ and $D^{T}=D\succeq0$, we obtain
\begin{align}
\operatorname{Re}\widehat K(s) =(\operatorname{Re}s)||\bm{x}_{s}||^{2}+\bm{x}_{s}^{\dagger}D\bm{x}_{s}\geq 0, \quad \operatorname{Re}s>0.
\end{align}
Since $\widehat{K}(s)$ is analytic for $\operatorname{Re}s>0$, this shows that $\widehat{K}(s)$ is a positive-real function.

Moreover, since $A$ is Hurwitz, $i\omega I_{n}-A$ is invertible for $\omega\in\mathbb{R}$. By defining 
\begin{align}
    \bm{x}_{\omega}:=(i\omega I_{n}-A)^{-1}\bm{c} \label{sup: xomega},
\end{align}
a similar argument leads to
\begin{align}
    \operatorname{Re}\widehat{K}(i\omega)=\bm{x}_{\omega}^{\dagger}D\bm{x}_{\omega}\geq 0 , \quad \omega \in \mathbb{R}. \label{sup: Komega positive}
\end{align}
 
Therefore, a finite-dimensional thermal linear Gaussian embedding considered in Sec.~\ref{sec:Markov_embedding_Langevin} generates a memory kernel whose Laplace transform $\widehat{K}(s)$ is real, strictly proper,
stable rational function, and satisfies $\operatorname{Re}\widehat{K}(i\omega)\geq 0$ for all $\omega\in\mathbb{R}$.

\subsection{Thermal realization of stable rational memory kernels}

We now show the converse direction. For a given memory kernel whose Laplace transform $\widehat{K}(s)$ is a real, strictly proper, stable rational function, and satisfies $J(\omega)/\omega=\operatorname{Re}\widehat K(i\omega)\geq0$ for all
$\omega\geq 0$, we can obtain a finite-dimensional thermal realization
of the form~\eqref{eq: note K Laplace}. 
In the following, we show this claim. 

Here, note that because $\widehat{K}(s)$ is real rational, we have $\widehat{K}(-i\omega)=\overline{\widehat{K}(i\omega)}$, and thus $\operatorname{Re}\widehat{K}(-i\omega)=\operatorname{Re}\widehat{K}(i\omega)$. Therefore, the above assumption implies $\operatorname{Re}\widehat K(i\omega)\geq0$ for all
$\omega\in \mathbb{R}$.

\subsubsection{Minimal realization}

Since $\widehat{K}(s)$ is a real, strictly proper rational function, it admits a real minimal realization
\begin{align}
    \widehat K(s)
    =
    C_{\rm m}(sI_{n_{{\rm m}}}-A_{\rm m})^{-1}B_{\rm m},
    \label{eq: sup minimal realization}
\end{align}
where $(A_{\rm m},B_{\rm m})$ is controllable
and $(C_{\rm m},A_{\rm m})$ is observable~\cite{ZhouDoyleGlover1996}. Note that the pair $(A_{\rm m},B_{\rm m})$ is controllable if and only if
\begin{align}
    \operatorname{rank}
    \begin{pmatrix}
        B_{\rm m} & A_{\rm m} B_{\rm m} & \cdots & A_{\rm m}^{n_{{\rm m}}-1}B_{\rm m}
    \end{pmatrix}
    =n_{{\rm m}}, \label{eq:controllability_matrix}
\end{align}
and the pair $(C_{\rm m},A_{\rm m})$ is observable if and only if
\begin{align}
    \operatorname{rank}
    \begin{pmatrix}
        C_{\rm m}\\
        C_{\rm m}A_{\rm m}\\
        \vdots\\
        C_{\rm m}A_{\rm m}^{n_{\rm m}-1}
    \end{pmatrix}
    =n_{{\rm m}}. \label{eq:observability_matrix}
\end{align}

Since the realization is minimal, $p$ is a pole of $\widehat{K}(s)$ if and only if it is an eigenvalue of $A_{\rm m}$~\cite{ZhouDoyleGlover1996}. Since $\widehat{K}(s)$ is stable, every pole $p$ satisfies $\operatorname{Re}p<0$.
Therefore, $A_{\rm m}$ is Hurwitz.  

\subsubsection{The non-strict KYP lemma}
\label{sec: sup KYP lemma}


We use the non-strict continuous-time
Kalman--Yakubovich--Popov (KYP) lemma (Corollary~24 of Ref.~\cite{Scherer2022}). Let us consider the minimal realization $\widehat{K}(s)=C_{{\rm m}}(sI_{n_{{\rm m}}}-A_{{\rm m}})^{-1}B_{{\rm m}}$ given in Eq.~\eqref{eq: sup minimal realization}. 
Since the realization is minimal, $(A_{{\rm m}},B_{{\rm m}})$ is controllable. Moreover, since $A_{{\rm m}}$ is Hurwitz, $\widehat{K}(s)$ has no poles on the imaginary axis. Then, for fixed $\rho\geq 0$, the following two statements are equivalent.

(i) For the real symmetric matrix \begin{align}
    P_{\rho}:=\begin{pmatrix}
        2\rho & -1 \\ -1 & 0
    \end{pmatrix},
\end{align}
the inequality
\begin{align}
    \begin{pmatrix}
        \widehat{K}(i\omega)\\1
    \end{pmatrix}^{\dagger}
    P_{\rho}
    \begin{pmatrix}
        \widehat{K}(i\omega)\\1
    \end{pmatrix} = 2\rho |\widehat{K}(i\omega)|^{2}-2\operatorname{Re}\widehat{K}(i\omega)
    \leq0
    \label{eq: sup KYP general frequency inequality}
\end{align}
holds for all $\omega\in\mathbb R$.

(ii) There exists a real symmetric matrix
$X_{\rho}=X_{\rho}^T$ satisfying
\begin{align}
   &\begin{pmatrix}
        A_{\rm m}  & B_{\rm m}  \\
        I_{n_{\rm m}} &0
    \end{pmatrix}^{T}\begin{pmatrix}
        0  & X_{\rho}  \\
        X_{\rho} &0
    \end{pmatrix}\begin{pmatrix}
        A_{\rm m}  & B_{\rm m}  \\
        I_{n_{\rm m}} &0
    \end{pmatrix}+ \begin{pmatrix}
        C_{\rm m}  & 0  \\
        0 & 1
    \end{pmatrix}^{T}P_{\rho}\begin{pmatrix}
        C_{\rm m}  & 0  \\
        0 & 1
    \end{pmatrix} \nonumber \\ 
    &=  \begin{pmatrix}
        A_{\rm m}^TX_{\rho}+X_{\rho}A_{\rm m}+2\rho C_{\rm m}^{T}C_{\rm m} &X_{\rho} B_{\rm m} - C_{\rm m}^{T} \\
        B_{\rm m}^TX_{\rho}-C_{\rm m}&0
    \end{pmatrix}
    \preceq0.
    \label{eq: sup KYP general matrix inequality}
\end{align}

For $\rho=0$, condition (i) follows directly from assumption $\operatorname{Re}\widehat{K}(i\omega)\geq 0$ for all $\omega \geq 0$. 
For $0<\mathcal{S}_{K}<\infty$, we will later show that condition also holds for $\rho=\rho_{*}:=1/\mathcal{S}_{K}$ in  Sec.~\ref{sec:construction}. We therefore derive the consequences of condition (ii) for fixed $\rho\geq 0$ for which condition (i) holds.  

\subsubsection{Realization of the thermal linear Gaussian embedding}
\label{sup sec: consequences when the statement}

In what follows, we fix $\rho\geq 0$ and discuss the consequences when Eq.~\eqref{eq: sup KYP general matrix inequality} holds.

We first note that a symmetric block matrix of the form
\begin{align}
    \begin{pmatrix}
        Q&R\\
        R^{T}&0
    \end{pmatrix}
    \preceq0
\end{align}
is equivalent to $R=0$ and $Q\preceq 0$. Therefore, Eq.~\eqref{eq: sup KYP general matrix inequality} is equivalent to  
\begin{align}
    & X_{\rho}B_{\rm m}=C_{\rm m}^{T} \label{sup: XBC relation} \\
    &  Y_{\rho}:=-(A_{\rm m}^{T}X_{\rho}+X_{\rho}A_{\rm m}+2\rho C_{\rm m}^{T}C_{\rm m}) \succeq 0. \label{sup: Yrho relation}
\end{align}
We now use the relation
\begin{align}
    \frac{d}{dt}(e^{A_{\rm m}^{T}t}X_{\rho}e^{A_{\rm m}t})=-e^{A_{\rm m}^{T}t}(Y_{\rho}+2\rho C_{\rm m}^{T}C_{\rm m})e^{A_{\rm m}t}. \label{sup: AXA relation}
\end{align}
Since $A_{\rm m}$ is Hurwitz, we have $e^{A_{\rm m}t}\rightarrow 0$ as $t\rightarrow \infty$. Then, integrating Eq.~\eqref{sup: AXA relation} from zero to infinity gives
\begin{align}
    X_{\rho}=\int^{\infty}_{0}dt e^{A_{\rm m}^{T}t}(Y_{\rho}+2\rho C_{\rm m}^{T}C_{\rm m} )e^{A_{\rm m}t} \succeq 0.
\end{align}
Next, suppose that $\bm{v}^{T}X_{\rho}\bm{v}=0$. Then, we have $(e^{A_{\rm m}t}\bm{v})^{T}(Y_{\rho}+2\rho C_{\rm m}^{T}C_{\rm m})e^{A_{\rm m}t}\bm{v}=0$ for all $t\geq 0$. Moreover, we have
\begin{align}
    (e^{A_{\rm m}s}\bm{v})^{T}X_{\rho}e^{A_{\rm m}s}\bm{v}=\int^{\infty}_{0}dt (e^{A_{\rm m}(t+s)}\bm{v})^{T}(Y_{\rho}+2\rho C_{\rm m}^{T}C_{\rm m})e^{A_{\rm m}(t+s)}\bm{v}=0\quad \forall s\geq 0,
\end{align}
which implies $X_{\rho}e^{A_{\rm m}s}\bm{v}=0$. Then, we use Eq.~\eqref{sup: XBC relation} and obtain 
\begin{align}
C_{\rm m}e^{A_{\rm m}s}\bm{v}=B_{\rm m}^{T}X_{\rho}e^{A_{\rm m}s}\bm{v}=0 \quad \forall s\geq 0.    \label{sup: CAv relation}
\end{align}
From Eq.~\eqref{sup: CAv relation}, we have $(d^{k}/ds^{k})(C_{\rm m}e^{A_{\rm m}s}\bm{v})|_{s=0}=C_{\rm m}A_{\rm m}^{k}\bm{v}=0$ for $k=0,1,\cdots, n_{\rm m}-1$, and therefore
\begin{align}
    \begin{pmatrix}
        C_{\rm m} \\
        C_{\rm m}A_{\rm m}\\
        \vdots \\
        C_{\rm m}A_{\rm m}^{n_{\rm m}-1}
    \end{pmatrix}\bm{v}=0.
\end{align}
We then use the observability of $(C_{\rm m},A_{\rm m})$ [Eq.~\eqref{eq:observability_matrix}] and find that $\bm{v}=0$. Therefore, we have
\begin{align}
    X_{\rho}\succ 0.
\end{align}
We now define a real symmetric matrix 
\begin{align}
    T_{\rho}:=X_{\rho}^{1/2}\succ 0,
\end{align}
which is invertible. Then, we define
\begin{align}
    A_{\rho}:=T_{\rho}A_{\rm m}T_{\rho}^{-1}, \quad \bm{c}_{\rho}:=T_{\rho}B_{\rm m}.
\end{align}
Using Eq.~\eqref{sup: XBC relation}, we also obtain $C_{\rm m}T_{\rho}^{-1}=B_{\rm m}^{T}T_{\rho}=\bm{c}_{\rho}^{T}$. 
This allows us to write
\begin{align}
    \widehat{K}(s)=C_{\rm m}(sI_{n_{\rm m}}-A_{\rm m})^{-1}B_{\rm m} = \bm{c}_{\rho}^{T}(sI_{n_{\rm m}}-A_{\rho})^{-1}\bm{c}_{\rho}, 
\end{align}
which takes the form of Eq.~\eqref{eq: note K Laplace}. 

We then define 
\begin{align}
    J_{\rho}:=\frac{A_{\rho}-A_{\rho}^{T}}{2}, \quad D_{\rho}:=-\frac{A_{\rho}+A_{\rho}^{T}}{2}.
\end{align}
From the definition, we have $J_{\rho}^{T}=-J_{\rho}$, $D_{\rho}^{T}=D_{\rho}$, and $A_{\rho}=J_{\rho}-D_{\rho}$. Using Eq.~\eqref{sup: Yrho relation}, we also have $A_{\rho}^{T}+A_{\rho}+2\rho \bm{c}_{\rho}\bm{c}_{\rho}^{T} \preceq 0$. Moreover, for $\rho\geq 0$, the following relation holds:
\begin{align}
    D_{\rho}\succeq \rho \bm{c}_{\rho}\bm{c}_{\rho}^{T}\succeq 0  .
\end{align}
We finally note that $A_{\rm m}$ and $A_{\rho}$ are similar matrices and share the same eigenvalues. Therefore, $A_{\rho}$ is also Hurwitz.

Therefore, whenever condition (i) holds for a fixed $\rho\geq 0$, the above construction provides a thermal linear Gaussian embedding of the type discussed in Sec.~\ref{sec:Markov_embedding_Langevin}.

\subsubsection{Construction of the thermal embedding}
\label{sup sec: construction of thermal embedding}

We now apply the above construction to $\rho=0$. In this case, Eq.~\eqref{eq: sup KYP general frequency inequality} reduces to $\operatorname{Re}\widehat{K}(i\omega)\geq 0$ for all $\omega\in \mathbb{R}$, which holds by assumption. Therefore, the construction discussed in Sec.~\ref{sup sec: consequences when the statement} yields
\begin{align}
    \widehat{K}(s)=\bm{c}_{0}^{T}(sI_{n_{\rm m}}-A_{0})^{-1}\bm{c}_{0},
\end{align}
with $A_{0}=J_{0}-D_{0}$, $J_{0}^{T}=-J_{0}$, $D_{0}=D_{0}^{T}\succeq 0$, and $A_{0}$ is Hurwitz.

\subsection{Simple-pole form and special cases of the spectral density}

If all poles are simple in Eq.~\eqref{eq: note K Laplace}, the memory kernel takes the form 
\begin{align}
    K(t)
    ={}&
    \sum_{k}
    e^{-\Gamma_k|t|}
    \left[
        a_k\cos(\Omega_k t)
        +
        b_k\sin(\Omega_k|t|)
    \right],
    \label{eq: note simple pole kernel}
\end{align}
with $\Gamma_k>0, \Omega_{k}\geq 0$ and $a_k,b_k\in\mathbb R$.

The spectral density $J(\omega)/\omega
=
\int_0^\infty dt\,K(t)\cos(\omega t)$ corresponding to Eq.~\eqref{eq: note simple pole kernel} is given by
\begin{align}
    J(\omega)
    =
    \frac{\omega}{2}
    \sum_{k=1}^{M}
    \Biggl[
    \frac{
        a_k\Gamma_k+b_k(\Omega_k-\omega)
    }{
        \Gamma_k^2+(\Omega_k-\omega)^2
    }
    +
    \frac{
        a_k\Gamma_k+b_k(\Omega_k+\omega)
    }{
        \Gamma_k^2+(\Omega_k+\omega)^2
    }
    \Biggr].
    \label{eq: sup spectral density simple pole}
\end{align}
We list several special choices of the coefficients in
Eq.~\eqref{eq: note simple pole kernel} and the corresponding
spectral densities.

\begin{itemize}

\item[(i)] \textit{Underdamped Brownian oscillator.}
Choosing $b_k=a_k\Gamma_k/\Omega_k$ with $\Omega_{k}>0$ gives
\begin{align}
    J^{\rm ub}(\omega)
    =
    \sum_k
    \frac{
        2a_k\Gamma_k(\Omega_k^2+\Gamma_k^2)\omega
    }{
        [\omega^2-(\Omega_k^2+\Gamma_k^2)]^2
        +4\Gamma_k^2\omega^2
    }.
\end{align}
\item[(ii)] \textit{Drude--Lorentz-type spectral density.}
Choosing $b_k=0$ gives
\begin{align}
    J^{\rm dl}(\omega)
    =
    \frac{\omega}{2}
    \sum_k a_k\Gamma_k
    \left[
        \frac{1}{
        \Gamma_k^2+(\Omega_k-\omega)^2}
        +
        \frac{1}{
        \Gamma_k^2+(\Omega_k+\omega)^2}
    \right].
\end{align}

\item[(iii)] \textit{Super-Ohmic spectral density at low frequencies.}
Choosing $b_k=-a_k\Gamma_k/\Omega_k$ with $\Omega_{k}>0$ gives
\begin{align}
    J^{\rm so}(\omega)
    =
    \sum_k
    \frac{
        2a_k\Gamma_k\omega^3
    }{
        [\omega^2-(\Omega_k^2+\Gamma_k^2)]^2
        +4\Gamma_k^2\omega^2
    }.
\end{align}

\end{itemize}

\subsection{Fokker-Planck equation}

The Fokker-Planck equation corresponding to Eq.~\eqref{eq: note z Langevin} in the unshifted coordinates $(X,P,\bm{y})$ is given by
\begin{align}
    \partial_t f_t^{SA}
    ={}&-\partial_X\left(\frac{P}{M}f_t^{SA}\right)
    -\partial_P\left[\left(-\partial_XV_{S}^{\lambda_t}
    +\bm c^T(\bm y-\bm cX)\right)f_t^{SA}\right]
    \nonumber\\
    &-\nabla_{\bm y}\cdot\left[J(\bm y-\bm cX)f_t^{SA}\right]
    +\nabla_{\bm y}\cdot D\left[(\bm y-\bm cX)f_t^{SA}
    +\frac{1}{\beta}\nabla_{\bm y}f_t^{SA}\right],
    \label{eq: note Fokker Planck expanded}
\end{align}
and $\pi_{\lambda_t}^{SA}\propto e^{-\beta H_{SA}^{\rm tot}}$ is an instantaneous stationary state of Eq.~\eqref{eq: note Fokker Planck expanded}. 

For arbitrary functions $F(X,P,\bm y)$ and $G(X,P,\bm y)$, we define the Poisson bracket by
\begin{align}
    \{F,G\}_{SA}
    :=\partial_XF\,\partial_PG-\partial_PF\,\partial_XG
    +(\nabla_{\bm y}F)^T J\nabla_{\bm y}G.
    \label{eq: note extended Poisson}
\end{align}
Note that when the auxiliary variables are expressed in canonical position and momentum coordinates, $\{\cdot,\cdot\}_{SA}$ reduces to the usual canonical Poisson bracket. 

Using Eq.~\eqref{eq: note extended Poisson}, Eq.~\eqref{eq: note Fokker Planck expanded} can be written in the compact form 
\begin{align}
    \partial_t f_t^{SA}
    =\{H_{SA}^{\rm tot},f_t^{SA}\}_{SA}+\mathcal D f_t^{SA},
    \label{eq: note compact FP}
\end{align}
where the dissipative part of the Fokker-Planck operator is
\begin{align}
    \mathcal D f
    :=\nabla_{\bm y}\cdot D\Bigl[(\nabla_{\bm y}H_{SA}^{\rm tot})f
    +\frac{1}{\beta}\nabla_{\bm y}f\Bigr]
    =\nabla_{\bm y}\cdot D\Bigl[(\bm y-\bm cX)f
    +\frac{1}{\beta}\nabla_{\bm y}f\Bigr].
    \label{eq: note dissipator}
\end{align}

\section{Finite-time approximation of memory kernels}
\label{sec: kernel approximation}

We show that the simple-pole subclass in
Eq.~\eqref{eq: note simple pole kernel} is already sufficient to approximate any continuous real, even, positive-definite memory kernel uniformly on any prescribed finite time interval. In fact, it suffices to use nonnegative damped-cosine components with a common decay rate. Note that below approximation scheme is closely related to the one discussed in Ref~\cite{PhysRevLett.127.250404}.

\begin{prop}[Finite-time approximation]
\label{prop:kernel_approximation}
Let $K_{\rm tar}:\mathbb R\to\mathbb R$ be a continuous, even,
positive-definite target kernel with $K_{\rm tar}(0)<\infty$.
For every $0<T<\infty$ and $\varepsilon>0$, there exist a finite
number of modes $N$, coefficients $a_k\geq0$, frequencies
$\Omega_k\geq0$, and a common decay rate $\gamma>0$ such that
\begin{align}
    K_N(t)
    :=
    \sum_{k=1}^{N}
    a_k e^{-\gamma|t|}\cos(\Omega_k t)
    \label{eq: approx damped cosine kernel}
\end{align}
satisfies
\begin{align}
    \sup_{|t|\leq T}
    |K_N(t)-K_{\rm tar}(t)|
    <\varepsilon.
    \label{eq: approx uniform accuracy}
\end{align}
Each $K_N$ is positive definite and admits a finite-dimensional thermal
linear Gaussian Markovian embedding discussed in Sec.~\ref{sec:Markov_embedding}. The number of modes and their parameters, including $\gamma$,
may depend on $K_{\rm tar}$, $T$, and $\varepsilon$.
\end{prop}

\noindent\textit{Proof.}
Because $K_{\rm tar}(t)$ is real and even, Bochner's theorem~\cite{Rudin1962} can be expressed in the cosine form. Thus, the target kernel admits
the spectral representation
\begin{align}
    K_{\rm tar}(t)
    =
    \int_{[0,\infty)}
    \cos(\omega t)\,\nu(d\omega),
    \qquad
    \nu([0,\infty))=K_{\rm tar}(0),
    \label{eq: approx Bochner representation}
\end{align}
where $\nu$ is a finite nonnegative Borel measure. 
When the spectral measure has a density, it corresponds to the spectral density introduced in the main text as
\begin{align}
    \nu(d\omega)
    =
    \frac{2}{\pi}
    \frac{J_{\rm tar}(\omega)}{\omega}\,d\omega.
\end{align}
The construction below does not require the existence of such
a density and also allows discrete spectral contributions, e.g., $\nu(d\omega)=\sum_{j}q_{j}\delta_{\omega_{j}}(d\omega)$ and $q_{j}\geq 0$. Here $\delta_{\omega_{j}}$ is defined by $\int f(\omega)\delta_{\omega_{j}}(d\omega)=f(\omega_{j})$ for any suitable test function $f$.

If $K_{\rm tar}(0)=0$, Eq.~\eqref{eq: approx Bochner representation}
implies $K_{\rm tar}\equiv0$, and the claim is trivially satisfied by taking all
$a_k=0$. We therefore assume $K_{\rm tar}(0)>0$ below.

First, choose a frequency cutoff $R>0$ and define
\begin{align}
    K_R(t)
    :=
    \int_{[0,R]}
    \cos(\omega t)\,\nu(d\omega).
\end{align}
Since $|\cos(\omega t)|\leq1$, the truncation error satisfies
\begin{align}
    \sup_{t\in\mathbb R}
    |K_{\rm tar}(t)-K_R(t)|
    \leq
    \nu((R,\infty)).
    \label{eq: approx truncation error}
\end{align}
The right-hand side can be made arbitrarily small by taking sufficiently large $R$ because
$\nu$ is finite.

Next, we partition $[0,R]$ into finitely many disjoint intervals
$I_1,\ldots,I_N$, and denote
\begin{align}
    h:=\max_{1\leq k\leq N}\operatorname{diam}(I_k), \label{sup: eq diam}
\end{align}
where $\operatorname{diam}(I_{k}):=\sup_{\omega,\omega'\in I_{k}}|\omega-\omega'|$. 
We choose a representative frequency $\Omega_k\in I_k$ and set
\begin{align}
    a_k:=\nu(I_k)\geq0,
    \quad
    K_N^{(0)}(t)
    :=
    \sum_{k=1}^{N}a_k\cos(\Omega_k t). 
\end{align}
Then, we find that
\begin{align}
    K_{R}(t)-K_{N}^{(0)}(t)=\sum_{k=1}^{N}\int_{I_{k}}[\cos(\omega t)-\cos(\Omega_{k}t)]\nu(d\omega). \label{sup: KR-KN}
\end{align}
For $|t|\leq T$ and $\omega\in I_k$, we have
\begin{align}
    |\cos(\omega t)-\cos(\Omega_k t)|
    \leq
    |t|\,|\omega-\Omega_k|
    \leq Th,
\end{align}
and substituting this into Eq.~\eqref{sup: KR-KN} yields
\begin{align}
    \sup_{|t|\leq T}
    |K_R(t)-K_N^{(0)}(t)|
    &\leq
    Th\sum_{k=1}^{N}\nu(I_k)
    = Th\nu([0,R])\leq 
    ThK_{\rm tar}(0).
    \label{eq: approx discretization error}
\end{align}

Finally, introduce a common positive decay rate $\gamma$ and
define
\begin{align}
    K_N(t)
    :=
    e^{-\gamma|t|}K_N^{(0)}(t)=e^{-\gamma|t|}\sum_{k=1}^{N}a_{k}\cos(\Omega_{k}t).
\end{align}
Since
$|K_N^{(0)}(t)|\leq\sum_k a_k\leq K_{\rm tar}(0)$,
we have
\begin{align}
    \sup_{|t|\leq T}
    |K_N^{(0)}(t)-K_N(t)|
    \leq
    K_{\rm tar}(0)\left(1-e^{-\gamma T}\right).
    \label{eq: approx damping error}
\end{align}
Combining Eqs.~\eqref{eq: approx truncation error},
\eqref{eq: approx discretization error}, and
\eqref{eq: approx damping error} gives
\begin{align}
    \sup_{|t|\leq T}
    |K_{\rm tar}(t)-K_N(t)|
    \leq
    \nu((R,\infty))
    +K_{\rm tar}(0)Th + K_{\rm tar}(0)
    \left[
        1-e^{-\gamma T}
    \right].
    \label{eq: approx total error}
\end{align}
We now choose $R$ to be sufficiently large, $h$ to be sufficiently small,
and $\gamma>0$ to be sufficiently small so that the three error
contributions are each smaller than $\varepsilon/3$.
This proves Eq.~\eqref{eq: approx uniform accuracy}.

We note that the corresponding spectral density $J_{N}(\omega)/\omega=\int^{\infty}_{0}dt K_{N}(t)\cos(\omega t)$ reads
\begin{align}
    \frac{J_{N}(\omega)}{\omega}=\sum_{k=1}^{N}\frac{a_{k}\gamma}{2}\Bigl[ \frac{1}{(\omega-\Omega_{k})^{2}+\gamma^{2}} + \frac{1}{(\omega+\Omega_{k})^{2}+\gamma^{2}}\Bigr]\geq 0, \quad \omega\in \mathbb{R},
\end{align}
and therefore $K_{N}$ is positive-definite. 

The corresponding thermal linear Gaussian Markovian embedding can be constructed as follows. For each $k$, we choose
\begin{align}
    J_k&=
    \begin{pmatrix}
        0&\Omega_k\\
        -\Omega_k&0
    \end{pmatrix},
    \quad
    D_k=\gamma I_2,
    \quad
    A_k=J_k-D_k, \quad
    \bm c_k=
    \sqrt{a_k}
    \begin{pmatrix}1\\0\end{pmatrix},
\end{align}
and obtain
\begin{align}
    K_{N}(t)=\sum_{k=1}^{N}\bm{c}_{k}^{T}e^{A_{k}|t|}\bm{c}_{k}.
\end{align}

\section{Hierarchy of entropy production and a lower bound on the correlation contribution}

In this section, we derive a hierarchy between the entropy production for the non-Markovian system and that for an embedded system discussed in Sec.~\ref{sec:Markov_embedding}. Then, we derive an embedding-independent lower bound on the correlation contribution to the entropy production.

\subsection{\label{sec:Gaussian_Hierarchy}Embedded entropy-production and hierarchy relation}

The heat current from the residual Markovian bath into $SA$ is defined as
\begin{align}
    \dot Q_{\rm emb}
    :=\int dX\,dP\,d\bm y\,H_{SA}^{\rm tot}\,\mathcal D f_t^{SA},
    \label{eq: note Qemb}
\end{align}
and the entropy production rate is defined as $\dot{\Sigma}_{\rm emb}=\dot{S}_{SA}-\beta \dot{Q}_{\rm emb}$. Using Eq.~\eqref{eq: note dissipator}, we obtain
\begin{align}
    \dot\Sigma_{\rm emb}
    ={}&\beta\int dX\,dP\,d\bm y\,f_t^{SA}\,
    \bm u_t^T D\bm u_t \nonumber 
    \\
    ={}&\beta\int dX\,dP\, f_{t}^{S}(X,P) \int d\bm y\,f_t^{A|S}(\bm{y}|X,P)\,
    \bm u_t^T D\bm u_t \geq0, \label{eq: note Sigma emb quadratic} \\
    \bm u_t
    :={}&(\bm y-\bm cX)
    +\frac{1}{\beta}\nabla_{\bm y}\ln f_t^{SA}=(\bm y-\bm cX)
    +\frac{1}{\beta}\nabla_{\bm y}\ln f_t^{A|S}(\bm{y}|X,P),
    \label{eq: note thermodynamic force}
\end{align}
where $f^{SA}_{t}=f^{S}_{t}(X,P)f^{A|S}_{t}(\bm{y}|X,P)$. 

On the other hand, the non-Markovian entropy production reads
\begin{align}
    \Sigma
    :=\Delta S_S-\beta Q_{\rm sys},
\end{align}
where
\begin{align}
    \dot Q_{\rm sys}
    :=\frac{d}{dt}\langle H_S^{\lambda_t}\rangle
    -\dot\lambda_t\langle\partial_{\lambda_t}H_S^{\lambda_t}\rangle.
\end{align}
Then, we have
\begin{align}
    \dot Q_{\rm sys}
    =\dot Q_{\rm emb}
    -\frac{d}{dt}\langle H_A+H_{\rm int}\rangle= \dot{Q}_{\rm emb}+\beta^{-1}\frac{d}{dt}\langle \ln \pi^{A|S}\rangle .
    \label{eq: sup Qsys Qemb fixed}
\end{align}
This shows
\begin{align}
    \Sigma
    ={}&\Sigma_{\rm emb}
    +D(f_\tau^{SA}\Vert f_\tau^S\pi^{A|S})
    -D(f_0^{SA}\Vert f_0^S\pi^{A|S}).
\end{align}
Under the conditional thermal initial state $f_0^{SA}=f_0^S\pi^{A|S}$, we obtain
\begin{align}
    \Sigma
    =\Sigma_{\rm emb}
    +\Sigma_{\rm cor}
    \geq\Sigma_{\rm emb},
    \label{eq: sup EP decomp}
\end{align}
where 
\begin{align}
    \Sigma_{\rm cor}:=D(f_\tau^{SA}\Vert f_\tau^S\pi^{A|S}). \label{eq:Sigma_mem_sup}
\end{align}
These explicitly show the hierarchy and decomposition of the entropy production presented in the main text for an embedded system discussed in Sec.~\ref{sec:Markov_embedding}.

\subsection{Embedding-independent lower bound on the correlation contribution}\label{sec: embedding-independent lower bound}
We next show an embedding-independent lower bound on the contribution $\Sigma_{\rm cor}$ presented in the main text by assuming $K(0)>0$:
\begin{align}
    \Sigma_{\rm cor}
    \geq
    \Sigma_{F^{\rm NM}}:=\frac{\beta}{2K(0)}
    \int dX\,dP\,
    \left[F_{\tau}^{\rm NM}(X,P)\right]^2
    f_{\tau}^{S}(X,P),
    \label{eq:lower-bound-sigma-mem}
\end{align}

Here, we define the force from the auxiliary modes
to the system as
\begin{align}
    \mathcal{F}_{A}(X,\bm{y})
    &:=
    -\partial_X H_{\rm int}=
    \bm{c}^{T}(\bm{y}-\bm{c}X)=\bm{c}^{T}\bm{z}.
    \label{eq:microscopic-auxiliary-force}
\end{align}
The conditional mean force is defined by
\begin{align}
    F_t^{\rm NM}(X,P)
    &:=
    \int d\bm{y}\,
    \mathcal{F}_{A}(X,\bm{y})
    f_t^{A|S}(\bm{y}|X,P)=
    -\int d\bm{y}\,
    (\partial_XH_{\rm int})
    f_t^{A|S}(\bm{y}|X,P).
    \label{eq:conditional-memory-force}
\end{align}
With this definition, we integrate over $\bm{y}$ in Eq.~\eqref{eq: note Fokker Planck expanded} and obtain the time-evolution equation of the reduced system probability distribution as
\begin{align}
    \partial_t f_t^S(X,P)
    =
    \{H_S^{\lambda_t},f_t^S\}
    -
    \partial_P
    \left[
        F_t^{\rm NM}(X,P)f_t^S(X,P)
    \right].
    \label{eq:reduced-FP-memory-force}
\end{align}

\subsubsection{Equivalent representation directly from the generalized Langevin equation}
We next show that the conditional mean force
$F_t^{\rm NM}(X,P)$ can be defined directly from the generalized
Langevin equation, without referring to a particular Markovian
embedding. This allows us to explicitly show that $\Sigma_{F^{\rm NM}}$ is independent of embedding. We introduce
\begin{align}
    \mathcal{F}_{t}^{\rm GLE}
    &:=
    -\int_{0}^{t}ds\,
    K(t-s)\frac{P(s)}{M}
    +\eta_t 
    \label{eq:GLE-stochastic-bath-force}
\end{align}
and define the conditional mean of the bath force as
\begin{align}
    F_t^{\rm NM}(X,P)
    &:=
    \mathbb E\!\left[
    \left.
    \mathcal{F}_{t}^{\rm GLE}
    \right|
    X(t)=X,\,
    P(t)=P
    \right]
    \nonumber\\
    &=
    \mathbb E\!\left[
    \left.
    -\int_{0}^{t}ds\,
    K(t-s)\frac{P(s)}{M}
    +\eta_t
    \right|
    X(t)=X,\,
    P(t)=P
    \right].
    \label{eq:FNM-GLE-conditional}
\end{align}
Equivalently, using
\begin{align}
    f_t^S(X,P)
    =
    \mathbb E\!\left[
    \delta(X-X(t))\delta(P-P(t))
    \right],
\end{align}
Eq.~\eqref{eq:FNM-GLE-conditional} can be written as
\begin{align}
    F_t^{\rm NM}(X,P)f_t^S(X,P)
    =
    \mathbb E\!\left[
    \mathcal{F}_{t}^{\rm GLE}
    \delta(X-X(t))\delta(P-P(t))
    \right].
    \label{eq:FNM-GLE-delta}
\end{align}
Taking the time derivative of $f_t^S$ and using the generalized Langevin equation~\eqref{sup: GLE}, we obtain
\begin{align}
    \partial_t f_t^S(X,P)
    ={}&
    -\partial_X
    \mathbb E\!\left[
    \dot X(t)
    \delta(X-X(t))\delta(P-P(t))
    \right]-
    \partial_P
    \mathbb E\!\left[
    \dot P(t)
    \delta(X-X(t))\delta(P-P(t))
    \right]
    \nonumber\\
    ={}&
    \{H_S^{\lambda_t},f_t^S\}
    -
    \partial_P
    \left[
    F_t^{\rm NM}(X,P)f_t^S(X,P)
    \right],
    \label{eq:reduced-FP-from-GLE}
\end{align}
which reproduces Eq.~\eqref{eq:reduced-FP-memory-force}.

We finally show explicitly that
Eq.~\eqref{eq:FNM-GLE-conditional} coincides with the definition
based on the Markovian embedding.  From Eqs.~\eqref{eq: note formal z},~\eqref{eq: note K general}, and
\eqref{eq: note eta general}, we obtain
\begin{align}
    \bm c^T\bm z(t)
    &=
    -\int_0^t ds\,
    K(t-s)\frac{P(s)}{M}
    +\eta_t=
    \mathcal{F}_{t}^{\rm GLE}.
    \label{eq:auxiliary-force-equals-GLE-force}
\end{align}
Taking the conditional
average of Eq.~\eqref{eq:auxiliary-force-equals-GLE-force} gives
\begin{align}
    F_t^{\rm NM}(X,P)
    &=
    \mathbb E\!\left[
    \left.
    \mathcal{F}_{t}^{\rm GLE}
    \right|
    X(t)=X,P(t)=P
    \right]
    \nonumber\\
    &=
    \mathbb E\!\left[
    \left.
    \bm c^T\bm z(t)
    \right|
    X(t)=X,P(t)=P
    \right]
    \nonumber\\
    &=
    \int d\bm z\,
    \bm c^T\bm z\,
    f_t^{A|S}(\bm z|X,P),
    \label{eq:FNM-GLE-embedding-equivalence}
\end{align}
which is precisely the conditional mean force defined in
Eq.~\eqref{eq:conditional-memory-force}.

\subsubsection{Derivation}
In what follows, we derive Eq.~\eqref{eq:lower-bound-sigma-mem}. We apply the classical variational inequality for relative entropy~\eqref{Petz ineq1 classical},
\begin{align}
    D(p\Vert q)
    \geq
    \langle h\rangle_p
    -
    \ln\langle e^h\rangle_q, \label{eq:classical_variational_ineq}
\end{align}
with $p=f^{SA}_{\tau}$ and $q=f^{S}_{\tau}\pi^{A|S}$, and
\begin{align}
    h_a(X,P,\bm{y})
    &:=
    aF_\tau^{\rm NM}(X,P)
    \mathcal{F}_{A}(X,\bm{y})    -
    \frac{a^2K(0)}{2\beta}
    \left[F_\tau^{\rm NM}(X,P)\right]^2,
    \label{eq:test-function-memory-bound}
\end{align}
where $a\in\mathbb{R}$. 

By noting that $K(0)=\bm{c}^T\bm{c}$, we complete the square
\begin{align}
-\beta(H_{\rm int}+H_A)
+aF_\tau^{\rm NM}\mathcal{F}_{A}
-\frac{a^2K(0)}{2\beta}(F_\tau^{\rm NM})^2
=
-\frac{\beta}{2}
\left(
    \bm{y}-\bm{c}X
    -
    \frac{a}{\beta}
    F_\tau^{\rm NM}\bm{c}
\right)^T
\left(
    \bm{y}-\bm{c}X
    -
    \frac{a}{\beta}
    F_\tau^{\rm NM}\bm{c}
\right),
\end{align}
and obtain
\begin{align}
    \left\langle e^{h_a}\right\rangle_
    {f_\tau^S\pi^{A|S}}
    =1,
\end{align}
where we use $    \pi^{A|S}(\bm{y}|X)= \exp[
        -\frac{\beta}{2}
        (\bm{y}-\bm{c}X)^{T}(\bm{y}-\bm{c}X)]/Z_{A}=\exp[-\beta(H_{\rm int}+H_{A})]/Z_{A}$.
Moreover, using the definition of the conditional mean force, we have the relation
\begin{align}
    \left\langle
        F_\tau^{\rm NM}\mathcal{F}_{A}
    \right\rangle_{f_\tau^{SA}}
    =
    \int dX\,dP\,
    \left[F_\tau^{\rm NM}(X,P)\right]^2
    f_\tau^S(X,P).
\end{align}
Therefore,  Eq.~\eqref{eq:classical_variational_ineq}  reads
\begin{align}
\Sigma_{\rm cor}=D(f_\tau^{SA}\Vert f_\tau^S\pi^{A|S})
\geq
\left(
    a-\frac{a^2K(0)}{2\beta}
\right)
\int dX\,dP\,
\left[F_\tau^{\rm NM}(X,P)\right]^2
f_\tau^S(X,P). \label{eq:Lowerbound_sigma_mem1}
\end{align}
The right-hand side is maximized at $    a^*=\frac{\beta}{K(0)}$. Therefore, substituting $a=a^{*}$ in Eq.~\eqref{eq:Lowerbound_sigma_mem1} gives
\begin{align}
    \Sigma_{\rm cor}
    \geq
    \frac{\beta}{2K(0)}
    \int dX\,dP\,
    \left[F_\tau^{\rm NM}(X,P)\right]^2
    f_\tau^S(X,P),
\end{align}
which completes the proof of
Eq.~\eqref{eq:lower-bound-sigma-mem}.

\subsubsection{Continuous Classical Derivation of Petz Variational Formula}

The relation~\eqref{eq:classical_variational_ineq} is essentially a classical version of the Petz variational formula~\cite{petz1988variational}. 
For concreteness, we show its classical version
\begin{equation}
    D(p\Vert q)
    \geq
    \int dx\, p(x) h(x)
    -
    \ln \int dx\, q(x)e^{h(x)} .
    \label{Petz ineq1 classical}
\end{equation}
Here, $p(x)$ and $q(x)$ are normalized probability
distributions and $h(x)$ is an arbitrary real-valued function such that $\int dx\, q(x)e^{h(x)}$ is finite.

Let us define $Z_q[h]
    :=
    \int dx\, q(x)e^{h(x)}$ 
and the normalized probability distribution
\begin{align}
    q_h(x)
    :=
    \frac{q(x)e^{h(x)}}{Z_q[h]}.
\end{align}
We then evaluate $D(p\Vert q_h)$ and obtain:
\begin{align}
    D(p\Vert q_h)
    &=
    D(p\Vert q)
    -
    \int dx\,p(x)h(x)
    +
    \ln Z_q[h] \geq 0,
\end{align}
where the last inequality follows from the non-negativity of the relative entropy, and we obtain Eq.~\eqref{Petz ineq1 classical}. From this derivation, we find that the equality of Eq.~\eqref{Petz ineq1 classical} is attained for
$p(x)=q_h(x)$.

\section{Derivation of the entropic bound}
\label{appendix:kernel_only_entropic_bound}

In this section, we derive the entropic bound presented in the main text:
\begin{align}
    \frac{1}{\Theta \Sigma}\Bigl( \int^{\tau}_{0}dt |\mathcal{J}_{\mathcal{O}}(t)| \Bigr)^{2} &\leq 1-\frac{\Sigma_{F^{\rm NM}}}{\Sigma} \leq 1   ,\label{sup: time-dep TUR} 
\end{align}
where 
\begin{align}
    \Theta := \frac{\mathcal{S}_{K}}{\beta}\int^{\tau}_{0}dt \int dXdP (\partial_{P}\mathcal{O}_{t})^{2} f_{t}^{S}(X,P) \label{sup: bound theta}
\end{align}
and
\begin{align}
    \mathcal{S}_{K}
    :=\sup_{\omega\geq 0}\frac{|\widehat K(i\omega)|^2}
    {\operatorname{Re}\widehat K(i\omega)}=\sup_{\omega\in \mathbb{R}}\frac{|\widehat K(i\omega)|^2}
    {\operatorname{Re}\widehat K(i\omega)}.
    \label{eq: sup SK definition}
\end{align}
Here, we adopt the convention that $|\widehat K(i\omega)|^2/\operatorname{Re}\widehat K(i\omega)=0$ when $\widehat K(i\omega)=0$, and $|\widehat K(i\omega)|^2/\operatorname{Re}\widehat K(i\omega)=+\infty$ when $\operatorname{Re}\widehat K(i\omega)=0$ but $\widehat K(i\omega)\neq 0$. 

Note that by using the relation $\widehat K(i\omega)=\int^{\infty}_{0}dt K(t)\cos\omega t - i \int^{\infty}_{0}dt K(t)\sin\omega t$ and $\int^{\infty}_{0}dt K(t)\cos\omega t=J(\omega)/\omega$, we can rewrite Eq.~\eqref{eq: sup SK definition} to obtain
\begin{align}
    \mathcal{S}_{K}
    =\sup_{\omega\geq0}
    \frac{
    [J(\omega)/\omega]^2
    +\left[\int_0^\infty dt\,K(t)\sin(\omega t)\right]^2
    }
    {J(\omega)/\omega}.
    \label{eq: sup SK spectral density}
\end{align}

\subsection{Finiteness of the coefficient and a special case}

We now show a sufficient condition for $\mathcal{S}_{K}<\infty$. Let us assume 
\begin{align}
    \frac{J(\omega)}{\omega}>0 \quad \text{for every finite } \omega \geq 0, \label{sup: Jomega_cond}
\end{align}
and 
\begin{align}
    0<\lim_{\omega\rightarrow\infty }\omega J(\omega)<\infty. \label{sup: assump omegaJ}
\end{align}

We first note that because $\widehat K(s)$ is a stable rational function, it has no poles on the imaginary axis. Therefore, $\widehat K(i\omega)$ is continuous as a function of $\omega\in\mathbb R$. 
Then,  $A(\omega):=J(\omega)/\omega=\operatorname{Re}\widehat{K}(i\omega)$ and $B(\omega):=\int^{\infty}_{0}dt K(t)\sin\omega t=-\operatorname{Im}\widehat{K}(i\omega)$ are continuous functions. We also define
\begin{align}
    R(\omega):= \frac{A(\omega)^{2}+B(\omega)^{2}}{A(\omega)}, 
\end{align}
such that $\mathcal{S}_{K}=\sup_{\omega\geq 0}R(\omega)$. 

Now, for any finite $\Omega>0$, the continuity of $A(\omega)$ and $B(\omega)$ with  $A(\omega)>0$ on $[0,\Omega]$ implies
\begin{align}
    \sup_{0\leq \omega\leq \Omega}R(\omega)<\infty. \label{sup: supR bound}
\end{align}  

Next, we consider the  high-frequency expansion of $\widehat{K}(i\omega)$ as 
\begin{align}
    \widehat K(i\omega)
    &=\bm{c}^{T}(i\omega I-A)^{-1}\bm{c} \nonumber \\
    &= -i\frac{\bm{c}^{T}\bm{c}}{\omega}-\frac{\bm{c}^{T}A\bm{c}}{\omega^{2}}+ i\frac{\bm{c}^{T}A^{2}\bm{c}}{\omega^{3}}
    +O(\omega^{-4})\nonumber \\
    &= -i\frac{K(0)}{\omega}-\frac{\dot{K}(0^{+})}{\omega^{2}}+\frac{i\ddot{K}(0^{+})}{\omega^{3}}+O(\omega^{-4}).
\end{align}
Then, we have $A(\omega)=-\dot{K}(0^{+})/\omega^{2}+O(\omega^{-4})$ and $B(\omega)=K(0)/\omega + O(\omega^{-3})$. 
We use the relation $-\dot{K}(0^{+})=\lim_{\omega\rightarrow\infty}\omega J(\omega)$ and Eq.~\eqref{sup: assump omegaJ} and find that $0<-\dot{K}(0^{+})<\infty$. 
This shows that 
\begin{align}
   \lim_{\omega\rightarrow\infty} R(\omega)= \frac{K(0)^{2}}{-\dot{K}(0^{+})}<\infty . \label{sup: supR inf}
\end{align}
Combining Eqs.~\eqref{sup: supR bound} and~\eqref{sup: supR inf} yields $\mathcal{S}_{K}=\sup_{\omega\geq 0}R(\omega)<\infty$.

We next consider the lower bound
\begin{align}
    \mathcal{S}_{K}\geq \widehat K(0)=\lim_{\omega\rightarrow 0}\frac{J(\omega)}{\omega}=\int^{\infty}_{0}dt K(t). \label{sup: lower bound S od}
\end{align}
Equality is attained by the overdamped Drude-Lorentz memory kernel $K^{\rm od}(t)=\sum_k a_ke^{-\Gamma_k|t|}$, $a_k>0, \, \Gamma_k>0$, where $\int^{\infty}_{0}dt K^{\rm od}(t)=\sum_{k}(a_{k}/\Gamma_{k})$ and $\widehat K^{\rm od}(i\omega)
    =
    \sum_k a_k/(\Gamma_k+i\omega)$. 
The Cauchy--Schwarz inequality gives
\begin{align}
    |\widehat K^{\rm od}(i\omega)|^2
    &\leq
    \left(\sum_k\frac{a_k}{\Gamma_k}\right)
    \left(\sum_k
        \frac{a_k\Gamma_k}{\Gamma_k^2+\omega^2}
    \right)=
    \left(\sum_k\frac{a_k}{\Gamma_k}\right)
    \operatorname{Re}\widehat K^{\rm od}(i\omega),
\end{align}
and equality is attained at $\omega=0$. Hence, the equality of~\eqref{sup: lower bound S od} is attained. 


\subsection{\label{appendix: proof thm 2}Fixed-embedding entropic bound}

Let $\mathfrak E(K)$ denote the set of Markovian embeddings $\mathcal E=(n,J,D,\bm c)$ satisfying the Langevin equation discussed in Sec.~\ref{sec:Markov_embedding_Langevin} and reproducing the prescribed kernel, $K_{\mathcal E}=K$. First, we derive an entropic bound with the coefficient $\mathcal{S}(\mathcal E)$ for a given embedding $\mathcal E$.

We start by deriving a lower bound on $\dot{\Sigma}_{\rm emb}$. We introduce 
\begin{align}
    \bm m_t(X,P)
    :=\int d\bm z\,\bm z\,f_t^{A|S}(\bm z|X,P)
    \label{eq: sup m conditional}
\end{align}
as the conditional mean displacement of the auxiliary coordinates.  

Applying Jensen's inequality to Eq.~\eqref{eq: note Sigma emb quadratic} and using $\bm{z}=\bm{y}-\bm{c}X$, we obtain
\begin{align}
    \dot\Sigma_{\rm emb}\geq
    \beta\int dX\,dP\,
    \bm m_t^TD\bm m_t\,f_t^S(X,P),
    \label{eq: sup EP Jensen general}
\end{align}
by noting that $\int d\bm z\,f_t^{A|S}\nabla_{\bm z}\ln f^{A|S}_t=0$.

The non-Markovian force appearing in the reduced Fokker-Planck equation can be expressed using Eq.~\eqref{eq: sup m conditional} as
\begin{align}
    F_t^{\rm NM}(X,P)
    =\bm c^T\bm m_t(X,P).
    \label{eq: sup FNM general}
\end{align}
When $\bm c\in\operatorname{Ran}D$, the Cauchy-Schwarz inequality associated with the positive semidefinite matrix $D$ yields
\begin{align}
    \left(F_t^{\rm NM}\right)^2
    \leq
    \left(\bm c^TD^{+}\bm c\right)
    \left(\bm m_t^TD\bm m_t\right),
\end{align}
where $D^{+}$ denotes the Moore-Penrose pseudoinverse of $D$. 
Combining this with Eq.~\eqref{eq: sup EP Jensen general} gives
\begin{align}
    \int dX\,dP\,
    (F_t^{\rm NM})^2f_t^S
    \leq
    \frac{\mathcal{S}(\mathcal E)}{\beta}
    \dot\Sigma_{\rm emb},
    \label{eq: sup force EP general}
\end{align}
where 
\begin{align}
    \mathcal{S}(\mathcal E)
    :=
    \begin{cases}
        \bm c^TD^+\bm c,
        &\bm c\in\operatorname{Ran}D,\\[1mm]
        +\infty,
        &\bm c\notin\operatorname{Ran}D.
    \end{cases}
\end{align}

For an observable $\mathcal O_t(X,P)$, we define the bath-induced current as
\begin{align}
    \mathcal J_{\mathcal O}(t)
    &:=-\int dX\,dP\,\mathcal O_t
    \partial_P\left(F_t^{\rm NM}f_t^S\right)
    =\int dX\,dP\,
    (\partial_P\mathcal O_t)F_t^{\rm NM}f_t^S.
    \label{eq: sup bath induced current general}
\end{align}
Using the Cauchy-Schwarz inequality and Eq.~\eqref{eq: sup force EP general}, we obtain
\begin{align}
    |\mathcal J_{\mathcal O}(t)|^2
    \leq
    \frac{\mathcal{S}(\mathcal E)}{\beta}
    \left\langle(\partial_P\mathcal O_t)^2\right\rangle_t
    \dot\Sigma_{\rm emb}(t).
    \label{eq: sup instantaneous fixed embedding}
\end{align}
We then apply a second Cauchy-Schwarz inequality in time and obtain
\begin{align}
    \left(\int_0^\tau dt\,|\mathcal J_{\mathcal O}(t)|\right)^2
    &\leq
    \frac{\mathcal{S}(\mathcal E)}{\beta}
    \left[\int_0^\tau dt\,
    \left\langle(\partial_P\mathcal O_t)^2\right\rangle_t\right]
    \Sigma_{\rm emb}
    \nonumber\\
    &=\frac{\mathcal{S}(\mathcal E)}{\beta}
    \left[\int_0^\tau dt\,
    \left\langle(\partial_P\mathcal O_t)^2\right\rangle_t\right]
    (\Sigma-\Sigma_{\rm cor})
    \nonumber\\
    &\leq
    \frac{\mathcal{S}(\mathcal E)}{\beta}
    \left[\int_0^\tau dt\,
    \left\langle(\partial_P\mathcal O_t)^2\right\rangle_t\right]
    (\Sigma-\Sigma_{F^{\rm NM}}) \nonumber \\
    &\leq
    \frac{\mathcal{S}(\mathcal E)}{\beta}
    \left[\int_0^\tau dt\,
    \left\langle(\partial_P\mathcal O_t)^2\right\rangle_t\right]
    \Sigma. 
    \label{eq: sup fixed embedding entropic bound}
\end{align}

\subsection{Lower bound on the embedding coefficient}

Next, we show an embedding-independent lower bound on $\mathcal{S}(\mathcal E)$. 
For $\omega\in \mathbb{R}$, we use $\bm{x}_{\omega}$ defined in Eq.~\eqref{sup: xomega} and express $\widehat{K}(i\omega)=\bm{c}^{T}\bm{x}_{\omega}$. 
If $\bm{c}\in \mathrm{Ran}D$, we apply the Cauchy-Schwarz inequality and obtain
\begin{align}
    |\widehat K(i\omega)|^2
    \leq
    (\bm c^TD^{+}\bm c)
    (\bm x_\omega^\dagger D\bm x_\omega)
    =\mathcal{S}(\mathcal E)
    \operatorname{Re}\widehat K(i\omega), \label{eq: K frequency}
\end{align}
where we use Eq.~\eqref{sup: Komega positive}. 
Because the above inequality holds for arbitrary $\omega\in\mathbb{R}$, we obtain 
\begin{align}
    \mathcal{S}(\mathcal E)\geq\mathcal{S}_{K}
    \qquad \forall\mathcal E\in\mathfrak E(K),
    \label{eq: sup universal lower SK}
\end{align}
where we used the definition~\eqref{eq: sup SK definition}. If $\bm c\notin\operatorname{Ran}D$, Eq.~\eqref{eq: sup universal lower SK} holds trivially because $\mathcal{S}(\mathcal E)=+\infty$.
From Eq.~\eqref{eq: sup universal lower SK}, we also obtain 
\begin{align}
    \inf_{\mathcal E\in\mathfrak E(K)}\mathcal{S}(\mathcal E)
    \geq\mathcal{S}_{K}.
\end{align}

\subsection{Attainability of the lower bound}
\label{sec:construction}

We next show that the lower bound~\eqref{eq: sup universal lower SK}
is attainable within the class of Markovian embeddings
$\mathfrak E(K)$ considered in Sec.~\ref{sec:Markov_embedding_Langevin}.
We assume
$0<\mathcal{S}_{K}<\infty$ and define $\rho_{*}=1/\mathcal{S}_{K}$.
From the definition of $\mathcal{S}_{K}$, the following inequality holds:
\begin{align}
    \operatorname{Re}\widehat K(i\omega)
    -\rho_{*}|\widehat K(i\omega)|^2\geq0,
    \qquad \forall\omega\in\mathbb R.
    \label{eq: sup K frequency inequality}
\end{align}
Therefore, condition (i) in Sec.~\ref{sec: sup KYP lemma} holds for $\rho=\rho_{*}$. Applying the construction in Sec.~\ref{sup sec: consequences when the statement}, we obtain a thermal linear Gaussian embedding $\mathcal{E}_{*}=(n_{\rm m},J_{\rho_{*}},D_{\rho_{*}},\bm{c}_{\rho_{*}})\in \mathfrak E(K)$, where 
\begin{align}
    \widehat K(s)
    &=\bm c_{\rho_{*}}^T(sI_{n_{\rm m}}-A_{\rho_{*}})^{-1}\bm c_{\rho_{*}},
    \label{eq: sup optimal collocated realization}
\end{align}
with $A_{\rho_{*}}=J_{\rho_{*}}-D_{\rho_{*}}$, $J_{\rho_{*}}^{T}=-J_{\rho_{*}}$, $D_{\rho_{*}}^{T}=D_{\rho_{*}}\succeq \rho_{*}\bm{c}_{\rho_{*}}\bm{c}_{\rho_{*}}^{T}\succeq 0$, and $A_{\rho_{*}}$ is Hurwitz. 

We finally show that the embedding $\mathcal{E}_{*}$ attains the lower bound in Eq.~\eqref{eq: sup universal lower SK}.

We start by noting that for any $\bm v\in\ker D_{\rho_{*}}$, we have $0=\bm{v}^{T}D_{\rho_{*}}\bm{v}\geq \rho_{*}(\bm{v}^{T}\bm{c}_{\rho_{*}})^{2} $. Since $\rho_{*}>0$, we have $\bm{v}^{T}\bm{c}_{\rho_{*}}=0$. Therefore, $\bm{c}_{\rho_{*}}$ is orthogonal to $\ker D_{\rho_{*}}$.
Since $D_{\rho_{*}}$ is real and symmetric,
$(\ker D_{\rho_{*}})^\perp=\operatorname{Ran}D_{\rho_{*}}$, and hence
\begin{align}
\bm{c}_{\rho_{*}}\in \operatorname{Ran}D_{\rho_{*}}. \label{sup: crho Ran}
\end{align}

We next define $\bm u:=D^{+}_{\rho_{*}}\bm{c}_{\rho_{*}}$. Then, we use Eq.~\eqref{sup: crho Ran} and obtain $D_{\rho_{*}}\bm u=\bm{c}_{\rho_{*}}$. This gives
\begin{align}
    \bm{c}^{T}_{\rho_{*}}D^{+}_{\rho_{*}}\bm{c}_{\rho_{*}}
    &=\bm u^TD_{\rho_{*}}\bm u
    \geq\rho_{*}(\bm u^T\bm{c}_{\rho_{*}})^2
    =\rho_{*}\left(\bm c^T_{\rho_{*}}D^{+}_{\rho_{*}}\bm c_{\rho_{*}}\right)^2. \label{eq: CTDc}
\end{align}
This shows
\begin{align}
    \mathcal{S}(\mathcal{E}_{*})=\bm{c}_{\rho_{*}}^{T}D^{+}_{\rho_{*}}\bm{c}_{\rho_{*}} \leq \frac{1}{\rho_{*}}=\mathcal{S}_{K}.
\end{align}
On the other hand, Eq.~\eqref{eq: sup universal lower SK}  gives
$\mathcal{S}(\mathcal E_*)\geq\mathcal{S}_{K}$.
Combining the two inequalities, we obtain
\begin{align}
        \mathcal{S}(\mathcal E_*)
        =\min_{\mathcal E\in\mathfrak E(K)}\mathcal{S}(\mathcal E)
        =\mathcal{S}_{K}.
    \label{eq: sup optimal coefficient}
\end{align}
Applying the bound~\eqref{eq: sup fixed embedding entropic bound} for $\mathcal{E}_{*}$ and using $\mathcal{S}(\mathcal{E}_{*})=\mathcal{S}_{K}$ completes the derivation of Eq.~\eqref{sup: time-dep TUR}.

\section{Extension by coefficient-preserving approximation}
\label{sec: coefficient preserving extension}

In this section, we extend the entropic-bound framework in Sec.~\ref{appendix:kernel_only_entropic_bound} beyond the finite-dimensional rational class by constructing a sequence of finite-dimensional thermal linear Gaussian embeddings whose memory kernels preserve the coefficient $\mathcal{S}_{K}$.
In the following, we state and prove this result.

\begin{prop}[Coefficient-preserving approximation]
\label{prop: coefficient preserving approximation}
Let $K:\mathbb R\to\mathbb R$ be continuous, even, and positive
definite, with
\begin{align}
    K\in L^1(0,\infty),
    \quad 0<K(0)<\infty,
    \quad 0<\mathcal{S}_{K}<\infty.
    \label{eq: cp kernel assumptions}
\end{align}
There exists a sequence of positive-definite kernels $K_N$
admitting finite-dimensional thermal linear Gaussian Markovian
embeddings of the form introduced in
Sec.~\ref{sec:Markov_embedding_Langevin} such that
\begin{align}
    &K_N(0)=K(0),
    \quad
    \mathcal{S}_{K_N}=\mathcal{S}_{K}
    \quad \text{for every }N,
    \label{eq: cp approximation statement coefficient}
    \\
    &\lim_{N\to\infty}
    \sup_{|t|\leq T}|K_N(t)-K(t)|=0
    \quad \text{for every }0<T<\infty.
    \label{eq: cp approximation statement convergence}
\end{align}
\end{prop}

For every $N$, the entropic bound applies to the kernel $K_N$
with exactly the same coefficient,
$\mathcal{S}_{K_N}=\mathcal{S}_{K}$.
Moreover, the sequence $K_N$ approximates the target memory kernel
$K$ with arbitrary accuracy on any fixed finite time interval.
Thus, Proposition~\ref{prop: coefficient preserving approximation} provides a coefficient-preserving finite-dimensional approximation for kernels beyond the finite-dimensional rational class.

Note that to transfer the corresponding entropic bound to the dynamics generated by the limiting kernel $K$, one may additionally need appropriate conditions ensuring convergence of the generalized Langevin dynamics and of the quantities entering the bound.

\subsection{Positive-real representation}
We set $G(s):=\widehat K(s)$ and $\rho_*:=1/\mathcal S_K$.
Since $K\in L^1(0,\infty)$, its Laplace transform $G(s)=\int_{0}^{\infty}dt e^{-st}K(t)$ is bounded and analytic for $\operatorname{Re}s>0$.
Next, let $s_{n}\to i\omega$ with $\operatorname{Re}s_{n}\geq 0$. Then, we have $e^{-s_{n}t}K(t)\to e^{-i\omega t}K(t)$ and $|e^{-s_{n}t}K(t)|\leq |K(t)|$. Because $K\in L^1(0,\infty)$, we can apply the dominated convergence theorem and obtain $G(s_{n})\rightarrow \int_{0}^{\infty}dt e^{-i\omega t}K(t)=: G(i\omega)$. Therefore, $G$ extends continuously to the imaginary axis. 
We also note that, because $K$ is real, we have $G(\bar s)=\overline{G(s)}$.

Since $K$ is positive definite, Bochner's theorem gives  $\operatorname{Re}G(i\omega)\geq 0$.
The definition of $\mathcal S_K$ then gives $    \operatorname{Re}G(i\omega)    -\rho_*|G(i\omega)|^2    \geq0$ for all $   \omega\in\mathbb R$.
If we define
\begin{align}
    \Phi(s):=1-2\rho_*G(s),
\end{align}
we have $|\Phi(i\omega)|^{2}=1-4\rho_{*}[\operatorname{Re}G(i\omega)-\rho_{*}|G(i\omega)|^{2}]$. Therefore, the above condition is equivalent to
\begin{align}
    |\Phi(i\omega)|\leq1. \label{eq: boundary phi bound}
\end{align}
Since $G$ is bounded in the right half-plane, $\Phi$ is also bounded there. 
Therefore, $\Phi$ is bounded and analytic in the right half-plane, continuously extends to the imaginary axis, and satisfies Eq.~\eqref{eq: boundary phi bound}. For $\epsilon>0$, we apply the maximum modulus theorem to $\Phi(s)e^{-\epsilon\sqrt{s}}$ on a right half-disk $D_{R}:=\{s:\operatorname{Re}s>0,|s|<R\}$. Taking first $R\rightarrow\infty$ and then $\epsilon\to 0^{+}$, we obtain  
\begin{align}
    |\Phi(s)|\leq1,
    \qquad
    \operatorname{Re}s>0.
    \label{eq: cp Phi bounded}
\end{align}

Using the change of variable $u=rt$, we have
\begin{align}
    rG(r)
    &=
    \int_0^\infty du\,e^{-u}K(u/r).
\end{align}
By taking $r\to\infty$, we have $u/r\to 0$ for fixed $u$ and $K(u/r)\to K(0)$. 
Noting that positive definiteness of $K$ implies
$|K(t)|\leq K(0)$, we therefore obtain $|e^{-u}K(u/r)|\leq e^{-u}K(0)$, where the right-hand side is integrable on $[0,\infty)$. We now apply the dominated convergence theorem and obtain
\begin{align}
\lim_{r\to\infty}rG(r)
    &=
    \int_0^\infty du\,e^{-u} \lim_{r\to\infty}K(u/r)=  K(0)>0.
    \label{eq: cp large r G}
\end{align}
Equation~\eqref{eq: cp large r G} implies that $\lim_{r\to\infty}G(r)=0$. Therefore, if $G$ is constant, $G\equiv 0$. However, this contradicts Eq.~\eqref{eq: cp large r G}.
Therefore, $G$ is nonconstant. Since $\rho_{*}>0$, this also implies that $\Phi$ is nonconstant.

Now suppose that $|\Phi(s_{0})|=1$ for some $s_{0}$ with $\operatorname{Re}s_{0}>0$. Together with Eq.~\eqref{eq: cp Phi bounded}, this means that $|\Phi|$ attains its maximum value at an interior point. 
The maximum modulus theorem then implies that $\Phi$ is constant, which contradicts the fact that $\Phi$ is nonconstant. Therefore, we conclude that
\begin{align}
    |\Phi(s)|<1,
    \qquad \operatorname{Re}s>0.
\end{align}
This means $1-\rho_*G(s)=(1+\Phi(s))/2\neq0$ throughout the right half-plane.

We can thus define
\begin{align}
    G_{\rm c}(s)
    :=
    \frac{G(s)}{1-\rho_*G(s)}.
    \label{eq: cp conservative transform}
\end{align}
It is analytic in the right half-plane and satisfies
\begin{align}
    \operatorname{Re}G_{\rm c}(s)
    &=
    \frac{
    \operatorname{Re}G(s)-\rho_*|G(s)|^2
    }{
    |1-\rho_*G(s)|^2
    }
    \geq0,\quad 
    G_{\rm c}(\bar s)
    =
    \overline{G_{\rm c}(s)},
    \quad
    \lim_{r\to\infty}rG_{\rm c}(r)
    =
    K(0).
    \label{eq: cp positive real transform}
\end{align}

We now use Cauer's representation for positive-real functions [Appendix A, Eq.~(A3) of Ref.~\cite{PhysRevB.110.024307}. Since the integrand is even in the spectral variable, we write it equivalently with a nonnegative Borel measure $\nu$ on $[0,\infty)$]. 
Therefore, if $F$ is analytic in the right half-plane, satisfies $\operatorname{Re}F(s)\geq0$, and obeys the real symmetry condition $F(\bar{s})=\overline{F(s)}$, then
\begin{align}
    F(s)=bs + \int_{[0,\infty)}\frac{s}{s^{2}+\Omega^{2}}\nu(d\Omega), \label{eq: cp Herglotz representation2}
\end{align}
where $b\geq 0$ and $\nu$ is a nonnegative Borel measure satisfying $\int_{[0,\infty)}\nu(d\Omega)/(1+\Omega^{2})<\infty$.

We now apply Eq.~\eqref{eq: cp Herglotz representation2} to $F=G_{\rm c}$.
Note that $G_{\rm c}(\bar{s})=\overline{G_{\rm c}(s)}$ and $\operatorname{Re}G_{\rm c}(s)\geq0$ from Eq.~\eqref{eq: cp positive real transform}. 
Taking $s=r>0$ in Eq.~\eqref{eq: cp Herglotz representation2} and multiplying by $r$ gives
\begin{align}
    rG_{\rm c}(r)=    br^{2}+\int_{[0,\infty)}
\frac{r^{2}}{r^{2}+\Omega^{2}}
\nu(d\Omega),
    \label{eq: cp Herglotz positive axis}
\end{align}
where both terms on the right-hand side are nonnegative. Since $\lim_{r\to\infty}rG_{\rm c}(r)= K(0)<\infty$, the finite limit implies $b=0$. We further note that $r^{2}/(r^{2}+\Omega^{2})\uparrow1 \ (r\rightarrow \infty)$ for every $\Omega\geq 0$. Therefore, we can apply the monotone convergence theorem and obtain
\begin{align}
\nu([0,\infty))
=\int_{[0,\infty)} \lim_{r\to\infty}\frac{r^{2}}{r^{2}+\Omega^{2}}\nu(d\Omega) = \lim_{r\to\infty}\int_{[0,\infty)}\frac{r^{2}}{r^{2}+\Omega^{2}}\nu(d\Omega)=\lim_{r\to\infty}rG_{\rm c}(r)=K(0).
\label{eq: cp spectral measure mass}
\end{align}
Therefore, $\nu$ is a finite nonnegative measure. 
Equation
\eqref{eq: cp Herglotz representation2} therefore reduces to
\begin{align}
G_{\rm c}(s)
=
\int_{[0,\infty)}
\frac{s}{s^{2}+\Omega^{2}}
\nu(d\Omega),
\qquad
\nu([0,\infty))=K(0).
\label{eq: cp finite spectral measure}
\end{align}

We define the corresponding kernel by
\begin{align}
K_{\rm c}(t)
:=
\int_{[0,\infty)}
\cos(\Omega t)
\nu(d\Omega).
\label{eq: cp conservative kernel}
\end{align}
Since $\nu$ is a finite nonnegative measure, $K_{\rm c}$ is continuous, real, even, and positive definite by Bochner's theorem.
Moreover, we have $|K_{\rm c}(t)|\leq
\nu([0,\infty))=K_{\rm c}(0)=K(0)$.
For $\operatorname{Re}s>0$, the Laplace transform of $K_{\rm c}$ is given by $\widehat{K}_{\rm c}(s)=\int_0^\infty  e^{-st} \Bigl[ \int_{[0,\infty)}
    \nu(d\Omega)\cos(\Omega t)\Bigr] dt$. This double integral is  absolutely convergent because
\begin{align}
\int_{[0,\infty)}\nu(d\Omega)\int^{\infty}_{0}dt|e^{-st} \cos(\Omega t)| \leq \int_{[0,\infty)}\nu(d\Omega)\int^{\infty}_{0}dt e^{-t\operatorname{Re}s}=\frac{\nu([0,\infty))}{\operatorname{Re}s}<\infty.
\end{align}
Therefore, Fubini's theorem gives
\begin{align}
    \widehat K_{\rm c}(s)
    &=
    \int_{[0,\infty)}
    \nu(d\Omega)
    \int_0^\infty dt\,
    e^{-st}\cos(\Omega t)=
    \int_{[0,\infty)}
    \frac{s}{s^2+\Omega^2}\,
    \nu(d\Omega)
    =
    G_{\rm c}(s).
\end{align}

\subsection{Finite-dimensional spectral discretization}

We now approximate $K_{\rm c}(t)$ by $K_{{\rm c},N}(t)$.

For each $N$, we choose a frequency cutoff $R_N>0$ and partition $(0,R_N]$ into finitely many disjoint intervals $I_{j,N}$ with $h_N=\max_{j}\operatorname{diam}(I_{j,N})$ with $\operatorname{diam}(I)=\sup_{\omega,\omega'\in I}|\omega-\omega'|$.
For each interval of nonzero $\nu$-measure, we define
\begin{align}
    a_{j,N}:=\nu(I_{j,N})>0
\end{align}
and choose a representative frequency
$\Omega_{j,N}\in I_{j,N}$.
We also define the zero-frequency weight
\begin{align}
    a_{0,N}
    :=
    \nu(\{0\})+\nu((R_N,\infty)).
\end{align}
We then define
\begin{align}
    K_{{\rm c},N}(t)
    :=
    a_{0,N}
    +\sum_j a_{j,N}\cos(\Omega_{j,N}t),
    \qquad
    a_{0,N}+\sum_j a_{j,N}=K(0),
\end{align}
which approximates $K_{\rm c}$. 
Now let us wirte the difference between $K_{{\rm c},N}(t)$ and $K_{\rm c}(t)$ as
\begin{align}
    K_{{\rm c},N}(t)-K_{\rm c}(t)=\sum_{j}\int_{I_{j,N}}[\cos(\Omega_{j,N}t)-\cos(\omega t)]\nu(d\omega) + \int_{(R_{N},\infty)}[1-\cos(\omega t)]\nu(d\omega).
\end{align}
For $\omega\in I_{j,N}$ and $|t|\leq T$,
we have $|\omega-\Omega_{j,N}|\leq h_N$, and therefore $|\cos(\Omega_{j,N}t)-\cos(\omega t)|\leq |t||\Omega_{j,N}-\omega|\leq Th_{N}$. Combined with $\sum_{j}\nu(I_{j,N})\leq \nu((0,R_{N}])\leq K(0)$ and $\int_{(R_{N},\infty)}[1-\cos(\omega t)]\nu(d\omega)\leq 2 \nu((R_{N},\infty))$, we obtain
\begin{align}
    \epsilon_N(T)
    &:=
    \sup_{|t|\leq T}
    |K_{{\rm c},N}(t)-K_{\rm c}(t)|
    \leq
    2\nu((R_N,\infty))
    +K(0)Th_N.
    \label{eq: cp conservative approximation}
\end{align}
We choose the approximation scheme such that
\begin{align}
    R_N\longrightarrow\infty, \qquad h_N\longrightarrow 0. \label{eq: cp refinement conditions}
\end{align}
Then, 
\begin{align}
    \epsilon_{N}(T) \longrightarrow 0 \qquad (N\to\infty)
\end{align}
for every fixed $T<\infty$. 

\subsection{Finite-dimensional thermal realization and exact coefficient preservation}

We now convert the kernel $K_{{\rm c},N}$ into a finite-dimensional thermal kernel~\eqref{eq: note K Laplace}.

For each positive frequency $\Omega_{j,N}$, we define
\begin{align}
    J_{j,N}
    :=\begin{pmatrix}0&\Omega_{j,N}\\-\Omega_{j,N}&0\end{pmatrix},
    \qquad
    \bm c_{j,N}
    :=\sqrt{a_{j,N}}\begin{pmatrix}1\\0\end{pmatrix}.
\end{align}
If $a_{0,N}>0$, we represent the zero-frequency contribution
by the one-dimensional block
\begin{align}
    J_{0,N}:=0,
    \qquad
    c_{0,N}:=\sqrt{a_{0,N}}.
\end{align}
We then set
\begin{align}
    J_N
    &:=
    J_{0,N}\oplus\bigoplus_j J_{j,N},
    \qquad
    \bm c_N
    :=
    \operatorname{col}
    \left(
        c_{0,N},
        \{\bm c_{j,N}\}_j
    \right),
\end{align}
where the zero-frequency block and the corresponding entry
$c_{0,N}$ are omitted when $a_{0,N}=0$.
We further define
\begin{align}
    D_N:=\rho_*\bm c_N\bm c_N^T,
    \qquad
    A_N:=J_N-D_N,
    \qquad
    \bm c_N^T\bm c_N=K(0),
    \label{eq: cp rank one damping}
\end{align}
where the matrix $A_N$ is Hurwitz. Indeed, if
$A_N\bm v=\lambda\bm v$, then $   \operatorname{Re}\lambda\,\|\bm v\|^2
    =
    -\rho_*|\bm c_N^T\bm v|^2
    \leq0$. 
Now, suppose $\bm c_N^T\bm v=0$. Then, we have
$J_N\bm v=\lambda\bm v$.
Since the positive frequencies are distinct, an eigenvector of
$J_N$ with eigenvalue $\pm i\Omega_{j,N}$ is supported on a
single frequency block. Within that block, it is proportional to
$(1,\pm i)^T$, and therefore
\begin{align}
    \bm c_{j,N}^T
    \begin{pmatrix}
        1\\
        \pm i
    \end{pmatrix}
    =
    \sqrt{a_{j,N}}
    \neq0.
\end{align}
If the zero-frequency block is present, it is one dimensional and
$c_{0,N}=\sqrt{a_{0,N}}\neq0$.
Hence every nonzero eigenvector of $J_N$ has nonzero overlap with
$\bm c_N$, contradicting $\bm c_N^T\bm v=0$.
Thus, $\operatorname{Re}\lambda<0$ holds for every eigenvalue of $A_N$.

We now define the final approximation kernel as
\begin{align}
    K_N(t):=\bm c_N^Te^{A_N|t|}\bm c_N,
\end{align}
Its one-sided Laplace transform is given by 
\begin{align}
    G_N(s)
    &:=\bm c_N^T(sI_{n_N}-A_N)^{-1}\bm c_N
    =\frac{G_{{\rm c},N}(s)}{1+\rho_*G_{{\rm c},N}(s)},
    \nonumber\\
     G_{{\rm c},N}(s)
    &=\bm c_N^T(sI_{n_N}-J_N)^{-1}\bm c_N
    =
    \frac{a_{0,N}}{s}
    +
    \sum_j a_{j,N}\frac{s}{s^2+\Omega_{j,N}^2},
    \label{eq: cp finite transfer}
\end{align}
where $n_N$ denotes the state dimension. Therefore, $K_N$ belongs to the
finite-dimensional rational kernel class generated by the thermal
linear Gaussian embeddings of
Sec.~\ref{sec:Markov_embedding_Langevin}.
Since $A_N$ is Hurwitz, $K_N$ is integrable.
With $\bm x_{N,\omega}:=(i\omega I_{n_N}-A_N)^{-1}\bm c_N$,
Eq.~\eqref{sup: Komega positive} gives
\begin{align}
    \operatorname{Re}G_N(i\omega)
    =\bm x_{N,\omega}^{\dagger}D_N\bm x_{N,\omega}
    =\rho_*|G_N(i\omega)|^2\geq0.
    \label{eq: cp exact frequency identity}
\end{align}
Since $K_N$ is real and even, its Fourier transform is
$2\operatorname{Re}G_N(i\omega)\geq0$.
Hence, by Bochner's theorem, $K_N$ is positive definite.
Also, note that $K_N(0)=\bm c_N^T\bm c_N=K(0)>0$, so $G_N$ is not
identically zero. Equation~\eqref{eq: cp exact frequency identity}
then shows that the quotient defining $\mathcal{S}_{K_N}$ equals
$\rho_*^{-1}$ whenever $G_N(i\omega)\neq0$, and equals zero at
their simultaneous zeros. Therefore,
\begin{align}
    \mathcal{S}_{K_N}
    =
    \rho_*^{-1}
    =
    \mathcal{S}_{K}.
    \label{eq: cp coefficient preservation}
\end{align}

\subsection{Uniform convergence of the memory kernel}

It remains to prove that the finite-dimensional kernels $K_{N}$ converge to the target kernel $K$ on every prescribed finite time interval. 

Using Eqs.~\eqref{eq: cp conservative transform}
and~\eqref{eq: cp finite transfer} yields
\begin{align}
    1-\rho_{*}G(s)=\frac{1}{1+\rho_{*}G_{\rm c}(s)}, \quad 1-\rho_{*}G_{N}(s)=\frac{1}{1+\rho_{*}G_{{\rm c},N}(s)}.
\end{align}
This leads to 
\begin{align}
    G_{N}(s)-G(s)&=(1-\rho_{*}G_{N}(s))(1-\rho_{*}G(s))(G_{{\rm c},N}(s)-G_{\rm c}(s)) \nonumber \\
    &= \widehat{d}_{N}(s)-\rho_{*}(G_{N}(s)+G(s))\widehat{d}_{N}(s)+ \rho_{*}^{2}G_{N}(s)G(s)\widehat{d}_{N}(s), \label{eq: GN-G relation}
\end{align}
where $\widehat{d}_{N}(s):=G_{{\rm c},N}(s)-G_{\rm c}(s)$. 
Taking the inverse Laplace transform of Eq.~\eqref{eq: GN-G relation}, we obtain
\begin{align}
    K_{N}-K=d_{N}-\rho_{*}(K_{N}+K)*d_{N}+\rho_{*}^{2}K_{N}*K*d_{N}, \label{eq: KN-K relation}
\end{align}
where $(f*g)(t):=\int_0^t ds\,f(t-s)g(s)$ is the convolution and $d_N:=K_{{\rm c},N}-K_{\rm c}$. We now evaluate each term that appears in Eq.~\eqref{eq: KN-K relation}. Using Eq.~\eqref{eq: cp conservative approximation}, $|K(t)|\leq K(0)$, and $|K_{N}(t)|\leq K(0)$ yields 
\begin{align}
    |d_{N}(t)|\leq \epsilon_{N}(T), \quad |((K_{N}+K)*d_{N})(t)|\leq 2tK(0)\epsilon_{N}(T), \quad |(K_{N}*K*d_{N})(t)|\leq \frac{t^{2}}{2}K(0)^{2}\epsilon_{N}(T).
\end{align}
We then obtain
\begin{align}
    |K_{N}(t)-K(t)|\leq \epsilon_{N}(T)\Bigl[ 1+2t\rho_{*}K(0)+\frac{t^{2}\rho_{*}^{2}}{2}K(0)^{2}\Bigr].
\end{align}
Using the evenness of $K_{N}$ and $K$, we finally obtain 
\begin{align}
    \sup_{|t|\leq T}|K_{N}(t)-K(t)|\leq \epsilon_{N}(T)\Bigl[ 1+2T\rho_{*}K(0)+\frac{T^{2}\rho_{*}^{2}}{2}K(0)^{2}\Bigr]\rightarrow 0,
\end{align}
where we use Eqs.~\eqref{eq: cp conservative approximation} and~\eqref{eq: cp refinement conditions}.
Together with $K_N(0)=K(0)$ and
Eq.~\eqref{eq: cp coefficient preservation}, this proves
Proposition~\ref{prop: coefficient preserving approximation}.


\section{Details of the entropic bound and multi-bath heat-current bound}

In this section, we generalize the entropic bound to multiple baths
and time-dependent couplings. We use the same thermal linear
Gaussian embeddings as in Sec.~\ref{sec:Markov_embedding}, applying
the construction separately to each bath.

\subsection{Generalized Langevin equation for multiple baths and time-dependent couplings}
\label{appendix: GLE}

In this section, we consider generalized Langevin equation with multiple baths and time-dependent couplings. We label each bath by $n$ and denote the corresponding coupling and bath inverse temperature as $g_{t}^{n}$ and $\beta_{n}$, respectively. 

For each $n$, we consider an auxiliary mode $A_{n}$ with its Hamiltonian given by  
\begin{align}
    H_{A_n}+H_{\rm int}^{g_t^n}
    =\frac{1}{2}(\bm y_{n}-g_t^n\bm c_n X)^T(\bm y_n -g_t^n \bm c_n X).
    \label{eq: sup time dependent whitened Hamiltonian}
\end{align}
By introducing $\bm{z}_{n}:=\bm{y}_{n}-g_{t}^{n}\bm{c}_{n}X$, the auxiliary equation becomes
\begin{align}
    \dot{\bm z}_n
    =A_n \bm z_n 
    -\bm c_n \left(g_t^n\frac{P}{M}+\dot g_t^n X\right)
    +\sqrt{\frac{2D_n }{\beta_{n}}}\,\bm\xi_n (t).
    \label{eq: sup time dependent z dynamics}
\end{align}
Eliminating the auxiliary variables gives the generalized Langevin equation
\begin{align}
    \dot X
    &=
    \frac{P}{M},
    \nonumber\\
    \dot P
    &=
    -\partial_XV_S^{\lambda_t}(X)
    +\sum_n g_t^n
    \left[
        \eta_n(t)
        -\int_0^t ds\,K_n(t-s)
        \left(g_s^n\frac{P(s)}{M}+\dot g_s^nX(s)\right)
    \right],
    \label{eq: GLE_general_nonstationary}
\end{align}
where
\begin{align}
    \mathbb{E}[\eta_{n}(t)\eta_{m}(s)]=\delta_{n,m}\frac{1}{\beta_{n}}K_{n}(t-s).
\end{align}

The bath force $\mathcal{F}^{\rm GLE}$ can be defined for the $n$-th
non-Markovian bath in a similar manner as
\begin{align}
    \mathcal{F}^{\rm GLE}_{n}(t)
    &:=
    \eta_n(t)
    -
    \int_0^t ds\,K_n(t-s)
    \left(
        g_s^n\frac{P(s)}{M}
        +
        \dot g_s^n X(s)
    \right)  =\bm c_n^T\bm z_n(t) . 
    \label{eq: sup GLE bath force multibath}
\end{align}
We then define its conditional mean for a given instantaneous system
state as
\begin{align}
    F_{t,n}^{\rm NM}(X,P)
    &:=
    \mathbb E\left[
    \left.
    \mathcal{F}^{\rm GLE}_{n}(t)
    \right|
    X(t)=X,\,
    P(t)=P
    \right]=\mathbb E\left[
    \left.
    \bm c_n^T\bm z_n(t)
    \right|
    X(t)=X,\,
    P(t)=P
    \right].
    \label{eq: sup FNM multibath}
\end{align}
The reduced system dynamics is similarly generalized to
\begin{align}
    \partial_t f_t^S(X,P)
    =
    \{H_S^{\lambda_t},f_t^S\}
    -
    \sum_{n}g_{t}^{n}\partial_P
    \left[
        F_{t,n}^{\rm NM}(X,P)f_t^S(X,P)
    \right].
    \label{eq:reduced-FP-memory-force-multibath}
\end{align}

\subsection{Work and heat current}

We first give the expression of work in the case of time-dependent
couplings $g_t^n$. The work rate is defined as 
\begin{align}
    \dot W
    ={}&
    \left\langle
        \dot\lambda_t\partial_{\lambda_t}H_S^{\lambda_t}
    \right\rangle
    -\sum_n\dot g_t^n
    \left\langle
        X(t)
        \left[
            \eta_n(t)
            -\int_0^t ds\,K_n(t-s)
            \left(g_s^n\frac{P(s)}{M}+\dot g_s^nX(s)\right)
        \right]
    \right\rangle.
    \label{eq: sup Work}
\end{align}
Note that this expression can be obtained from the time-derivative of the total Hamiltonian as
\begin{align}
    \dot W
    =
    \left\langle
        \dot\lambda_t\partial_{\lambda_t}H_S^{\lambda_t}
    \right\rangle
    +\sum_n
    \left\langle
        \dot g_t^n\partial_{g_t^n}H_{\rm int}^{g_t^n}
    \right\rangle.
\end{align}

For multiple heat baths, we decompose the system heat current
$\dot Q_{\rm sys}$ into contributions from each bath labeled by $n$,
which we denote as $\dot Q_{\rm sys}^n$.
We start from the definition of $\dot Q_{\rm sys}$ and use the
generalized Langevin equation~\eqref{eq: GLE_general_nonstationary}
to obtain
\begin{align}
    \dot Q_{\rm sys}
    &:=
    \frac{d}{dt}\langle H_S^{\lambda_t}\rangle-\dot W
    \nonumber\\
    &=
    \sum_n
    \left\langle
        \left(g_t^n\frac{P(t)}{M}+\dot g_t^nX(t)\right)
        \left[
            \eta_n(t)
            -\int_0^t ds\,K_n(t-s)
            \left(g_s^n\frac{P(s)}{M}+\dot g_s^nX(s)\right)
        \right]
    \right\rangle.
    \label{eq: sup Q expression}
\end{align}
From this expression, we decompose the heat into contributions
from the bath labeled by $n$ as
$\dot Q_{\rm sys}=\sum_n\dot Q_{\rm sys}^n$, where
$\dot Q_{\rm sys}^n$ is defined as
\begin{align}
    \dot Q_{\rm sys}^n
    =
    \left\langle
        \left(g_t^n\frac{P(t)}{M}+\dot g_t^nX(t)\right)
        \left[
            \eta_n(t)
            -\int_0^t ds\,K_n(t-s)
            \left(g_s^n\frac{P(s)}{M}+\dot g_s^nX(s)\right)
        \right]
    \right\rangle.
    \label{eq: sup Qa expression}
\end{align}

\subsection{Hierarchy relation}
For multiple baths, the conditional thermal state is defined as
\begin{align}
    \pi^{A|S}_{\bm{g}_{t}}:=\prod_{n}\pi^{A_{n}|S}_{g_{t}^{n}} = \prod_{n}\frac{ \exp( - \beta_{n} (H_{A_{n}}+H_{\rm int}^{g_{t}^{n}}))}{Z_{A_{n}}}.
\end{align}
We define the entropy productions as
\begin{align}
    \Sigma:=\Delta S_{S}-\sum_{n}\beta_{n}Q^{n}_{\rm sys}, \quad \Sigma_{\rm emb}:=\Delta S_{SA}-\sum_{n}\beta_{n}Q^{n}_{\rm emb},
\end{align}
where $Q^{n}_{\rm emb}$ follows the standard definition. Under the initial condition $f^{SA}_{0}=f^{S}_{0}\pi^{A|S}_{\bm{g}_{0}}$, we obtain 
\begin{align}
    \Sigma = \Sigma_{\rm emb}+\Sigma_{\rm cor}, \quad \Sigma_{\rm cor}:=D(f^{SA}_{\tau}||f^{S}_{\tau}\pi^{A|S}_{\bm{g}_{\tau}})\geq 0,
\end{align}
which follows from an argument similar to the one in Sec.~\ref{sec:Gaussian_Hierarchy}.  

\subsection{\label{appendix: proof multiple bath}Thermodynamic trade-off relation for multiple baths}

Using Eq.~\eqref{eq:reduced-FP-memory-force-multibath}, the heat current is expressed as
\begin{align}
    \dot{Q}_{\rm sys}^{n}=\int dXdP \Bigl( g_{t}^{n}\frac{P}{M}+\dot{g}_{t}^{n}X\Bigr) F_{t,n}^{\rm NM}(X,P)f_{t}^{S}(X,P). 
\end{align}
Using this expression, we can apply an argument similar to that in Sec.~\ref{appendix:kernel_only_entropic_bound} and obtain
\begin{align}
    \Bigl(\int^{\tau}_{0}dt \sum_{n}|\dot{Q}_{\rm sys}^{n}|\Bigr)^{2} 
    &\leq \Theta_{\rm m} \Sigma, \label{eq: sup multibath CD tradeoff}
\end{align}
where 
\begin{align}
    \Theta_{\rm m}&:=\sum_{n}\frac{\mathcal{S}_{K_{n}}}{\beta_{n}}\int^{\tau}_{0}dt\int dXdP \Bigl( \frac{g_{t}^{n}P}{M}+\dot{g}_{t}^{n}X\Bigr)^{2}f_{t}^{S}(X,P),
\end{align}
and
\begin{align}
    \mathcal{S}_{K_{n}}
    =\sup_{\omega\in\mathbb R}\frac{|\widehat K_{n}(i\omega)|^2}
    {\operatorname{Re}\widehat K_{n}(i\omega)}.
\end{align}

\subsection{Derivation of power-efficiency trade-off relation}

Finally, we derive the power-efficiency trade-off relation presented in the main text by following Ref.~\cite{PhysRevLett.117.190601}: 
   \begin{align}
    \mathcal{P} \leq 
    \beta_{C} \bar{\Theta}_{\rm m} \eta (\eta_{\rm Car}-\eta). \label{eq: sup power-efficiency trade-off}
\end{align}

We denote by $Q^{H}_{\rm sys}$ and $Q^{C}_{\rm sys}$ the heat from the hot and cold baths. 
When the engine cycle is stationary, i.e., $f_{0}^{SA}=f_{\tau}^{SA}$ and $\bm{g}_{\tau}=\bm{g}_{0}$, and satisfies the condition under which it operates as a heat engine, i.e., $W_{\rm ext}\geq 0, Q^{H}_{\rm sys}\geq 0, Q^{C}_{\rm sys}\leq 0$, the power and efficiency of the heat engine are defined as $\mathcal{P}:=W_{\rm ext}/\tau$ and $\eta:=W_{\rm ext}/Q^{H}_{\rm sys}$. Here, $W_{\rm ext}:=-W=Q^{H}_{\rm sys}+Q^{C}_{\rm sys}$ is the extractable work and $\tau$ is the duration of one engine cycle. 

From Eq.~\eqref{eq: sup multibath CD tradeoff} and $|Q_{\rm sys}^{n}|\leq \int^{\tau}_{0}dt |\dot{Q}^{n}_{\rm sys}|$, we obtain 
\begin{align}
    (Q^{H}_{\rm sys})^{2}\leq(|Q^{H}_{\rm sys}|+|Q^{C}_{\rm sys}|)^{2}\leq \Theta_{\rm m} \Sigma . \label{eq: sup PE derivation1}
\end{align}
Combining with the relation 
\begin{align}
    \Sigma =-\beta_{H}Q^{H}_{\rm sys}-\beta_{C}Q^{C}_{\rm sys}= \beta_{C}Q^{H}_{\rm sys}(\eta_{\rm Car}-\eta),
\end{align}
we obtain
\begin{align}
   (Q_{\rm sys}^{H})^{2} \leq \Theta_{\rm m} \beta_{C}(\eta_{\rm Car}-\eta) Q_{\rm sys}^{H}.
\end{align}
We further use $\mathcal{P}=W_{\rm ext}/\tau$, $\bar{\Theta}_{\rm m}=\Theta_{\rm m}/\tau$, and $\eta=W_{\rm ext}/Q_{\rm sys}^{H}$, and obtain Eq.~\eqref{eq: sup power-efficiency trade-off}.

\section{Overdamped regime}

In this section, we consider the overdamped generalized Langevin dynamics~\eqref{eq: sup overdamped GLE} and derive the hierarchy relation for entropy production. Because Eq.~\eqref{eq: sup overdamped GLE} is obtained by taking the small-mass limit of the generalized Langevin equation, the relation~\eqref{eq: sup EP decomp} is still valid. In the following, we explicitly show this relation. Moreover, we derive an extended hierarchy relation:
\begin{align}
    \tilde{\Sigma}_{\rm tp}
    \leq
    \tilde{\Sigma}_{S}
    \leq
    \Sigma_{\rm emb}
    =
    \Sigma-\Sigma_{\rm cor}
    \leq
    \Sigma-\Sigma_{F^{\rm NM}}
    \leq
    \Sigma
    =
    \Sigma_{\rm M}+\Sigma_{\rm NM}
    \leq
    \Sigma_{\rm M},
    \label{eq: sup EP hierarchy overdamped}
\end{align}
where $\Sigma_{F^{\rm NM}}$ is the embedding-independent lower bound
on $\Sigma_{\rm cor}$ defined in Eq.~\eqref{eq: sup overdamped Sigma FNM}, and  $\tilde{\Sigma}_{S}$ is the partial entropy production of the system defined in Eq.~\eqref{eq: sup partial EP S}. Finally, we give details of the thermodynamic trade-off relations presented in the main text.

\subsection{Overdamped generalized Langevin equation}

In the following, we take the small-mass limit of the generalized Langevin equation to obtain the overdamped generalized Langevin equation
\begin{align}
    \frac{1}{\mu}\dot{X}(t) &= -\partial_{X}V_{S}^{\lambda_{t}}  - \int^{t}_{0}ds K^{\rm od}(t-s)\dot{X}(s) + \eta^{\rm od}(t) + \frac{1}{\mu} \xi(t) , \label{eq: sup overdamped GLE}
\end{align}
where $\mu$ is the mobility, $\xi(t)$ and $\eta^{\rm od}(t)$ are uncorrelated Gaussian white and colored noises, respectively satisfying
\begin{align}
    \mathbb{E}[\xi(t)]=\mathbb{E}[\eta^{\rm od}(t)]=0, \quad \mathbb{E}[\xi(t)\xi(s)]=\frac{2\mu}{\beta}\delta(t-s), \quad \mathbb{E}[\eta^{\rm od}(t)\eta^{\rm od}(s)]=\frac{1}{\beta}K^{\rm od}(|t-s|), \label{eq: sup property Gaussian noise overdamped}
\end{align}
and we define the memory kernel $K^{\rm od}(t) =
\sum_{k}a_{k}e^{-\Gamma_{k}|t|}$ with $a_{k}\geq 0$, corresponding to the overdamped Drude-Lorentz spectral density. 


For simplicity, let us start from the underdamped generalized Langevin equation with a single heat bath and constant coupling $g_{t}=1$:
\begin{align}
\frac{dX}{dt}&= \frac{P}{M} \nonumber \\
    \frac{dP}{dt}&= -\partial_{X}V_{S}^{\lambda_{t}} - \int^{t}_{0}ds K(t-s)\frac{P(s)}{M} + \eta(t). \label{eq: sup underdamped GLE 2}
\end{align}
We first take the following form of the memory kernel $K(t)$ and the noise $\eta(t)$: 
\begin{align}
    K(t-s)&=K^{\rm od}(t-s)+\frac{2}{\mu}  \delta(t-s), \nonumber \\
    \eta(t)&=\eta^{\rm od}(t)+\frac{1}{\mu} \xi(t).
\end{align}
Then, the generalized Langevin equation~\eqref{eq: sup underdamped GLE 2} takes the form 
\begin{align}
     \frac{dP}{dt}&= -\partial_{X}V_{S}^{\lambda_{t}} - \frac{1}{\mu} \dot{X}(t) - \int^{t}_{0}ds K^{\rm od}(t-s)\dot{X}(s) + \eta^{\rm od}(t) + \frac{1}{\mu} \xi(t). \label{eq: sup underdamped GLE Markov Non}
\end{align}
We now formally take the small-mass limit $M\rightarrow 0$ to obtain~\cite{nguyen2018small} 
\begin{align}
    0 = -\partial_{X}V_{S}^{\lambda_{t}} - \frac{1}{\mu}\dot{X}(t) - \int^{t}_{0}ds K^{\rm od}(t-s)\dot{X}(s) + \eta^{\rm od}(t) + \frac{1}{\mu} \xi(t) ,
\end{align}
which gives Eq.~\eqref{eq: sup overdamped GLE}. 

Next, we consider a different route to obtain Eq.~\eqref{eq: sup overdamped GLE} via Markovian embedding.
We start from the generalized Langevin equation with underdamped Brownian oscillator spectral density. 
The Langevin equation of the joint system reads
\begin{align}
    \frac{dX}{dt}&=\frac{P}{M}, \nonumber\\
    \frac{dP}{dt}&=-\partial_{X}V_{S}^{\lambda_{t}}+\sum_k c_k\left(x_k-\frac{c_k}{\kappa_{k}}X\right) - \frac{1}{\mu}\dot{X}+\frac{1}{\mu}\xi , \nonumber\\
    \frac{1}{\mu_{k}}\frac{dx_k}{dt}&=- \kappa_{k}\left(x_k-\frac{c_k}{\kappa_{k}}X\right)+ \frac{1}{\mu_{k}} \xi_{k}, \label{eq: sup underdamped LE 2}
\end{align}
where $\mathbb{E}[\xi_{k}(t)\xi_{l}(s)]=(2\mu_{k}/\beta)\delta(t-s)\delta_{k,l}$. 

We now take the small-mass limit of Eq.~\eqref{eq: sup underdamped LE 2} and obtain coupled overdamped Langevin dynamics for $(X,\bm{x})$:
\begin{align}
    \frac{1}{\mu}\frac{dX}{dt}&=-\partial_{X}V_{S}^{\lambda_{t}}+\sum_k c_k\left(x_k-\frac{c_k}{\kappa_{k}}X\right) +\frac{1}{\mu}\xi , \nonumber\\
    \frac{1}{\mu_{k}}\frac{dx_k}{dt}&=- \kappa_{k}\left(x_k-\frac{c_k}{\kappa_{k}}X\right)+ \frac{1}{\mu_{k}} \xi_{k}. \label{eq: sup overdamped Markov embedding}
\end{align}
We formally solve Eq.~\eqref{eq: sup overdamped Markov embedding} and obtain
\begin{align}
    x_{k}(t)-\frac{c_{k}}{\kappa_{k}}X(t)=e^{-\mu_{k}\kappa_{k}t} \Bigl(  x_{k}(0)-\frac{c_{k}}{\kappa_{k}}X(0)\Bigr)  + \int^{t}_{0}ds e^{-\mu_{k}\kappa_{k}(t-s)} \Bigl( \xi_{k}(s)-\frac{c_{k}}{\kappa_{k}}\dot{X}(s) \Bigr),
\end{align}
which gives
\begin{align}
    \sum_{k}c_{k}\Bigl(  x_{k}(t)-\frac{c_{k}}{\kappa_{k}}X(t)\Bigr) = - \int^{t}_{0}ds K^{\rm od}(t-s)\dot{X}(s) + \eta^{\rm od}(t), \label{eq: sup underdamped LE 4}
\end{align}
where 
\begin{align}
    \eta^{\rm od}(t):=\sum_{k}c_{k}e^{-\mu_{k}\kappa_{k}t}\Bigl(  x_{k}(0)-\frac{c_{k}}{\kappa_{k}}X(0)\Bigr)+\int^{t}_{0}ds \sum_{k}c_{k}e^{-\mu_{k}\kappa_{k}(t-s)}\xi_{k}(s), \label{eq: sup zeta Markov def}
\end{align}
and we identified the parameters as $a_{k}=c_{k}^{2}/\kappa_{k}$ and $\Gamma_{k}=\mu_{k}\kappa_{k}$. 
We assume the initial state of the form 
\begin{align}
    f^{SA}_{0}(X,\bm{x})=f_{0}^{S}(X)\pi^{A|S}(\bm{x}|X), \label{eq: sup overdamped initial cond}
\end{align}
where $\pi^{A|S}(\bm{x}|X)$ is the auxiliary thermal state conditioned on $X$, given by
\begin{align}
    \pi^{A|S}(\bm{x}|X):= \frac{e^{-\beta(H_{\rm int}+V_{A}) }}{Z_{A}}, \label{eq: sup overdamped initial cond2}
\end{align}
with $Z_{A}=\int d\bm{x}\exp(-\beta V_{A})$ and the potentials read
\begin{align}
    H_{\rm int} &:= -\sum_{k}c_{k}x_{k}X + \sum_{k}\frac{c_{k}^{2}}{2\kappa_{k}}X^{2}, \quad 
    V_{A} :=\sum_{k}\frac{\kappa_{k}}{2}x_{k}^{2} .
\end{align}
Note that under the initial condition~\eqref{eq: sup overdamped initial cond}, the noise $\eta^{\rm od}(t)$~\eqref{eq: sup zeta Markov def} satisfies the property given by Eq.~\eqref{eq: sup property Gaussian noise overdamped}. 

Finally, substituting Eq.~\eqref{eq: sup underdamped LE 4} into Eq.~\eqref{eq: sup overdamped Markov embedding} yields the overdamped generalized Langevin equation~\eqref{eq: sup overdamped GLE}.

\subsection{Fokker-Planck equation and entropy production for embedded system}

From the Langevin equation~\eqref{eq: sup overdamped Markov embedding} for the joint system $SA$, the Fokker-Planck equation for the Markovian embedded model reads
\begin{align}
    \partial_{t}f^{SA}_{t}(X,\bm{x})=-\partial_{X}(\nu_{X}f_{t}^{SA}) -\sum_{k}\partial_{x_{k}}(\nu_{x_{k}}f_{t}^{SA}), \label{eq: sup overdamped FP}
\end{align}
where the mean local velocities are
\begin{align}
    \nu_{X}=\mu (-\partial_{X}H_{SA}^{\rm tot} -\frac{1}{\beta}\partial_{X}\ln f_{t}^{SA}), \quad \nu_{x_{k}}=\mu_{k}(- \partial_{x_{k}}H_{SA}^{\rm tot} - \frac{1}{\beta}\partial_{x_{k}}\ln f_{t}^{SA}). 
\end{align}
Here, 
\begin{align}
    H^{\rm tot}_{SA}&=V_{S}^{\lambda_{t}}+H_{\rm int} +V_{A} \nonumber \\
    &=V_{S}^{\lambda_{t}}(X)+\sum_{k}\frac{\kappa_{k}}{2}\Bigl( x_{k}-\frac{c_{k}}{\kappa_{k}}X\Bigr)^{2}
\end{align}
is the potential of the joint system. 
The steady state of Eq.~\eqref{eq: sup overdamped FP} is given by
\begin{align}
    \pi^{SA}_{\lambda_{t}}(X,\bm{x})=\frac{e^{-\beta H_{SA}^{\rm tot}(X,\bm{x})}}{Z}.
\end{align}

For the joint system $SA$, the standard Markovian entropy production rate is defined as
\begin{align}
    \dot{\Sigma}_{\rm emb}&:=\dot{S}_{SA}-\beta \dot{Q}_{\rm emb} \nonumber \\
    &= \frac{\beta}{\mu}\int dXd\bm{x} \nu_{X}^{2}f_{t}^{SA} + \sum_{k}\frac{\beta}{\mu_{k}}\int dXd\bm{x} \nu_{x_{k}}^{2}f_{t}^{SA} \nonumber \\
    &= \dot{\tilde{\Sigma}}_{S} + \dot{\tilde{\Sigma}}_{A} \geq 0, \label{eq: EP def overdamped}
\end{align}
where the heat is defined as
\begin{align}
    \dot{Q}_{\rm emb}&:= \int dXd\bm{x} H_{SA}^{\rm tot}\partial_{t}f_{t}^{SA}, \label{eq: sup overdamped def heat aux}
\end{align}
and $\dot{S}_{SA}$ is the Shannon entropy change of the joint system $SA$. We also introduce the partial entropy production of the subsystems (see, for example Ref.~\cite{PhysRevResearch.3.043093}):
\begin{align}
    \tilde{\Sigma}_{S}&:=\int^{\tau}_{0}dt \frac{\beta}{\mu}\int dXd\bm{x} \nu_{X}^{2}f_{t}^{SA} \geq 0, \label{eq: sup partial EP S} \\
    \tilde{\Sigma}_{A}&:=\int^{\tau}_{0}dt\sum_{k}\frac{\beta}{\mu_{k}}\int dXd\bm{x} \nu_{x_{k}}^{2}f_{t}^{SA} \geq 0. \label{eq: sup partial EP A}
\end{align}
From Eq.~\eqref{eq: EP def overdamped} and the nonnegativity of the partial entropy productions, we have
\begin{align}
    \tilde{\Sigma}_{S} \leq \Sigma_{\rm emb},
\end{align}
which shows the second inequality of Eq.~\eqref{eq: sup EP hierarchy overdamped}.

\subsection{Hierarchy of entropy production}

\subsubsection{\texorpdfstring{$\Sigma_{\rm emb}\leq \Sigma$}{Sigma\_emb leq Sigma}}
Now, we define the system heat flux through the change in the system energy (note that we take constant coupling $g_{t}=1$ for simplicity): 
\begin{align}
    \dot{Q}_{\rm sys}&:= \int dXd\bm{x} \ V_{S}^{\lambda_{t}}(X)\partial_{t}f^{SA}_{t}, \label{eq: sup overdamped def heat sys}
\end{align}
and define the non-Markovian entropy production as
\begin{align}
    \Sigma = \Delta S_{S}-\beta \int^{\tau}_{0}dt \dot{Q}_{\rm sys}.
\end{align}

By comparing Eqs.~\eqref{eq: sup overdamped def heat aux} and \eqref{eq: sup overdamped def heat sys}, the relation between $\dot{Q}_{\rm emb}$ and $\dot{Q}_{\rm sys}$ reads
\begin{align}
    \dot{Q}_{\rm sys}&=\dot{Q}_{\rm emb} - \int dXd\bm{x} (H_{\rm int}+V_{A})\partial_{t}f_{t}^{SA} \nonumber \\
    &= \dot{Q}_{\rm emb} +\frac{1}{\beta} \int dXd\bm{x} (\ln \pi^{A|S}) \partial_{t}f_{t}^{SA} .
\end{align}
Then, we find that
\begin{align}
    \Sigma &= \Delta S_{S} -\beta \int^{\tau}_{0}dt \dot{Q}_{\rm emb} - \int dXd\bm{x} f_{\tau}^{SA} \ln \pi^{A|S} + \int dXd\bm{x} f_{0}^{SA}\ln\pi^{A|S} \nonumber \\
    &= \Sigma_{\rm emb} + D(f_{\tau}^{SA}\Vert f_{\tau}^{S}\pi^{A|S})-D(f_{0}^{SA}\Vert f_{0}^{S}\pi^{A|S}) \nonumber \\
    &= \Sigma_{\rm emb} + D(f_{\tau}^{SA}\Vert f_{\tau}^{S}\pi^{A|S})\nonumber \\
    &\geq \Sigma_{\rm emb}, \label{eq: EP relation between aux and nonMarkov}
\end{align}
where the third line follows from the initial condition~\eqref{eq: sup overdamped initial cond}. This expression~\eqref{eq: EP relation between aux and nonMarkov} shows the entropy-production decomposition used in Eq.~\eqref{eq: sup EP hierarchy overdamped}.

\subsubsection{Embedding-independent lower bound on the correlation contribution}

We next derive an embedding-independent lower bound on
$\Sigma_{\rm cor}$ in the overdamped regime. We define the force
exerted by the auxiliary modes on the system as
\begin{align}
    \mathcal{F}_{A}(X,\bm{x})
    &:=
    -\partial_X H_{\rm int}
    =
    \sum_k c_k
    \left(
        x_k-\frac{c_k}{\kappa_k}X
    \right).
    \label{eq: sup overdamped auxiliary force}
\end{align}
Its conditional mean is given by
\begin{align}
    F_t^{\rm NM}(X)
    &:=
    \int d\bm{x}\,
    \mathcal{F}_{A}(X,\bm{x})
    f_t(\bm{x}|X)
    =
    \sum_k c_k v_k^A(X)
    =
    \frac{1}{\mu}\nu_t^{\rm NM}(X).
    \label{eq: sup overdamped FNM}
\end{align}

This quantity can also be defined directly from the overdamped
generalized Langevin equation. Introducing the stochastic force
associated with the non-Markovian bath as
\begin{align}
    \mathcal{F}^{\rm GLE,od}_{t}
    &:=
    -\int_0^t ds\,
    K^{\rm od}(t-s)\dot X(s)
    +\eta_t^{\rm od},
    \label{eq: sup overdamped GLE force}
\end{align}
we have
\begin{align}
    F_t^{\rm NM}(X)
    =
    \mathbb E\left[
    \left.
    \mathcal{F}^{\rm GLE,od}_{t}
    \right|
    X(t)=X
    \right].
    \label{eq: sup overdamped FNM GLE}
\end{align}
Indeed, Eq.~\eqref{eq: sup underdamped LE 4} shows that
$\mathcal{F}^{\rm GLE,od}_{t}
=\mathcal{F}_{A}(X(t),\bm{x}(t))$.
Therefore, $F_t^{\rm NM}(X)$ is determined by the reduced overdamped
generalized Langevin dynamics and is independent of the specific
Markovian embedding.

Using the above definitions, we can prove the following lower bound by following an argument similar to that given in Sec.~\ref{sec: embedding-independent lower bound}:
\begin{align}
    \Sigma_{\rm cor}
    \geq
    \Sigma_{F^{\rm NM}}
    &:=
    \frac{\beta}{2K^{\rm od}(0)}
    \int dX\,
    \left[F_\tau^{\rm NM}(X)\right]^2
    f_\tau^S(X),
    \label{eq: sup overdamped Sigma FNM}
\end{align}
for $K^{\rm od}(0)>0$.

\subsubsection{\texorpdfstring{$\Sigma_{\rm M}$ and $\Sigma_{\rm NM}$}{Sigma\_M and Sigma\_NM}}
We further decompose $\Sigma$ into two contributions. First, let us  write down the continuity equation for the marginal distribution of the system $f_{t}^{S}(X)$ by integrating out auxiliary degrees of freedom from Eq.~\eqref{eq: sup overdamped FP} as
\begin{align}
    \partial_{t}f_{t}^{S}(X)&=-\partial_{X}J_{t}^{S}(X),
\end{align}
with current
\begin{align}
    J_{t}^{S}(X):= [\nu_{t}^{\rm M}(X) + \nu_{t}^{\rm NM}(X) ]  f_{t}^{S}(X), \label{eq: sup cont}
\end{align}
where 
\begin{align}
    \nu_{t}^{\rm M}(X):=-\mu V_{\lambda_{t}}'(X)-\frac{\mu}{\beta}\partial_{X}\ln f_{t}^{S}(X).
\end{align}
is the Markovian mean local velocity and
\begin{align}
    \nu^{\rm NM}_{t}(X)&:=  \mu \sum_{k}c_{k}v_{k}^{A}(X) = -\mu \int d\bm{x} (\partial_{X}H_{\rm int})f_{t}(\bm{x}|X),
\end{align}
is the non-Markovian correction term, with
\begin{align}
        v^{A}_{k}(X)&:= \int d\bm{x} \delta x_{k} f_{t}(\bm{x}|X), \quad \delta x_{k}:= x_{k}-\frac{c_{k}}{\kappa_{k}}X ,
\end{align}
and $f_{t}(\bm{x}|X)=f_{t}^{SA}(X,\bm{x})/f_{t}^{S}(X)$.

Then, we decompose the entropy production into
\begin{align}
    \Sigma = \frac{\beta}{\mu} \int^{\tau}_{0}dt \int dX \nu_{t}^{\rm M}(X)[\nu_{t}^{\rm M}(X)+\nu_{t}^{\rm NM}(X)]f_{t}^{S}(X) =\Sigma_{\rm M}+\Sigma_{\rm NM},
\end{align}
with
\begin{align}
    \Sigma_{\rm M}&:= \frac{\beta}{\mu} \int^{\tau}_{0}dt \int dX [\nu_{t}^{\rm M}(X)]^{2}f_{t}^{S}(X)
\end{align}
is the usual expression of the entropy production for Markovian overdamped systems, and
\begin{align}
    \Sigma_{\rm NM}:= \frac{\beta}{\mu} \int^{\tau}_{0}dt \int dX \nu_{t}^{\rm M}(X)\nu_{t}^{\rm NM}(X)f_{t}^{S}(X)
\end{align}
quantifies the non-Markovian correction. 

In what follows, we show the hierarchy relation
\begin{align}
    \Sigma_{\rm NM}\leq 0, \quad \Sigma_{\rm M}\geq \Sigma. \label{eq: sup Sigma M relation}
\end{align}
To show Eq.~\eqref{eq: sup Sigma M relation}, we first use the relation
\begin{align}
    \tilde{\Sigma}_{S}&= \frac{\beta}{\mu}  \int^{\tau}_{0}dt \int dX d\bm{x} \Bigl[ \nu_{t}^{\rm M}(X) - \frac{\mu}{\beta}\partial_{X}\ln \frac{f_{t}(\bm{x}|X)}{\pi^{A|S}(\bm{x}|X)}\Bigr]^{2} f_{t}(\bm{x}|X)f_{t}^{S}(X) \nonumber \\
    &= \Sigma_{\rm M} + 2\Sigma_{\rm NM} + I_{X},
\end{align}
where
\begin{align}
    I_{X}:= \frac{\mu}{\beta} \int^{\tau}_{0}dt \int dX d\bm{x} \Bigl[ \partial_{X}\ln \frac{f_{t}(\bm{x}|X)}{\pi^{A|S}(\bm{x}|X)}\Bigr]^{2} f_{t}^{SA}(X,\bm{x})\geq 0,
\end{align}
and we use the relation
\begin{align}
    \int d\bm{x} f_{t}(\bm{x}|X) \frac{\mu}{\beta} \partial_{X}\ln \frac{f_{t}(\bm{x}|X)}{\pi^{A|S}(\bm{x}|X)} = -\nu_{t}^{\rm NM}(X).
\end{align}
Finally, we combine with Eq.~\eqref{eq: EP relation between aux and nonMarkov} and obtain
\begin{align}
    \Sigma &= \Sigma_{\rm emb}+D(f_{\tau}^{SA}\Vert f^{S}_{\tau}\pi^{A|S}) \nonumber \\
    &=\tilde{\Sigma}_{S}+\tilde{\Sigma}_{A} +D(f_{\tau}^{SA}\Vert f^{S}_{\tau}\pi^{A|S}) \nonumber \\
    &= \tilde{\Sigma}_{A} +D(f_{\tau}^{SA}\Vert f^{S}_{\tau}\pi^{A|S}) + \Sigma + \Sigma_{\rm NM}+I_{X},
\end{align}
which leads to
\begin{align}
    \Sigma_{\rm NM}= - \tilde{\Sigma}_{A} -D(f_{\tau}^{SA}\Vert f^{S}_{\tau}\pi^{A|S}) - I_{X} \leq 0.
\end{align}

\subsubsection{\texorpdfstring{$\tilde{\Sigma}_{\rm tp}$}{tilde{Sigma}\_tp}}
We introduce a transport cost associated with the continuity equation~\eqref{eq: sup cont} as
\begin{align}
    \tilde{\Sigma}_{\rm tp}:= \frac{\beta}{\mu}\int^{\tau}_{0}dt \int dX \frac{[J_{t}^{S}(X)]^{2}}{f_{t}^{S}(X)}\geq 0. 
\end{align}
This quantity is related to the entropy production as
\begin{align}
     \tilde{\Sigma}_{\rm tp}&= \tilde{\Sigma}_{S} + \tilde{I}_{X}-I_{X} \leq \tilde{\Sigma}_{S},
\end{align}
where
\begin{align}
    \tilde I_X&:=\frac{\beta}{\mu}\int_0^\tau d t\int d X\, [\nu_{t}^{\rm NM}(X)]^{2} f_{t}^{S}(X),
\end{align}
which satisfies $I_{X}\geq \tilde{I}_{X}$. 

We also obtain a relation
\begin{align}
    \tilde{\Sigma}_{\rm tp} = \Sigma+\Sigma_{\rm NM} + \tilde{I}_{X} \leq \Sigma ,
\end{align}
where 
\begin{align}
    \Sigma_{\rm NM} + \tilde{I}_{X} = -\tilde{\Sigma}_{A}-D(f_{\tau}^{SA}\Vert f^{S}_{\tau}\pi^{A|S}) - (I_{X}-\tilde{I}_{X}) \leq 0.
\end{align}

\subsection{Thermodynamic trade-off relations in the overdamped regime}

\subsubsection{Thermodynamic speed limit}
We now derive the thermodynamic speed limit for overdamped generalized Langevin systems described by Eq.~\eqref{eq: sup overdamped GLE}:
\begin{align}
    \frac{\mathcal{W}_{2}(f^{S}_{0},f^{S}_{\tau})^{2}}{\tau}&\leq \frac{\mu}{\beta} \Sigma_{\rm emb}=\frac{\mu}{\beta}(\Sigma-\Sigma_{\rm cor}) \leq \frac{\mu}{\beta}(\Sigma-\Sigma_{F^{\rm NM}})\leq \frac{\mu}{\beta} \Sigma . \label{eq: sup TSL} 
\end{align}
To show Eq.~\eqref{eq: sup TSL}, we use the continuity equation~\eqref{eq: sup cont} and the Benamou-Brenier formula~\cite{benamou2000computational}
\begin{align}
    \frac{\mathcal{W}_{2}(f^{S}_{0},f^{S}_{\tau})^{2}}{\tau} & \leq \frac{\mu}{\beta} \tilde{\Sigma}_{\rm tp}  .\label{eq: sup TSL Nakazato}
\end{align}
We further combine Eq.~\eqref{eq: sup TSL Nakazato} with the hierarchy of the entropy production~\eqref{eq: sup EP hierarchy overdamped} and obtain Eq.~\eqref{eq: sup TSL}.

\subsubsection{TUR}

Let us discuss details of the TUR shown in the main text:
\begin{align}
     2\frac{ \left[\langle \mathcal{J}_{\tau}^{S}\rangle +\Delta \langle \mathcal{J}_{\tau}^{S} \rangle\right]^{2} }{\Var[\mathcal{J}_{\tau}^{S}]} \leq \Sigma_{\rm emb}=\Sigma-\Sigma_{\rm cor}\leq (\Sigma-\Sigma_{F^{\rm NM}}) \leq \Sigma, \label{eq: sup TUR}
\end{align}
where $\Delta:=\tau\partial_{\tau}-v\partial_{v}$.

Here, the explicit form of the current $\mathcal{J}_{\tau}^{S}$ that we consider reads:
\begin{align}
    \mathcal{J}_{\tau}^{S}=\frac{1}{\tau}\int^{\tau}_{0} dt \psi[X(t),\lambda_{t}^{v}]\circ \dot{X}(t),
\end{align}
where $\psi$ is an arbitrary function and $\circ$ denotes the Stratonovich product. By further using the overdamped generalized Langevin equation~\eqref{eq: sup overdamped GLE}, the current is expressed as 
\begin{align}
    \mathcal{J}_{\tau}^{S} &= \frac{1}{\tau}\int^{\tau}_{0} dt \psi[X(t),\lambda_{t}^{v}]\circ \xi_{t} \nonumber \\
    &+ \frac{\mu}{\tau}\int^{\tau}_{0} dt \psi[X(t),\lambda_{t}^{v}]\Bigl( - \partial_{X}V_{S}^{\lambda_{t}^{v}}(X(t))-\int^{t}_{0}ds K^{\rm od}(t-s)\dot{X}(s)+\eta^{\rm od}_{t}\Bigr).  \label{eq: sup current GLE}
\end{align}
From this expression, the current $\mathcal{J}^{S}_{\tau}$ can be obtained by solving Eq.~\eqref{eq: sup overdamped GLE} and is independent of the specific form of the Markovian embedding. 

We now express Eq.~\eqref{eq: sup current GLE} using the Markovian embedding, which reads
\begin{align}
    \mathcal{J}_{\tau}^{S}
    &= \frac{1}{\tau}\int^{\tau}_{0} dt \psi[X(t),\lambda_{t}^{v}]\circ \xi_{t} \nonumber \\
    &+ \frac{1}{\tau}\int^{\tau}_{0} dt \psi[X(t),\lambda_{t}^{v}]\Bigl[ -\mu\partial_{X}V_{S}^{\lambda_{t}^{v}}(X(t)) + \mu\sum_{k} c_{k}\Bigl( x_{k}(t)-\frac{c_{k}}{\kappa_{k}}X(t)\Bigr) \Bigr] \nonumber \\
    &= \frac{1}{\tau}\int^{\tau}_{0} dt  \Bigl( \psi[X(t),\lambda_{t}^{v}]\circ \xi_{t} - \mu \psi[X(t),\lambda_{t}^{v}]\partial_{X}H^{\rm tot}_{SA}(X,\bm{x})\Bigr). \label{eq: sup current emb}
\end{align}
Therefore, using Eq.~\eqref{eq: sup current emb} for the Markovian embedded overdamped system allows direct application of Ref.~\cite{PhysRevLett.125.260604} to obtain 
\begin{align}
    2\frac{ \left[\langle \mathcal{J}_{\tau}^{S}\rangle +\Delta \langle \mathcal{J}_{\tau}^{S} \rangle\right]^{2} }{\Var[\mathcal{J_{\tau}^{S}}]} \leq \Sigma_{\rm emb}. \label{eq: sup TUR Koyuk Seifert}
\end{align}
By further applying the hierarchy of entropy production~\eqref{eq: sup EP hierarchy overdamped}, we obtain the non-Markovian TUR~\eqref{eq: sup TUR}. Note that the left-hand side of Eq.~\eqref{eq: sup TUR}, as well as $\Sigma$ and $\Sigma_{F^{\rm NM}}$, is independent of the specific choice of embedding.

\subsubsection{Entropic bound}
We consider a current of the form
\begin{align}
    \mathcal J_t^{\rm od}[\mathcal{O}]:=-\int dX\,\mathcal{O}_t(X)\,\partial_X J_t^S(X).
\end{align}
By using the Cauchy-Schwarz inequality, the entropic bound is obtained as:
\begin{align}
    \left(\int_0^\tau dt\,\bigl|\mathcal J_t^{\rm od}[\mathcal{O}]\bigr|\right)^2
    &\leq
    \frac{\mu}{\beta}\,\tilde\Sigma_{\rm tp}
    \int_0^\tau dt\int dX\,\bigl[\partial_X \mathcal{O}_t(X)\bigr]^2 f_t^S(X) \nonumber \\
    &\leq
    \frac{\mu}{\beta}\,\Sigma
    \int_0^\tau dt\int dX\,\bigl[\partial_X \mathcal{O}_t(X)\bigr]^2 f_t^S(X).
\end{align}

\section{Hierarchy of entropy production for the general bath case in the classical regime}

For completeness, we give the classical fixed-coupling specialization of the general hierarchy derived in the End Matter. 

We now show the hierarchy relation 
\begin{align}
    \Sigma = \Sigma_{\rm emb}+D(f^{SA}_{\tau} \Vert  f^{S}_{\tau}  \pi^{A|S}) \label{eq: sup general bath dec}
\end{align}
for the general bath case, where the total Hamiltonian is given by
\begin{align}
    H^{\rm tot}_{SA}(\bm{z}_{S},\bm{z}_{A})= H^{S}_{\lambda_{t}}(\bm{z}_{S})+H_{\rm int}(\bm{z}_{S},\bm{z}_{A}) + H_{A}(\bm{z}_{A}).
\end{align}
Here, we only allow $H^{S}_{\lambda_{t}}$ to depend on time via a control parameter $\lambda_{t}$.

Let us start from the definition of the heat current
\begin{align}
    \dot{Q}_{\rm sys}:=\frac{d}{dt}\langle \mathcal{E}_{S}\rangle - \dot{W},
\end{align}
where the internal energy reads
\begin{align}
    \mathcal{E}_{S}&:= H_{S}^{*} + \beta \partial_{\beta}H_{S}^{*} 
\end{align}
and the Hamiltonian of mean force is given as
\begin{align}
    H_{S}^{*}&:=H_{S}^{\lambda_{t}} -\beta^{-1} \ln \Bigl[ \int d\bm{z}_{A} e^{-\beta (H_{\rm int} + H_{A})}/Z_{A}\Bigr] \nonumber \\
    &= H_{SA}^{\rm tot} + \beta^{-1}\ln \pi^{A|S} +\beta^{-1}\ln Z_{A} , 
\end{align}
with the conditional thermal state of the auxiliary modes
\begin{align}
    \pi^{A|S}:=\frac{e^{-\beta(H_{\rm int}+H_{A})}}{\int d\bm{z}_{A}e^{-\beta(H_{\rm int}+H_{A})}}.
\end{align}

Now, using the relation 
\begin{align}
    \dot{Q}_{\rm emb}=\frac{d}{dt}\langle H_{SA}^{\rm tot}\rangle-\langle\dot{H}_{SA}^{\rm tot}\rangle = \frac{d}{dt}\langle H_{SA}^{\rm tot}\rangle - \dot{W},
\end{align}
we obtain
\begin{align}
    \dot{Q}_{\rm sys}&= \dot{Q}_{\rm emb} - \frac{d}{dt}\langle H^{\rm tot}_{SA}\rangle +\frac{d}{dt}\langle H^{*}_{S}\rangle + \frac{d}{dt}\langle \beta\partial_{\beta}H_{S}^{*}\rangle  \nonumber \\
    &=\dot{Q}_{\rm emb}+\frac{1}{\beta}\frac{d}{dt}\langle \ln \pi^{A|S}\rangle +\frac{d}{dt}\langle \beta\partial_{\beta}H_{S}^{*}\rangle,
\end{align}
by noting that $\ln Z_{A}$ is time-independent. 

Using the definitions of the entropy production
\begin{align}
    \Sigma&:=\Delta S^{*}_{S}-\beta \int^{\tau}_{0}dt \dot{Q}_{\rm sys} , \label{eq: sup EP strong coupling} \\
    \Delta S^{*}_{S} &:= \Delta S_{S} +\beta^{2}\Delta \langle \partial_{\beta}H_{S}^{*}\rangle ,\nonumber \\
    \Sigma_{\rm emb}&:=\Delta S_{SA} - \beta \int^{\tau}_{0}dt \dot{Q}_{\rm emb}, 
\end{align}
we find that
\begin{align}
    \Sigma 
    &= \Delta S^{*}_{S} - \beta \int^{\tau}_{0}dt \dot{Q}_{\rm emb} - \Delta \langle \ln \pi^{A|S}\rangle - \beta^{2} \Delta \langle \partial_{\beta}H_{S}^{*}\rangle   \nonumber \\
    &= \Sigma_{\rm emb} + \Delta S_{S} - \Delta S_{SA} - \Delta \langle \ln \pi^{A|S}\rangle \nonumber \\
    &=\Sigma_{\rm emb} + D(f^{SA}_{\tau} \Vert  f^{S}_{\tau}  \pi^{A|S})-D(f_{0}^{SA}\Vert f^{S}_{0}\pi^{A|S}), 
\end{align}
where the last term vanishes under the initial condition $f_{0}^{SA}=f_{0}^{S}\pi^{A|S}$, i.e., $D(f_{0}^{SA}\Vert f^{S}_{0}\pi^{A|S})=0$, which completes the derivation of Eq.~\eqref{eq: sup general bath dec}.

Let us now introduce the reduced thermal state of the system $\pi^{*}_{\lambda_{t}}$ as
\begin{align}
    \pi^{*}_{\lambda_{t}}(\bm{z}_{S}):= \int d\bm{z}_{A} \frac{e^{-\beta H_{SA}^{\rm tot}}}{Z_{SA}}, 
\end{align}
and express thermodynamic quantities using $\pi^{*}_{\lambda_{t}}$.

First, we express $\dot{Q}_{\rm sys}$ using $\pi^{*}_{\lambda_{t}}$. Note that using the relation
\begin{align}
    \mathcal{E}_{S}
    &= H_{S}^{\lambda_{t}} - \partial_{\beta}\ln \Bigl[ \int d\bm{z}_{A} e^{-\beta(H_{\rm int}+H_{A})}/Z_{A}\Bigr],
\end{align}
the heat current is expressed as
\begin{align}
    \dot{Q}_{\rm sys}&= \int d\bm{z}_{S}d\bm{z}_{A} \mathcal{E}_{S}\partial_{t}f^{SA}_{t} + \langle \dot{\mathcal{E}}_{S}\rangle - \dot{W} \nonumber \\
    &= \int d\bm{z}_{S}d\bm{z}_{A} \mathcal{E}_{S}\partial_{t}f^{SA}_{t}.
\end{align}
We further use the relation
\begin{align}
    H_{S}^{*}&= -\beta^{-1}\ln \Bigl[ \pi^{*}_{\lambda_{t}}Z_{SA}/Z_{A} \Bigr] ,\nonumber \\
    \mathcal{E}_{S} &:= -\partial_{\beta} \ln \pi^{*}_{\lambda_{t}} - \partial_{\beta}\ln \frac{Z_{SA}}{Z_{A}},
\end{align}
and obtain
\begin{align}
    \dot{Q}_{\rm sys}
    &= -\int d\bm{z}_{S}d\bm{z}_{A}  \Bigl( \partial_{\beta}\ln \pi^{*}_{\lambda_{t}} + \partial_{\beta}\ln \frac{Z_{SA}}{Z_{A}}\Bigr)\partial_{t}f^{SA}_{t} \nonumber \\
    &= -\int d\bm{z}_{S}  \partial_{\beta}\ln \pi^{*}_{\lambda_{t}}(\bm{z}_{S})\partial_{t}f^{S}_{t}(\bm{z}_{S}). \label{eq: sup qstar}
\end{align}

Next, let us express $\dot{S}^{*}_{S}$ using $\pi^{*}_{\lambda_{t}}$. We first note that $\partial_{\beta}H_{S}^{*}$ does not depend on $\lambda_{t}$. Therefore, we have
\begin{align}
    \dot{S}^{*}_{S} &= \dot{S}_{S} + \beta^{2}\int d\bm{z}_{S}d\bm{z}_{A}   \partial_{\beta}H_{S}^{*}   \partial_{t}f_{t}^{SA} \nonumber \\
    &= \dot{S}_{S} - \beta^{2}\int d\bm{z}_{S}d\bm{z}_{A}   \partial_{\beta}\Bigl( \frac{1}{\beta}\ln \pi^{*}_{\lambda_{t}} + \frac{1}{\beta}\ln \frac{Z_{SA}}{Z_{A}} \Bigr)   \partial_{t}f_{t}^{SA} \nonumber \\
    &=\dot{S}_{S}- \beta^{2}\int d\bm{z}_{S} \partial_{\beta}\Bigl( \frac{1}{\beta}\ln \pi^{*}_{\lambda_{t}}(\bm{z}_{S})\Bigr)\partial_{t}f_{t}^{S}(\bm{z}_{S}). \label{eq: sup Sstar}
\end{align}
Combining the expressions~\eqref{eq: sup qstar} and~\eqref{eq: sup Sstar}, we find that
\begin{align}
    \Sigma = - \int^{\tau}_{0}dt \int d\bm{z}_{S} \partial_{t}f^{S}_{t}(\bm{z}_{S}) \ln \frac{f_{t}^{S}(\bm{z}_{S})}{\pi^{*}_{\lambda_{t}}(\bm{z}_{S})}, \label{eq: sup EP strong coupling 2}
\end{align}
and the entropy production Eq.~\eqref{eq: sup EP strong coupling 2} depends only on the marginal probability distribution $f^{S}_{t}(\bm{z}_{S})$ and the reduced thermal state of the system $\pi^{*}_{\lambda_{t}}(\bm{z}_{S})$. Therefore, as long as the reduced system dynamics of $f^{S}_{t}$ and the thermal state $\pi^{*}_{\lambda_{t}}$ reproduce those of the original non-Markovian dynamics, the entropy production Eq.~\eqref{eq: sup EP strong coupling 2} does not depend on the specific choice of Markovian embedding.  

\section{Alternative definition of entropy production for product initial states}

In this section, we consider an alternative definition of the entropy production for a product initial state with non-vanishing interaction Hamiltonian for the quantum case and derive the hierarchy of entropy production.

\subsection{Unitary system-bath model}
The total Hamiltonian of the system-bath model reads
\begin{align}
    H^{\rm tot}_{SB}=H_{S}^{\lambda_{t}} +H_{\rm int}^{g_{t}} +H_{B},
\end{align}
and the time-evolution is described by the von Neumann equation:
\begin{align}
    \partial_{t}\rho^{SB}_{t}= -\frac{i}{\hbar}[H^{\rm tot}_{SB},\rho^{SB}_{t}]. 
\end{align}
We assume that the initial state is given by the product state of the form
\begin{align}
    \rho^{SB}_{0}=\rho^{S}_{0}\otimes \pi^{B},
\end{align}
where $\pi^{B}:=\exp(-\beta H_{B})/Z_{B}$ with $Z_{B}:=\Tr[\exp(-\beta H_{B})]$. The state at time $t=\tau$ is given by $\rho^{SB}_{\tau}=U_{SB}\rho^{SB}_{0} U_{SB}^{\dagger}$, where $U_{SB}=T\exp(-\frac{i}{\hbar}\int^{\tau}_{0}dt H_{SB}^{\rm tot})$ is the unitary time-evolution operator from $t=0$ to $t=\tau$. 

We define bath heat current as the change in the bath energy:
\begin{align}
    \dot{Q}_{\rm bath}&= -\frac{d}{dt}\langle H_{B}\rangle \nonumber \\
    &= \frac{i}{\hbar}\Tr[ H_{B}[H_{SB}^{\rm tot},\rho^{SB}_{t}]] \nonumber \\ 
    &= \Tr[ (H_{S}^{\lambda_{t}}+H_{\rm int}^{g_{t}})\partial_{t}\rho_{t}^{SB}]. \label{eq: sup Qb quantum}
\end{align}
Then, the entropy production is defined as~\cite{Esposito_2010}
\begin{align}
    \Sigma &=\Delta S_{S} -\beta Q_{\rm bath} \nonumber \\
    &=D(\rho^{SB}_{\tau}\Vert \rho^{S}_{\tau}\otimes \pi^{B})\geq 0, \label{eq: EP def second}
\end{align}
where $Q_{\rm bath}:=\int^{\tau}_{0}dt \dot{Q}_{\rm bath}$ and $\Delta S_{S}$ is the von Neumann entropy change of the system.

\subsection{\label{sec: sup: global GKSL}Markovian embedded model using a global GKSL master equation}

The total Hamiltonian of the Markovian embedded model reads
\begin{align}
    H_{SA}^{\rm tot}=H_{S}^{\lambda_{t}}+H_{\rm int}^{g_{t}}+H_{A},
\end{align}
and the initial state is given by
\begin{align}
    \rho^{SA}_{0}=\rho^{S}_{0}\otimes \pi^{A}. \label{eq: sup initial condition}
\end{align}
The time-evolution equation is assumed to take the (global) Gorini-Kossakowski-Sudarshan-Lindblad (GKSL) form
\begin{align}
    \partial_{t}\rho^{SA}_{t}=\mathcal{L}_{t}[\rho^{SA}_{t}]=-\frac{i}{\hbar}[H_{SA}^{\rm tot},\rho^{SA}_{t}] + \mathcal{D}_{t}^{SA}[\rho^{SA}_{t}],
\end{align}
with the steady state given by the thermal state $\pi^{SA}_{\lambda_{t},g_{t}}:=\exp(-\beta H_{SA}^{\rm tot})/Z_{SA}$, i.e., $\mathcal{L}_{t}[\pi^{SA}_{\lambda_{t},g_{t}}]=0$. 
Note that this type of global master equation is typically used in the reaction-coordinate approach~\cite{Nazir2018,strasberg2016nonequilibrium}. 
Then, the Markovian entropy production defined for the embedded model reads
\begin{align}
    \dot{\Sigma}_{\rm emb} = \dot{S}_{SA} - \beta \dot{Q}_{\rm emb} \geq 0,
\end{align}
where 
\begin{align}
    \dot{Q}_{\rm emb}= \Tr[ H_{SA}^{\rm tot}\partial_{t}\rho^{SA}_{t}].
\end{align}

We now define the bath heat current for the Markovian embedded model as 
\begin{align}
    \dot{Q}_{\rm bath}=\Tr[(H_{S}^{\lambda_{t}}+H_{\rm int}^{g_{t}})\partial_{t}\rho^{SA}_{t}]   . \label{eq: sup heat bath quantum Markov}
\end{align}
It should be noted that whether Eq.~\eqref{eq: sup heat bath quantum Markov} reproduces the bath heat current defined for the original non-Markovian dynamics~\eqref{eq: sup Qb quantum} requires careful analysis because of the dependence on the interaction Hamiltonian, and we leave this for future study. 

Using this expression~\eqref{eq: sup heat bath quantum Markov}, we obtain the relation between $\dot{Q}_{\rm bath}$ and $\dot{Q}_{\rm emb}$: 
\begin{align}
    \dot{Q}_{\rm bath}=\dot{Q}_{\rm emb}-\Tr[H_{A}\partial_{t}\rho^{SA}_{t}]. \label{eq: equivalence EP assumption two}
\end{align}
We now use Eq.~\eqref{eq: equivalence EP assumption two} and relate $\Sigma$ and $\Sigma_{\rm emb}$ as
\begin{align}
    \Sigma &= \Delta S_{S}-\beta Q_{\rm bath} \nonumber \\
    &= \Delta S_{S} - \beta Q_{\rm emb} - \int_{0}^{\tau} dt\Tr[\partial_{t}\rho^{SA}_{t} \ln \pi^{A}] \nonumber \\
    &= \Sigma_{\rm emb} + \Delta S_{S}-\Delta S_{SA} - \Tr[ (\rho^{SA}_{\tau}-\rho^{SA}_{0})\ln \pi^{A}] \nonumber \\
    &= \Sigma_{\rm emb} + D(\rho^{SA}_{\tau}\Vert \rho^{S}_{\tau}\otimes \pi^{A}) -D(\rho^{SA}_{0}\Vert \rho^{S}_{0}\otimes \pi^{A}), 
\end{align}
where the last term vanishes using the initial condition~\eqref{eq: sup initial condition}, i.e., $D(\rho^{SA}_{0}\Vert \rho^{S}_{0}\otimes \pi^{A})=0$. Hence, we finally show 
\begin{align}
    \Sigma = \Sigma_{\rm emb} + D(\rho^{SA}_{\tau}\Vert \rho^{S}_{\tau}\otimes \pi^{A})\geq \Sigma_{\rm emb}. \label{eq: sup decomp product}
\end{align}
Note that Eq.~\eqref{eq: sup decomp product} has been obtained in the case of the mesoscopic-leads approach in Ref.~\cite{PhysRevE.110.014125}. 
We also note that Eq.~\eqref{eq: sup decomp product} applies to the classical case with initial factorized states, i.e., $f_{0}^{SA}=f_{0}^{S}\pi^{A}$. 

\subsection{Markovian embedded model using a local GKSL master equation}

Suppose we instead consider the following embedding based on a local GKSL master equation.
The time-evolution equation is assumed to take the (local) Gorini-Kossakowski-Sudarshan-Lindblad (GKSL) form
\begin{align}
    \partial_{t}\rho^{SA}_{t}=\mathcal{L}^{\rm loc}_{t}[\rho^{SA}_{t}]=-\frac{i}{\hbar}[H_{SA}^{\rm tot},\rho^{SA}_{t}] + \mathcal{D}_{t}^{A}[\rho^{SA}_{t}], \label{eq: sup local ME}
\end{align}
where the jump operators only act on the auxiliary modes
\begin{align}
    \mathcal{D}_{t}^{A}[\rho_{t}^{SA}]:=\sum_{k}\gamma_{k}\Bigl( L^{A}_{k} \rho_{t}^{SA} (L^{A}_{k})^{\dagger} - \frac{1}{2}\{ (L^{A}_{k})^{\dagger}L_{k}^{A}, \rho_{t}^{SA} \} \Bigr),
\end{align}
and the dissipator satisfies $\mathcal{D}_{t}^{A}[\pi^{A}]=0$. Note that this master equation is the usual setup for the pseudo-modes~\cite{PhysRevA.55.2290,PhysRevLett.120.030402} (here, we assume real parameters) and mesoscopic-leads approach~\cite{PhysRevX.10.031040}. 

In this case, the entropy production for the embedded system $SA$ is defined as
\begin{align}
    \dot{\Sigma}_{\rm emb}^{\rm loc}:=\dot{S}_{SA} - \beta \dot{Q}_{\rm emb}^{\rm loc} \geq 0,
\end{align}
where 
\begin{align}
    \dot{Q}_{\rm emb}^{\rm loc}:= \Tr[ H_{A}\mathcal{D}_{t}^{A}[\rho_{t}^{SA}]]. 
\end{align}

By using the expression~\eqref{eq: sup heat bath quantum Markov} and the master equation~\eqref{eq: sup local ME}, the bath heat current reads
\begin{align}
    \dot{Q}_{\rm bath} &=\Tr[(H_{S}^{\lambda_{t}}+H_{\rm int}^{g_{t}})\partial_{t}\rho^{SA}_{t}]   \nonumber \\
    &= \frac{i}{\hbar}\Tr[ H_{A} [H_{SA}^{\rm tot},\rho_{t}^{SA}] ] + \Tr[ H_{\rm int}^{g_{t}}\mathcal{D}_{t}^{A}[\rho^{SA}_{t}]],
\end{align}
and therefore we obtain the relation
\begin{align}
    \dot{Q}_{\rm bath}=\dot{Q}_{\rm emb}^{\rm loc}+\beta^{-1}\Tr[ \ln \pi^{A} \partial_{t}\rho^{SA}_{t}]+\Tr[H_{\rm int}^{g_{t}}\mathcal{D}_{t}^{A}[\rho^{SA}_{t}]]. \label{eq: heat relation pseudomode}
\end{align}
Because of the last term $\Tr[H_{\rm int}^{g_{t}}\mathcal{D}_{t}^{A}[\rho^{SA}_{t}]]$, a hierarchy between $\Sigma$ and $\Sigma_{\rm emb}^{\rm loc}$ similar to Eq.~\eqref{eq: sup decomp product} does not hold in general. 

Note that in Ref.~\cite{PhysRevE.110.014125}, the authors use a different definition of the entropy production for the embedded system $SA$ as (here, we do not consider chemical potentials of the bath)
\begin{align}
    \dot{\tilde{\Sigma}}_{\rm emb}^{\rm loc}:=\dot{S}_{SA} - \beta \dot{\tilde{Q}}_{\rm emb} , \label{eq: sup mesoscopic-leads EP def}
\end{align}
with
\begin{align}
    \dot{\tilde{Q}}_{\rm emb}:= \Tr[H_{SA}^{\rm tot}\mathcal{D}_{t}^{A}[\rho^{SA}_{t}]],
\end{align}
and obtain a relation similar to Eq.~\eqref{eq: sup decomp product} as
\begin{align}
    \Sigma \geq \tilde{\Sigma}_{\rm emb}^{\rm loc}, \label{eq: sup decomp product 2}
\end{align}
but it should be noted that $\dot{\tilde{\Sigma}}_{\rm emb}^{\rm loc}$ (and also $\tilde{\Sigma}_{\rm emb}^{\rm loc}$) can take negative values, and only when the steady state of Eq.~\eqref{eq: sup local ME} is given by the thermal state of the joint system, i.e., $\mathcal{L}^{\rm loc}_{t}[\pi^{SA}_{\lambda_{t},g_{t}}]=0$, it is nonnegative. This condition is not always satisfied for the local master equation~\eqref{eq: sup local ME}, but as discussed in Ref.~\cite{PhysRevE.110.014125}, such a condition for the mesoscopic leads approach holds in the high-temperature limit with a large number of auxiliary modes.

\section{Details of the numerical calculation in the main text}

Here, we give details of the parameters used for the figure presented in the main text. 

\subsection{Figure 1 (c) and 2}

We consider an underdamped system coupled to a single
overdamped auxiliary coordinate $y$.
Using the notation in Sec.~\ref{sec:Markov_embedding_Langevin}, we take
\begin{align}
    H_{S}
    &=\frac{P^{2}}{2M}+\frac{\lambda X^{2}}{2}, \quad     H_{A}+H_{\rm int}
    =\frac{1}{2}(y-cX)^{2},
    \label{eq: sup case2 single H}
\end{align}
and choose $n=1$, $J=0$, and $D=\epsilon$. The memory kernel reads $K(t)=ae^{-\epsilon|t|}$ with $a=c^{2}$. 

For Fig.~1 (c) and Fig.~2, we use dimensionless units
and set $\beta=M=\lambda=c=1$ and
$\tau=2\pi/\sqrt{2}$.
The initial system distribution is assumed to be Gaussian, with
$\langle X\rangle_0=0$, $\langle P\rangle_0=1$,
$\operatorname{Var}[X]_0=\operatorname{Var}[P]_0=1$,
and $\operatorname{Cov}(X,P)_0=0$. The initial state of the auxiliary coordinate is given by the conditional thermal equilibrium state $\pi^{A|S}$. 

\subsection{Figure 3 and 4}
In Fig.~3 and Fig.~4, we assume a single exponential for the memory kernel $K^{\rm od}(t)=a_{1}e^{-\Gamma_{1}|t|}$. The parameters are $\beta=\mu=\kappa=1$, $\tau=2.5$, $v=5$, $\kappa_{1}=3$, and $\mu_{1}=5$. Using the coupling $c$ to the auxiliary mode, we have $a_{1}=c^{2}/\kappa_{1}$ and $\Gamma_{1}=\mu_{1}\kappa_{1}$. The initial distribution of the system is assumed to be Gaussian, with $\langle X\rangle =0$ and $\text{Var}[X]=0.3$.

\end{document}